\documentclass[aps,prb,twocolumn,superscriptaddress]{revtex4-1}

\usepackage{graphicx}
\usepackage{amsmath}
\usepackage{bbold}
\usepackage[T1]{fontenc}
\usepackage{hyperref}
\hypersetup{backref,pdfpagemode=FullScreen,colorlinks=true,allcolors=blue}

\newcommand*{\covdev}{\text{\dh}}

\begin{document}

\title{Quantum mechanics in an evolving Hilbert space}

\author{Emilio Artacho}
\affiliation{CIC Nanogune and DIPC, Tolosa Hiribidea 76, 
             20018 San Sebastian, Spain}
\affiliation{Ikerbasque, Basque Foundation for Science, 48011 Bilbao, Spain}
\affiliation{Theory of Condensed Matter,
             Cavendish Laboratory, University of Cambridge, 
             J. J. Thomson Ave, Cambridge CB3 0HE, United Kingdom}
             
\author{David D. O'Regan}
\affiliation{School of Physics, CRANN and AMBER, Trinity College Dublin, 
             Dublin 2, Ireland}
\affiliation{Theory of Condensed Matter,
             Cavendish Laboratory, University of Cambridge, 
             J. J. Thomson Ave, Cambridge CB3 0HE, United Kingdom}

\date{\today}

\begin{abstract}
  Many basis sets for electronic structure calculations evolve 
with varying external parameters, such as moving atoms in dynamic 
simulations, giving rise to extra derivative terms in the dynamical equations. 
  Here we revisit these derivatives in the context of differential geometry,
thereby obtaining a more transparent formalisation, and a geometrical 
perspective for better understanding the resulting equations.
  The effect of the evolution of the basis set within the spanned Hilbert
space separates explicitly from the effect of the turning of the space 
itself when moving in parameter space, as the tangent space turns when moving 
in a curved space.
  New insights are obtained using familiar 
concepts in that context such as the Riemann curvature.
  The differential geometry is not strictly that for curved spaces as in general 
relativity, a more adequate mathematical framework being provided by fibre 
bundles.
  The language used here, however, will be restricted to tensors and basic 
quantum mechanics.
  The local gauge implied by a smoothly varying basis set readily connects 
with Berry's formalism for geometric phases.
  Generalised expressions for the Berry connection and curvature are 
obtained for a parameter-dependent occupied Hilbert space spanned 
by non-orthogonal Wannier functions.
  The formalism is applicable to basis sets made of atomic-like 
orbitals and also more adaptative moving basis functions (such as 
in methods using Wannier functions as intermediate or support bases),
but should also apply to other situations in which non-orthogonal
functions or related projectors should arise.
  The formalism is applied to the time-dependent quantum evolution 
of electrons for moving atoms.
  The geometric insights provided here allow us to propose new 
finite-differences time integrators, and also better understand 
those already proposed. 
\end{abstract}

\pacs{}

\maketitle

\section{Introduction}

  Many electronic-structure methods use basis sets made of states
that move or change with atomic positions. 
  A very popular subset of these methods (most quantum-chemical\cite{Gaussian,
Turbomol,Gamess,ADF} and a significant fraction of solid-state 
methods\cite{Siesta,CP2K,Crystal,OpenMX,FHI-AIMS}) use atomic-like basis 
functions, 
composed
of the product of a radial function and a spherical harmonic, normally 
centered around atoms.
  In other cases, the localised basis is obtained dynamically, using a finer
auxiliary basis.\cite{Onetep,Conquest,BigDFT}
  The detail of the kind of functions is not important in this work,
what matters here is that such a basis is generally not orthonormal,
it spans a subspace of the Hilbert space (a finite basis is always used), 
and both the basis and the subspace change with the evolution of a set 
of external parameters such as the atomic positions.

  Non-orthogonal basis sets have been used since the early times of 
quantum mechanics, especially in the context of quantum chemistry.\cite{Lowdin1950} 
  The matrix representation of Schr\"odinger's equation (using Dirac notation)
$H | \psi \rangle = E | \psi \rangle$ in a basis $\{ | e_{\mu} \rangle , \mu = 1 \dots \cal{N} \} $
gives
\begin{equation*}  
\sum_{\nu} H_{\mu \nu} C_{\nu} = E \sum_{\nu} S_{\mu\nu} C_{\nu} \; ,
\end{equation*}
where 
\begin{equation}
\label{coeff-expansion}
| \psi \rangle = \sum_{\mu} | e_{\mu} \rangle \, C_{\mu} \,  ,
\end{equation}
$H_{\mu \nu}= \langle e_{\mu} | H | e_{\nu} \rangle$ and 
$S_{\mu \nu}= \langle e_{\mu} | e_{ \nu} \rangle$, the latter being the
overlap matrix.
 
  Similarly, there are electronic structure methods based on 
the integration of time-evolving quantum problems, most prominently
based on time-dependent density-functional 
theory.\cite{RungeGross,Yabana,Tsolakidis2002,Octopus,ONTDDFT,Bowler} 
  The time-dependent Kohn-Sham equation arising is 
analogous to the time dependent Schr\"odinger equation 
$H | \psi \rangle = i \,  \partial_t | \psi \rangle$ (using 
$\hbar = m_e = e = 1$), which, for a non-orthogonal 
basis set, becomes
\begin{equation*}
\sum_{\nu} H_{\mu \nu}  C_{\nu} = 
i \, \sum_{\nu} S_{\mu \nu} \, \partial_t C_{\nu} \; ,
\end{equation*}
in a situation in which the basis set is fixed.\cite{Tsolakidis2002}
  If the basis set moves in time (e.g., related to nuclear motion)
the equation becomes rather
\begin{equation}
\label{old-td-eq}
\sum_{\nu} (H_{\mu \nu} - i \, D_{\mu \nu} ) C_{\nu} = 
i \, \sum_{\nu} S_{\mu \nu} \, \partial_t C_{\nu} \; ,
\end{equation}
in which the new terms $D_{\mu \nu}= \langle e_{\mu} | \partial_t | e_{\nu} \rangle
= \langle e_{\mu} | \partial_t  e_{\nu} \rangle$ appear related to the
basis set evolution,\cite{Todorov2001,Rudiger2002,Kaxiras2015} 
although at sufficiently low nuclear velocities these terms can be
neglected.\cite{Kaxiras}
  Similar objects to the $D_{\mu\nu}$ matrix in Eq.~\ref{old-td-eq} can also 
be found in time-dependent methods using localised molecular 
orbitals.\cite{WeitaoYang2010}
  Extra terms related to derivatives also appear when calculating
the forces on atoms for geometry relaxation, ab initio molecular
dynamics calculations, or Ehrenfest dynamics simulations.\cite{Todorov2001,Rudiger2002} 
   There are terms arising, called Pulay forces,\cite{Pulay1969} which again involve 
basis vector derivatives. 

  The matrix representation of the quantum formalism used above has its limitations,
however, and a more general formalisation was introduced\cite{Vanderbilt1984,
Ballantine1986, Artacho1991, Head-Gordon1993} based on
tensors, which offers a better suited and more flexible framework for non-orthogonal
basis sets, corresponding to a description of magnitudes in an Euclidean space with
oblique axes. 
  It allowed, for instance, the formalisation of second quantisation based on 
non-orthogonal bases,\cite{Artacho1991} which was then used to
formulate many-body theories using such bases,\cite{Head-Gordon1993,Holomorphic}
and to formulate corrective methods such as DFT+$U$ for the non-orthogonal 
case.\cite{ORegan2011,Palacios2014,Jacob2015}
  It also connected naturally with non-Hermitian representations proposed
earlier for the better exploitation of localisation.\cite{Weeks,Bullett,Anderson}

  In this paper we use 
concepts of differential geometry 
to extend that oblique-axis formalism to the calculation of derivatives 
when the basis and the Hilbert space it spans change with parameters 
such as atomic positions or time. 
  This offers insights into the geometric interpretation of the dynamical equations 
arising with moving basis sets.
  In particular, the affine connection defined for the changing basis allows the
proposal of optimised propagators for the numerical integration of quantum 
time-evolving problems. 
  It should be noted that there have been previous works on derivatives 
in the tensorial formalism for non-orthogonal bases in electronic 
structure,\cite{MHG-gradients} including relaxations with curvy steps.\cite{MHG-curvy}
  They were, however, always addressing 
derivatives of a scalar, the total energy, 
a special case which allows for circumvention of the key concepts in this work, 
the affine connection and the covariant derivative.
  An alternative way of using differential geometry in electronic 
structure was initially explored in 
Ref.~\onlinecite{ORegan}
when calculating derivatives with respect to the basis functions themselves,
instead of external parameters as in this work.
  This is beyond the scope of the present paper.
  
  The ideas in this paper should be useful for time-dependent 
(or parameter dependent) methods involving basis functions, 
auxiliary support functions, or any kind of states, which move
during simulations, including atomic-like basis orbitals, support functions  
or generalised Wannier functions in large scale electronic-structure methods, or
even the projectors for the core-electron description in projected-augmented-wave 
(PAW) methods.  
  The connection is also made with the Berry formalism 
of geometric phases.
  In Section II the general formalism is presented, while Section III shows its
application in several contexts.
  Many derivations have been pushed to appendices 
with a view to attaining
a more concise exposition of the relevant ideas in the main text, 
while preserving a reasonably self-contained paper.

\section{Formulation}

\subsection{Tensorial representations}
\label{SectForm}

  In this work we will use tensorial representations as used 
in Ref.~\onlinecite{Artacho1991}.
  Here are the essentials before we get into derivatives.
  Consider a basis consisting on a set of linearly-independent, non-orthogonal
states,
\begin{equation*}
\{ | e_{\mu} \rangle, \; \mu = 1 \dots \cal{N} \}
\end{equation*}
spanning a subspace $\Omega$ of the relevant Hilbert space $\cal{H}$
for our quantum problem.
  We will use here the tensorial notation for oblique 
angles.\cite{Vanderbilt1984, Ballantine1986, Artacho1991, Head-Gordon1993}
  For this, the dual basis\cite{Algebra,Artacho1991} is defined as the set of vectors 
$\{ | e^{\mu} \rangle, \; \mu = 1 \dots \cal{N} \}$ in the same space $\Omega$ that 
fulfil
\begin{equation*}
\langle e^{\mu} | e_{\nu} \rangle = \delta^{\mu}_{\phantom{e}\nu} = 
\langle e_{\nu} | e^{\mu} \rangle = \delta_{\nu}^{\phantom{e}\mu} \; ,
\end{equation*}
where $\delta^{\mu}_{\phantom{e}\nu}=\delta_{\nu}^{\phantom{e}\mu}=
\delta^{\mu}_{\nu}$ is Kronecker's delta.
  They also fulfil\cite{Artacho1991}
\begin{equation*}
\sum_{\mu} | e_{\mu} \rangle \langle e^{\mu} | = \sum_{\mu} | e^{\mu} \rangle \langle 
e_{\mu} | = P_{\Omega} \; ,
\end{equation*}
where $P_{\Omega}$ is the projector onto the $\Omega$ Hilbert space.
  The metric tensors are given by the overlap, 
$S_{\mu\nu} = \langle e_{\mu} | e_{\nu} \rangle$, and its upper-indices counterpart 
$S^{\mu\nu} = \langle e^{\mu} | e^{\nu} \rangle$, which is the inverse
(in the matrix sense) of the overlap matrix. 

  In this paper we will mostly (always, unless explicitly stated) use the
{\it natural representation} as defined in Ref.~\onlinecite{Artacho1991}.
  A state $| \psi \rangle \in \Omega$ is represented by the contravariant 
first-rank tensor 
\begin{equation*}
\psi^{\mu}= \langle e^{\mu} | \psi \rangle \; ,
\end{equation*}
which corresponds to the coefficients $C_{\mu}$ of the vector expansion
in Eq.~\ref{coeff-expansion}, since, 
\begin{equation*}
| \psi \rangle = P_{\Omega} | \psi \rangle
= \sum_{\mu} | e_{\mu} \rangle \langle e^{\mu} | \psi \rangle
= \sum_{\mu} | e_{\mu} \rangle \psi^{\mu}
= | e_{\mu} \rangle \psi^{\mu} \; .
\end{equation*}
  The last identity just reflects the fact that from now on we will use 
Einstein's convention for tensors,\cite{Einstein1916} by which repeated 
indices imply a sum.
  A bra $\langle \psi | \in \Omega^\dagger$ will be represented by the 
equivalent covariant tensor,
\begin{equation*}
\psi_{\mu} = \langle \psi | e_{\mu} \rangle \; ,
\end{equation*}
coming from 
\begin{equation*}
\langle \psi | = \langle \psi | e_{\mu} \rangle \langle e^{\mu} | = 
\psi_{\mu} \langle e^{\mu} |  \; .
\end{equation*}
  Note that the representation of the bra is not the complex conjugate of
the representation of the ket (see Appendix~\ref{NotationAppendix}).
  The covariant and contravariant character of a tensor relates to the way
they transform under basis change (see Appendix~\ref{BasisChangeAppendix}
for the definition).
  
  An operator acting in $\Omega$ is represented by the second-rank tensor given by
\begin{equation*}
H^{\mu}_{\phantom{e}\nu} = \langle e^{\mu} | H | e_{\nu} \rangle \; ,
\end{equation*}
since
\begin{equation*}
P_{\Omega} H P_{\Omega} = \left ( |e_{\mu} \rangle \langle e^{\mu} | \right ) \, H \, 
\left ( | e_{\nu} \rangle \langle e^{\nu} | \right ) = |e_{\mu} \rangle 
H^{\mu}_{\phantom{e}\nu} \langle e^{\nu} | \; .
\end{equation*}
  Schr\"odinger's equation $H | \psi \rangle = E | \psi \rangle$ in this representation 
then becomes
\begin{equation*}
H^{\mu}_{\phantom{e}\nu} \, \psi^{\nu} = E \, \psi^{\mu}  \; .
\end{equation*}

  It should be noted at this point that the formalism described here is equally 
valid for $| \psi \rangle$ being a single-particle or a many-particle state.
  It just requires the $| e_{\mu} \rangle$ basis states (and their duals) to 
represent the same number of particles as $|\psi \rangle$.
  One can also use this formalism for many-particle systems building
on single-particle non-orthogonal basis states by using 
non-orthogonal second quantization.\cite{Artacho1991,Head-Gordon1993}
  In some parts below we will refer to single-particle (mean-field-like)
situations since they are frequently found in different contexts, such as
Kohn-Sham density-functional theory,\cite{Kohn-Sham} but the present
formalism is not limited to such situations.

  The other representation to be considered in this work is the traditional 
quantum-chemical representation (henceforth called the {\it matrix representation}), 
which uses $\psi^{\mu}$ and $H_{\mu\nu}= \langle e_{\mu} | H | e_{\nu} \rangle$ 
for the representation of states and operators, respectively, the Schr\"odinger 
equation now reading
\begin{equation*}
H_{\mu\nu} \, \psi^{\nu} = E \, S_{\mu\nu} \, \psi^{\nu} \; ,
\end{equation*}
as detailed in Ref.~\onlinecite{Artacho1991}.

  An Hermitian operator in the natural representation would fulfil
\begin{equation*}
H^{\mu}_{\phantom{e}\nu} = \left ( H_{\nu}^{\phantom{e}\mu} \right )^*
= \left ( S_{\nu\lambda} H^{\lambda}_{\phantom{e}\sigma} S^{\sigma\mu} \right )^* \; ,
\end{equation*}
where $H_{\nu}^{\phantom{e}\mu} = \langle e_{\nu} | H | e^{\mu}\rangle$. 
  In the matrix representation hemiticity is reflected just by 
$H_{\mu\nu}= \left ( H_{\nu\mu} \right )^*$.
  Further details on the tensorial notation used in this work are found in 
Appendix~\ref{NotationAppendix}.

  All of these magnitudes are tensors in the sense that they represent abstract 
objects that are defined independently of the basis set.
  Tensor components transform in a well defined fashion when
changing the basis set.
  Transformations under basis change of the different tensors in this paper 
are discussed in Appendix~\ref{BasisChangeAppendix}.

\subsection{Parameter vector space}
\label{ParameterSpace}

  In this work we provide a comprehensive formalisation of derivatives of
the quantities just defined with respect to any parameters that the basis
may depend on, including both the basis change within $\Omega$ and
the evolution of $\Omega$ itself.
  Such parameters will normally be nuclear positions as in a molecular 
dynamics or Ehrenfest simulations, or time, as when following the
dynamics governed by time-dependent Schr\"odinger equation (or the
time-dependent Kohn-Sham equation in time-dependent density-functional
theory, or analogous mean-field-like equation).

  We will then consider those parameters as defining a vector space $\Theta$
of dimension $N$, spanned by the basis
\begin{equation*}
\{ \mathbf{u}_i, \; i = 1 \dots N \} \; ,
\end{equation*}
such that any vector $\mathbf{R} \in \Theta$ is expanded as
\begin{equation*}
{\mathbf R} = R^i {\mathbf u}_i \: .
\end{equation*} 
  Please keep in mind that these $R^i$ variables represent any parameters
that a particular quantum problem may depend on, not necessarily nuclear
positions (in the Applications section below there will be examples for 
nuclear positions, but also for time as a single parameter in a one-dimensional
$\Theta$).
  We keep the tensor notation for the vectors in $\Theta$ for convenience.
This allows for oblique angles in this space as well if ever wanted.
  We will always use Greek letters as indices for the quantum (electronic)
components and Latin for the components of vectors in parameter space.
  Our electronic basis set and space do then depend on $\mathbf{R}$, i.e.
$\Omega=\Omega(\mathbf{R})$, $| e_{\mu} \rangle = | e_{\mu} (\mathbf{R}) \rangle$,
and so will the projector $P_{\Omega}$ and all the tensors defined above. 

  It is important to note that, although we will exploit analogies with the 
differential geometry defined for curved spaces as in general relativity, 
the situation described here may be discussed  more formally using the 
language of fibre bundles,\cite{FibreBundle} with $\Theta$ as the base space, 
and $\Omega(\mathbf{R})$ as a fibre for each {\bf R}. 
  When moving in the base space, the Hilbert space associated to each fibre
turns within the ambient Hilbert space ${\cal H}$, very much as the tangent space 
would turn for a curved space.
  Although both the base space $\Theta$ and each fiber $\Omega(\mathbf{R})$ 
are each flat Euclidean spaces, the overall bundle is curved.

  The parameter-dependent Hilbert space $\Omega (\mathbf{R})$ and its turning
is also the basis of the Berry formalism of geometric phases in quantum 
mechanics.\cite{Berry-orig, Berry} 
  In this case the relevant $\Omega$ space would 
be the one associated with the ground state, or, in a mean-field-like 
setting (as in e.g. density-functional theory), the space spanned by
the occupied single-particle states (occupied space),
although also larger spaces are considered, e.g. for
metallic systems, or disentangling bands.\cite{Marzari1997}
  We relate this work to the Berry formalism below.
  Finally, the change of basis between fibres can be regarded as a gauge transformation
given in principle by any $\cal{N} \times \cal{N}$ invertible matrix of complex numbers,
i.e., belonging to the general linear group GL$(\cal{N},\mathbb{C})$.
  In this paper, however, we prefer to introduce the formalism in an accessible, 
self-contained manner, using basic quantum mechanics, tensors, and simple 
manipulations therein.

   In what follows, we will investigate the rate of change of quantum states, 
and operators acting upon them, with respect to the evolving basis vectors as we 
navigate the parameter vector space. 
   We will find that the components of such change due to space preserving and 
space non-preserving basis function evolution must be separately considered, 
as follows.

\subsection{Differential geometry}

\subsubsection{Covariant derivative}

  The derivative of the $\psi^{\mu}$ components of a quantum state $| \psi \rangle$
with respect to $R^i$ will be indicated by 
\begin{equation*}
\partial_i \psi^{\mu} = {\partial \psi^{\mu} \over \partial R^i} \; .
\end{equation*}
  It is easy to show (see Appendix~\ref{BasisChangeAppendix}) that
such a derivative does not transform as a tensor under basis change.
  Using the conventional nomenclature: it is a non-tensor.
  
  Let us then define the covariant derivative as one that transforms 
as a tensor, which we can easily do as follows:
\begin{equation}
\label{covdevdef}
\covdev_i \psi^{\mu} \equiv \langle e^{\mu} | P_{\Omega} \partial_i 
\left \{ P_{\Omega} | \psi \rangle \right \} =
\langle e^{\mu} | \partial_i \left \{ P_{\Omega} | \psi \rangle \right \} \; .
\end{equation}
  The projector directly acting on $| \psi \rangle$ might appear redundant
if starting with $| \psi \rangle \in \Omega$.
  It is not so, however, the important point being that the derivative is calculated
for the state being projected on the {\it varying} $\Omega$ space, 
and therefore, even though $| \psi \rangle$ and $P_{\Omega} |\psi \rangle$ 
are equal at {\bf R}, they are not necessarily equal at any nearby point, 
$\mathbf{R} + \mathrm{d} R^i \mathbf{u}_i$.
  Put another way, the covariant derivative must be applied before 
projection onto $\Omega$, since both $|\psi \rangle$ and $P_{\Omega}$ 
may evolve in time.

  This definition gives a well-behaved tensor since
the defined $\covdev_i \psi^{\mu}$ is the tensor representation of the vector
$P_{\Omega} \partial_i \left \{ P_{\Omega} | \psi \rangle \right \}  \in \Omega$
(see also Appendix~\ref{BasisChangeAppendix}).
  The justification for this particular definition will become clear throughout 
this section. 
  We can already point to the fact that for $| \psi \rangle \in \Omega$
such that neither $|\psi \rangle$ nor $\Omega$ change with {\bf R},
and for a basis set that does change, the proposed covariant derivative is
zero (while the usual derivative is not), thereby indicating that it is the 
intrinsic rate of change of the state what is being measured, excluding
the basis set change. 
  This point will be proven more generally below.
  It is analogous to the definition of covariant derivatives in gauge-dependent
theories, as the physical derivative independent of change of local gauge.

  The 
relationship between
$\covdev_i \psi^{\mu}$ and $\partial_i \psi^{\mu}$ is
obtained as follows:
\begin{align*}
\covdev_i \psi^{\mu} &= \langle e^{\mu} | \partial_i \left \{ P_{\Omega} | \psi \rangle \right \} =
\langle e^{\mu} | \partial_i \left \{  | e_{\nu} \rangle \langle e^{\nu} | \psi \rangle \right \} \\
&= \langle e^{\mu} | \partial_i \left \{  | e_{\nu} \rangle \psi^{\nu} \right \} 
= \langle e^{\mu} | \partial_i | e_{\nu} \rangle \psi^{\nu} + \partial_i \psi^{\mu} \; .
\end{align*}
  This gives us an alternative definition of the covariant derivative,
expressed in objects all defined within $\Omega$, namely,
\begin{equation}
\label{covdevdef2}
\covdev_i \psi^{\mu} = \partial_i \psi^{\mu} + D^{\mu}_{\phantom{i} \nu i} \psi^{\nu} \; ,
\end{equation}
where we have used the following definition
\begin{equation}
\label{christoffel}
D^{\mu}_{\phantom{i} \nu i}  \equiv  \langle e^{\mu} | \partial_i | e_{\nu} \rangle
= \langle e^{\mu} | \partial_i e_{\nu} \rangle \; .
\end{equation}
  The $i$ index is located after the index of the state being differentiated.
  This choice can be remembered by thinking of it as $\partial_i  | e_{\mu} \rangle = 
\partial | e_{\mu} \rangle / \partial R^i$.
  This second expression of the covariant derivative (Eq.~\ref{covdevdef2})
can be shown to give a tensor within this 
formalism (see Appendix~\ref{BasisChangeAppendix}).

\subsubsection{Affine connection}

  The quantity defined in Eq.~\ref{christoffel} is also a non-tensor and 
plays the role of the Christoffel symbols of the second kind in the Levi-Civita 
connection of conventional differential geometry.
  Keeping the nomenclature, our Christoffel symbols as defined in 
Eq.~\ref{christoffel} thus define the affine connection relevant to our problem.
  Remember, however, that this is not differential geometry for curved
spaces, where the tangent space at one point directly relates to the 
overall manifold, but 
rather
for a rotating Hilbert space $\Omega$ within ${\cal H}$ 
when moving in parameter space $\Theta$.
  In the former, there is one metric tensor at every point, while in the latter
there are still two metrics, one for $\Omega$ and one for $\Theta$.
  Hence, 
we do not establish a relationship
between the defined Christoffel symbols 
and the electronic metric and its derivatives.

  The defined Christoffel symbols 
exhibit
other expected properties.
  They give, for instance, the expansion coefficients in $\Omega$ of the 
derivative of a basis vector:
\begin{equation*}
P_{\Omega} \, \partial_i | e_{\nu} \rangle = | e_{\mu} \rangle \langle e^{\mu} 
| \partial_i | e_{\nu} \rangle = | e_{\mu} \rangle D^{\mu}_{\phantom{i} \nu i} \, ,
\end{equation*}
so that when moving from one point $\mathbf{R} \in \Theta$ to another
infinitesimally close to it $\mathbf{R} + \mathrm{d}\mathbf{R}$, the basis
vectors transform as
\begin{equation}
\label{BasisProp}
| e_{\nu} (\mathbf{R} + \mathrm{d}\mathbf{R}) \rangle =
| e_{\nu} (\mathbf{R}) \rangle + | e_{\mu} (\mathbf{R}) \rangle 
D^{\mu}_{\phantom{e}\nu i} \mathrm{d} R^i \; ,
\end{equation}
to linear order for infinitesimal $\mathrm{d} \mathbf{R}$.
  The second term of the right hand side accounts for the fact that
not only the space is turning, but the basis set itself is changing when
displacing in $\Theta$.

  The turning of the Hilbert space $\Omega$ as we move in parameter
space $\Theta$ demands the definition of
the way the vector $|\psi (\mathbf{R}) \rangle$ in $\Omega (\mathbf{R})$
propagates when moving to the neighbouring point $\mathbf{R} + \mathrm{d}
\mathbf{R}$ into the slightly turned Hilbert space $\Omega(\mathbf{R}+\mathrm{d}
\mathbf{R})$. 
   The required propagation is given by 
\begin{equation}
\label{prop}
\psi^{\mu} (\mathbf{R} + \mathrm{d}\mathbf{R}) = 
\psi^{\mu} (\mathbf{R}) + \partial_i \psi^{\mu} (\mathbf{R}) \mathrm{d} R^i \; ,
\end{equation}
again for infinitesimal $\mathrm{d} \mathbf{R}$.
  This means that the vector is first propagated in $\Omega (\mathbf{R})$
and then projected into $\Omega (\mathbf{R} + \mathrm{d}\mathbf{R})$ by 
retaining the same vector components. 
  This is consistent 
to linear order
with the projective propagation of the state $|\psi\rangle$ when
moving from $\mathbf{R}$ to $\mathbf{R} + \mathrm{d}\mathbf{R}$, 
\begin{multline}
\label{abstractprop}
|\psi (\mathbf{R} + \mathrm{d}\mathbf{R}) \rangle =
P_{\Omega (\mathbf{R} + \mathrm{d}\mathbf{R})} \{
P_{\Omega (\mathbf{R})} |\psi (\mathbf{R}) \rangle \\ +
P_{\Omega (\mathbf{R})} \partial_i [ P_{\Omega (\mathbf{R})}
| \psi (\mathbf{R})\rangle ] \mathrm{d} R^i \} \, .
\end{multline}
  This last expression shows the propagation of a vector when moving in
$\Theta$, irrespective of basis set and basis set transformation.
  It involves the covariant derivative in its final term.
  The equivalence between Eqs.~\ref{prop} and \ref{abstractprop}
is shown in Appendix~\ref{PropAppendix}.

 Differently from canonical differential geometry of curved spaces,
in the formalization presented here the proposed Christoffel symbols
do not necessarily contract with the metric tensor following conventional
rules 
for lowering or raising indices. 
  The required equalities 
among these objects are presented in Appendix~\ref{ChristAppendix}.

\subsubsection{Covariant derivative for the bra representation}

  The natural representation of the bra of any state, $\langle \psi | $,
is $\psi_{\mu} = \langle \psi | e_{\mu} \rangle$.
  Its covariant derivative is defined in analogy to Eq.~\ref{covdevdef},
namely,
\begin{equation}
\label{covdevbra}
\covdev_i \psi{\mu} \equiv \, \partial_i \left \{ \langle \psi | P_{\Omega} \, \right \} 
| e_{\mu} \rangle \, ,
\end{equation}
which, again working in analogy with that which was
done for Eq.~\ref{covdevdef2}, gives
\begin{equation}
\label{covdevbra2}
\covdev_i \psi_{\mu} = \partial_i \psi_{\mu} + \psi_{\nu} D^{\nu}_{\phantom{i} i \mu} \; ,
\end{equation}
where we have defined the corresponding Christoffel symbol
\begin{equation}
\label{christoffelbra}
D^{\nu}_{\phantom{i} i \mu} \equiv  \langle \partial_i e^{\nu} | e_{\mu} \rangle 
\end{equation}
(see the different order of indices in the symbol as compared to the 
definition in Eq.~\ref{christoffel}).

  The Christoffel symbols of Eqs.~\ref{christoffel} and \ref{christoffelbra}
are easy to interrelate, since 
$\langle e^{\mu} | e_{\nu} \rangle = \delta^{\mu}_{\phantom{i}\nu} $ for all
$\mathbf{R} \in \Theta$, and, therefore,
\begin{equation*}
\partial_i \delta^{\mu}_{\phantom{i}\nu} = \langle \partial_i e^{\mu} | e_{\nu} \rangle +
\langle e^{\mu} | \partial_i e_{\nu} \rangle = D^{\mu}_{\phantom{i} i \nu} +
D^{\mu}_{\phantom{i} \nu i} = 0 \; ,
\end{equation*}
which leads to
\begin{equation}
\label{christoffelrelation}
D^{\mu}_{\phantom{i} i \nu} = - D^{\mu}_{\phantom{i} \nu i} \; 
\end{equation}
(general relations among Christoffel symbols can be found in 
Appendix~\ref{ChristAppendix}).
  The covariant derivative for the bra can then be written as
\begin{equation}
\label{covdevbra3}
\covdev_i \psi_{\mu} = \partial_i \psi_{\mu} - \psi_{\nu} D^{\nu}_{\phantom{i} \mu i}  \; .
\end{equation}

  This derivative transforms as the corresponding tensor, as can be easily checked
following any of the two procedures used above (and in 
Appendix~\ref{BasisChangeAppendix}) for the case of the ket representation.

\subsubsection{Covariant derivative for operators}

  Following the spirit of Eqs.~\ref{covdevdef} and \ref{covdevbra}, let us
define the covariant derivative of an operator $H$ in its natural representation
$H^{\mu}_{\phantom{e}\nu}$ as
\begin{equation}
\label{covdevop}
\covdev_i H^{\mu}_{\phantom{e}\nu} \equiv \langle e^{\mu} |  \partial_i \left \{
P_{\Omega} H P_{\Omega} \right \} | e_{\nu} \rangle \;,
\end{equation}
which becomes
\begin{equation}
\label{covdevop2}
\covdev_i H^{\mu}_{\phantom{e}\nu} = \partial_i H^{\mu}_{\phantom{e}\nu} + 
D^{\mu}_{\phantom{e}\lambda i} H^{\lambda}_{\phantom{e}\nu} - 
H^{\mu}_{\phantom{e}\lambda} D^{\lambda}_{\phantom{e}\nu i} \; .
\end{equation}
  For the last term we have used Eq.~\ref{christoffelrelation}.
  This last expression coincides with the usual definition of the 
covariant derivative of a second-rank tensor in other differential
geometry contexts, including general relativity, once the 
connection (the Christoffel symbols) has been defined.
  It can also be written as 
\begin{equation*}
\covdev_i H^{\mu}_{\phantom{e}\nu} = \partial_i H^{\mu}_{\phantom{e}\nu} 
+  [D,H]^{\mu}_{\phantom{e}\nu i} \; ,
\end{equation*}
where the conmutator in the last term is defined as the two last terms
in Eq.~\ref{covdevop2}.

  This definition gives a well-behaved tensor, transforming as such
under a basis set change. 
  It can be straightforwardly checked following either of
the two procedures used before for $\covdev_i \psi{^\mu}$: by noticing
from Eq.~\ref{covdevop} that it is a tensor representation of an operator 
within $\Omega$, and by following Appendix~\ref{BasisChangeAppendix}. 
  It is also a definition consistent with the previous ones. 
  The following Leibniz chain rule for a vector
\begin{equation}
\label{chain}
\covdev_i \left ( H^{\mu}_{\phantom{e}\nu} \psi^{\nu} \right ) =
\left ( \covdev_i  H^{\mu}_{\phantom{e}\nu} \right ) \psi^{\nu} +
H^{\mu}_{\phantom{e}\nu} \left ( \covdev_i \psi^{\nu} \right ) \; ,
\end{equation}
and the expected behavior of a scalar (zero-rank tensor)
\begin{equation}
\label{e_deriv}
\covdev_i E = 
\covdev_i \left ( \psi_{\mu} H^{\mu}_{\phantom{e}\nu} \psi^{\nu} \right ) = 
\partial_i \left ( \psi_{\mu} H^{\mu}_{\phantom{e}\nu} \psi^{\nu} \right ) = \partial_i E \; ,
\end{equation}
are proved in Appendix~\ref{ChainRuleAppendix}.

\subsubsection{Matrix representation}

  Let us extend the previous definitions to the matrix representation, 
which will be useful in the next section.  
  Since the ket is equally represented in the natural and matrix representations,
the covariant derivative of a ket is also equally defined, as in Eq.~\ref{covdevdef2}.
  The bra is, however, different, since $\langle \psi |$ is represented by
$\psi^{\mu *} = \langle \psi | e^{\mu} \rangle$ (see details of present notation
in Appendix~\ref{NotationAppendix}).
  Its covariant derivative is then defined as
\begin{equation*}
\covdev_i \psi^{\mu *} = \partial_i  \left \{ \langle \psi | P_{\Omega} \right \} | 
e^{\mu} \rangle \; ,
\end{equation*}
and, therefore, without resorting to the ambient Hilbert space ${\cal H}$, 
it is expressed as
\begin{equation*}
\covdev_i \psi^{\mu *} = \partial_i \psi^{\mu *} + \psi^{\nu *} 
D_{\nu i}^{\phantom{ei}\mu} \; ,
\end{equation*}
which is just the Hermitian conjugate of Eq.~\ref{covdevdef2}
(see Appendix~\ref{ChristAppendix} for relations among Christoffel symbols).
  For an operator in this representation, the covariant derivative is defined as
\begin{equation*}
\covdev_i H_{\mu\nu} = \langle e_{\mu} | \partial_i \left \{ P_{\Omega} H
P_{\Omega} \right \} | e_{\nu} \rangle \; ,
\end{equation*}
which becomes
\begin{equation}
\label{covdevHmatrix}
\covdev_i H_{\mu\nu} = \partial_i H_{\mu\nu} +
D_{\mu\phantom{e} i}^{\phantom{e}\sigma} H_{\sigma\nu} +
H_{\mu\sigma} D_{\phantom{e} i \nu}^{\sigma} \; .
\end{equation}

\subsubsection{Geometric interpretation of the affine connection}

  Eqs.~\ref{prop} and \ref{abstractprop} give the basis for a clearer geometric 
interpretation of the affine connection defined above, and the
corresponding covariant derivative.
  Closing Eq.~\ref{abstractprop} from the left with $\langle e^{\mu} (
\mathbf{R} + \mathrm{d}\mathbf{R}) |$, one obtains
\begin{equation}
\label{affinerot}
\psi^{\mu} ( \mathbf{R} + \mathrm{d}\mathbf{R})  =
A^{\mu}_{\phantom{e}\nu} (\mathbf{R} + \mathrm{d}\mathbf{R} : \mathbf{R})
\big \{ \psi^{\nu} + ( \covdev_i \psi^{\nu} ) \mathrm{d} R^i \big \} \, ,
\end{equation}
where the terms within the curly brackets are defined at {\bf R},
and 
\begin{equation}
\label{TimeOverlap}
A^{\mu}_{\phantom{e}\nu} (\mathbf{R} + \mathrm{d}\mathbf{R} : \mathbf{R})
\equiv \langle e^{\mu} (\mathbf{R} + \mathrm{d}\mathbf{R}) | 
e_{\nu} (\mathbf{R}) \rangle \, ,
\end{equation}
defines the basis transformation (as in Appendix~\ref{BasisChangeAppendix})
when moving from {\bf R} to $\mathbf{R} + \mathrm{d}\mathbf{R}$ 
(the linear equivalence between Eqs.~\ref{affinerot} and \ref{prop} is shown in 
Appendix~\ref{Geometric-Appendix}).
  Eq.~\ref{TimeOverlap} represents the local gauge transformation.
In principle any invertible matrix with a smooth behaviour with respect to {\bf R}
is allowed.

  In Eq.~\ref{affinerot}, the basis set transformation information is now carried by
the $A$ tensor instead of the Christoffel symbols.
  The Christoffel symbols would reappear in Eq.~\ref{affinerot}
if replacing the covariant derivative by its definition in terms of 
the regular derivative, Eq.~\ref{covdevdef2}.
  That would indeed defeat the purpose of Eq.~\ref{affinerot} (and would
bring us back to Eq.~\ref{prop}, as shown in the mentioned 
Appendix~\ref{Geometric-Appendix}). 
  The usefulness of  expressing the propagation as in Eq.~\ref{affinerot}
becomes evident when using it, for instance, for finite-differences time integrators 
for solving the time-dependent Schr\"odinger equation, as we will do in 
Section~\ref{quantumevolution} and Appendix~\ref{ModifiedCK-Appendix}, 
where we will essentially replace 
$\covdev_t \psi^{\mu}$ by $-i H^{\mu}_{\phantom{e}\nu} \psi^{\nu}$. 

  In addition to its utility for integrators, it is presented here because it conveys 
the geometric meaning of the affine connection quite clearly: the covariant derivative 
is the intrinsic one, independent of the basis set change, accounting for both the 
physical variation of the state and the turning of the Hilbert space $\Omega$, 
while the Christoffel symbols linearly account for the basis set transformation.

\subsubsection{Rotation versus deformation}

  So far we have talked about basis change or transformation in general.
  We make here the distinction between pure rotations of
the basis and what we will call basis {\it deformation} (in analogy with 
elasticity theory).
  They are defined as the ones for anti-Hermitian and Hermitian $D_{\mu\nu i}$,
repectively.
   A small arbitrary transformation will thus have a rotation and a deformation
 component, that can be obtained from $( D_{\mu\nu i} - D^{*}_{\nu\mu i}) / 2$,
 and $( D_{\mu\nu i} + D^{*}_{\nu\mu i} ) / 2$, respectively.

   Appendix~\ref{rotation-appendix} shows how a small unitary transformation of
 the basis, defined as one that keeps constant overlap (and therefore
corresponding to basis rotations in $\Omega$) has an associated anti-Hermitian
$D_{\mu\nu i}$ tensor.
  This consideration will be relevant in the Applications section below, in the
context of finite-differences integrators.

\subsubsection{Parallel transport and unitary propagation}

\label{ParallelTransport}

  The analog of the parallel transport in curved spaces would be the 
propagation of a state vector $|\psi \rangle \in \Omega (\mathbf{R})$, 
which, in itself, would not vary (it would be constant if $\Omega={\cal H}$) 
when moving in $\Theta$ away from point {\bf R}. 
  Such intrinsic constancy is reflected by a null covariant derivative, 
\begin{equation*}
\covdev_i \psi^{\mu} = \partial_i \psi^{\mu} + 
D^{\mu}_{\phantom{e}\nu i} \psi^{\nu} = 0 \, .
\end{equation*}
  Therefore, parallel transport of any vector along any line in $\Theta$
is then obtained from Eq.~\ref{prop}, propagating 
\begin{equation*}
\psi^{\mu} (\mathbf{R} + \mathrm{d}\mathbf{R}) = 
\psi^{\mu} (\mathbf{R}) - D^{\mu}_{\phantom{e}\nu i} (\mathbf{R}) 
\psi^{\nu} (\mathbf{R}) \mathrm{d} R^i 
\end{equation*}
(for infinitesimal d{\bf R})
along the corresponding line.

  Vectors that are orthogonal to each other at a given point would
propagate keeping their orthogonality. 
  Appendix~\ref{UnitaryAppendix} shows that any unitary propagation
(one such the condition $\covdev_i \{ \psi_{n\mu} \psi^{\mu}_{\phantom{e}m} \} 
= 0$ is preserved)
maintains the orthogonality of propagated vectors.
  Parallel transport is a special case, since $\covdev_i \psi^{\nu}_m
= \covdev_i \psi_{\mu n} = 0$.

  More generally, a set of vectors in parallel transport keep their mutual 
scalar products (this can also be seen following the reasoning of 
Appendix~\ref{UnitaryAppendix}).
  This also applies to the two metric tensors, which are composed 
precisely of the scalar products of basis vectors.
  Since such scalar products are preserved under parallel transport, 
then
\begin{equation*}
\covdev_i S_{\mu\nu}= \covdev_i S^{\mu\nu} = 0 \, ,
\end{equation*}
which reflects a fundamental property of the theory, that the covariant derivative 
conserves the metric.
  This result is also derived explicitly in Appendix~\ref{ChristAppendix}.

\subsubsection{Curvature}

  The Riemann-Christoffel curvature for a curved space characterises the
fact that if you take one vector in a vector field using the Levi-Civita connection 
along one direction, and then along another direction to reach a certain point, 
it gives a different result than if changing the order of directions in which you 
arrive to the same point (or analogously, if following a closed loop).
  This is locally quantified with the difference in changing the order of
second derivatives, by defining the curvature tensor
$R^{\mu}_{\phantom{e}i\nu j}$ such that
\begin{equation}
\label{curvdef}
R^{\mu}_{\phantom{e}i \nu j} \psi^{\nu} = 
\covdev_i \covdev_j \psi^{\mu} - \covdev_j \covdev_i \psi^{\mu} \; .
\end{equation}
  Using the covariant derivatives defined above, one obtains:
\begin{equation}
\label{curvexp}
R^{\mu}_{\phantom{e} i \nu j} = 
\partial_i D^{\mu}_{\phantom{e}\nu j} - \partial_j D^{\mu}_{\phantom{e}\nu i} + 
D^{\mu}_{\phantom{e}\lambda i} D^{\lambda}_{\phantom{e}\nu j} -
D^{\mu}_{\phantom{e}\lambda j} D^{\lambda}_{\phantom{e}\nu i} \, ,
\end{equation}
again in perfect analogy to the expression for curved spaces.

\subsubsection{Relation to Berry connection and curvature}

  In Subsection~\ref{ParameterSpace} the analogy with Berry's geometric phase
formalism\cite{Berry} was mentioned. 
  Indeed, for a quantum mechanical (single- or many-particle) state 
$| \Psi ( \mathbf{R} ) \rangle$,
the Berry connection is normally defined as 
\begin{equation*}
{\cal A}_{j} = i \, \langle \Psi | \partial_j \Psi \rangle
\end{equation*}
($i=\sqrt{-1}$),
which is nothing but ($i$ times) the connection defined in this work 
(Eq.~\ref{christoffel}) for a space $\Omega$ spanned by a single state.
  It generalises to the trace
\begin{equation}
\label{BerryConnection}
{\cal A}_{j} = i \, \sum_n^{occ} \langle \psi_n | \partial_j \psi_n \rangle \; ,
\end{equation}  
for a set of single-particle states $| \psi_n \rangle$ spanning the occupied space
in the context of a mean-field-like approach to the many-particle problem (as is the case 
for the Kohn-Sham states in density-functional theory).
  As expected from Berry's work, it is easy to see that 
\begin{equation}
\label{TraceConnection}
{\cal A}_j = i \, D^{\mu}_{\phantom{e}\mu j} \; ,
\end{equation}
where $D^{\mu}_{\phantom{e}\mu j}$ is 
the trace of the connection defined in 
Eq.~\ref{christoffel}
(bear in mind that the $| \psi_n \rangle$ states in this context play
the role of the basis of the relevant space, i.e., the $| e_{\mu}\rangle$ states
of the previous sections, and $\Omega$ refers here to the occupied space).
  The expression in Eq.~\ref{BerryConnection} assumes orthonormal states,
whereas Eq.~\ref{TraceConnection} is valid for any non-orthogonal basis of 
the relevant $\Omega$ space.
  More generally, the Berry connection matrix 
${\cal A}_{mnj} = i \, \langle \psi_m | \partial_j \psi_n \rangle$ corresponds to
($i$ times) this work's $D_{mn j}$ connection in the matrix representation, which
can be transformed to any other tensorial representation for
non-orthogonal states.

  Similarly to what happens to the connection, the curvature of this work and that 
of Berry are closely related.
  The Berry curvature is usually defined as 
\begin{equation*}
{\cal R}_{ij} = - 2 \; \mathrm{Im} \left \{ \langle \partial_i \Psi | \partial_j \Psi \rangle \right \} \; ,
\end{equation*}
for a quantum mechanical state $| \Psi \rangle$,
which generalises to 
\begin{equation}
\label{BerryCurv}
{\cal R}_{ij} = - 2 \; \mathrm{Im} \left \{ \sum_n^{occ} \langle \partial_i \psi_n | \partial_j \psi_n 
\rangle \right \} 
\end{equation}
for a set of single-particle states $| \psi_n \rangle$ spanning the occupied space.
  The curvatures in Eqs.~\ref{curvexp} and \ref{BerryCurv} are very closely 
interrelated when considering our $\Omega$ Hilbert space as the occupied space
(or any specific subspace that we are computing the curvature for).
  Starting with 
\begin{equation*}
\partial_i D^{\mu}_{\phantom{e}\nu j} = \langle \partial_i e^{\mu} | \partial_j e_{\nu} \rangle
- \langle e^{\mu} | \partial_i \partial_j e_{\nu} \rangle \; ,
\end{equation*}
we can easily see that
\begin{equation*}
\partial_i D^{\mu}_{\phantom{e}\nu j} - \partial_j D^{\mu}_{\phantom{e}\nu i} = 
\langle \partial_i e^{\mu} | \partial_j e_{\nu} \rangle -
\langle \partial_j e^{\mu} | \partial_i e_{\nu} \rangle \; .
\end{equation*}
If we now trace over the quantum variables, in analogy with the Ricci curvature
\begin{equation}
\label{RicciBerry}
{\cal R}_{ij} = R^{\mu}_{\phantom{e} i \mu j} = 
\langle \partial_i e^{\mu} | \partial_j e_{\mu} \rangle -
\langle \partial_j e^{\mu} | \partial_i e_{\mu} \rangle \; ,
\end{equation} 
since the following traces annihilate:
\begin{equation*}
D^{\mu}_{\phantom{e}\lambda i} D^{\lambda}_{\phantom{e}\mu j} -
D^{\mu}_{\phantom{e}\lambda j} D^{\lambda}_{\phantom{e}\mu i} = 0 \; .
\end{equation*}

  If the basis $\{ | e_{\mu} \rangle \}$ is invariably orthonormal, then
$\langle e^{\mu} | = \langle e_{\mu} |$, for any {\bf R}, and thus
$\langle \partial_i e^{\mu} | = \langle \partial_i e_{\mu} |$, and
\begin{equation*}
\langle \partial_j e_{\mu} | \partial_i e_{\mu} \rangle =
\langle \partial_i e_{\mu} | \partial_j e_{\mu} \rangle ^* \; ,
\end{equation*}
and, therefore, 
\begin{equation}
\label{BerryRicci}
{\cal R}_{ij} =   2 i \; \mathrm{Im} \left \{
\langle \partial_i e_{\mu} | \partial_j e_{\mu} \rangle \right \} \; ,
\end{equation} 
which is nothing but Eq.~\ref{BerryCurv} (times $i$) for the 
$| \psi_n \rangle$ states taken as an orthonormal basis of 
occupied space $\Omega$.
  Therefore, Berry's curvature is nothing but the Ricci curvature of 
our turning occupied space.

  This result is directly generalizable to any other orthonormal basis
of occupied space, e.g., a basis of Wannier functions, still 
under Eq.~\ref{BerryRicci}.
  If the basis is non-orthogonal (non-orthogonal Wannier functions), 
Eq.~\ref{RicciBerry} is then the relevant definition.
  If seeking an expression closer to Eq.~\ref{BerryRicci}, the definition in 
Eq.~\ref{RicciBerry} can also be re-expressed as
\begin{align}
\label{BerryRicciNon}
{\cal R}_{ij} =   & 2 i \; \mathrm{Im}  \left \{ S^{\mu\nu}
\langle \partial_i e_{\nu} | \partial_j e_{\mu} \rangle \right \}  \notag \\ 
& + ( \partial_i S^{\mu\nu} ) D_{\nu\mu j} - ( \partial_j S^{\mu\nu} ) D_{\nu\mu i} \; ,
\end{align}
which, in addition to the expected redefinition of the trace with the metric tensor 
in the first term, includes two additional terms related to the variation 
of the metric itself.

\subsubsection{The topology of $\Omega (\mathbf{R})$}

   When solving a quantum-mechanical problem using a finite basis
set that changes in parameter space, we will then have two relevant
fibre bundles, one within the other: the one spanned by the basis, $\Omega$,
and the one spanned by the occupied states. 
  The curvature and topology of $\Omega$({\bf R})
and its relation with the ones corresponding to the occupied space could have 
implications on the effect of the approximation implied by the basis.
  We will not explore this point further in this paper. 
  We can however, point to Ref.~\onlinecite{Mead1992} for the implications
of Berry concepts on the dependence of the occupied space on atomic positions
in molecular systems, including the effect of conical intersections, for instance.
  Rethinking such ideas considering a wider-than-occupied space could be
an avenue for future investigation.

\section{Applications}

\subsection{Quantum time evolution}

\label{quantumevolution}

\subsubsection{Basic equations}
  
  Using the present formalism, we can consider a one-dimensional 
parameter space with time as the only variable. 
 In the natural representation, the time dependent Schr\"odinger 
equation $H | \psi \rangle = i \partial_t | \psi \rangle$ becomes 
simply
\begin{equation}
\label{td}
H^{\mu}_{\phantom{e}\nu} \psi^{\nu} = i \, \covdev_t \, \psi^{\mu} \, ,
\end{equation}
where the time covariant derivative is defined as 
\begin{equation*}
\covdev_t    \, \psi^{\mu} = \partial_t  \, \psi^{\mu} + D^{\mu}_{\phantom{e}\nu t} 
\, \psi^{\nu} \, ,
\end{equation*}
and the corresponding temporal Christoffel symbol as
\begin{equation*}
D^{\mu}_{\phantom{e}\nu t}  = \langle e^{\mu} | \partial_t | e_{\nu} \rangle
= \langle e^{\mu} | \partial_t  e_{\nu} \rangle \, .
\end{equation*}
   Equation \ref{td} reflects the physics in a basis-set independent form, 
in the sense that the well-behaved tensors in the equation
all transform as in Appendix~\ref{BasisChangeAppendix}, 
and transmit the physics of the original Schr\"odinger
equation for an evolving basis set and Hilbert space.
  Eq.~\ref{td} can be obtained by minimising the action
for the following Lagrangian: 
\begin{equation*}
L = i \, \psi_{\mu} \covdev_t \psi^{\mu} - \psi_{\mu} H^{\mu}_{\phantom{e}\nu}
\psi^{\nu} \; ,
\end{equation*}
which is easily obtained using the ideas above from the standard
$L = i \, \langle \psi | \partial_t \psi \rangle - \langle \psi | H | \psi \rangle$.
  
  The matrix representation also allows a concise representation
of the physical equation, albeit less elegantly, carrying around
the metric tensors, as follows.
\begin{equation}
\label{matschroe}
H_{\mu\nu} \psi^{\nu} = i \, S_{\mu\nu} \,\covdev_t \, \psi^{\nu} \, ,
\end{equation}
which corresponds to Eq.~\ref{old-td-eq} of the Introduction, or
\begin{equation*}
S^{\mu\sigma} H_{\sigma\nu} \psi^{\nu} = i \,\covdev_t \, \psi^{\mu} \, .
\end{equation*}

  If following the propagation of the density matrix
instead of that of the wave functions, the dynamics is
defined by the Liouville - Von Neumann equation,
\begin{equation*}
i \, \partial_t \rho = [ H, \rho]  \; ,
\end{equation*}
where the density operator $\rho$ would be
\begin{equation*}
\rho (t) = | \Psi (t) \rangle \langle \Psi (t) |
\end{equation*}
for a pure quantum state, or 
\begin{equation*}
\rho (t) = \sum_n^{occ} | \psi_n (t) \rangle \langle \psi_n (t) |
\end{equation*}
for the set of occupied states in a mean-field setting. 
  (It can also be generalised to statistical mixtures in general,
including thermal). 

   Again, for the evolving Hilbert space $\Omega$, the
expression of this equation in terms of the corresponding
tensors is obtained by closing it with $\langle e^{\mu} |$ 
from the left and $ | e_{\nu} \rangle $ from the right, and
substituting $\rho$ and $H$ by 
$P_{\Omega} \rho P_{\Omega}$ and 
$P_{\Omega} H P_{\Omega}$, giving
\begin{equation*}
i \, \covdev_t \rho^{\mu}_{\phantom{e}\nu} = 
[ H, \rho]^{\mu}_{\phantom{e}\nu} = 
H^{\mu}_{\phantom{e}\sigma} \rho^{\sigma}_{\phantom{e}\nu} -
\rho^{\mu}_{\phantom{e}\sigma} H^{\sigma}_{\phantom{e}\nu}  \; 
\end{equation*}
in its natural representation.
  The matrix representation of this equation
is much less elegant.
  The conventional definition of  density matrix in a typical quantum chemistry
setting is $\sum_n^{occ} C_{\mu n} C_{\nu n}^*$, in the language 
of Eq.~\ref{coeff-expansion} of the Introduction. 
  This is nothing but  
\begin{equation*}
\rho^{\mu\nu} = \langle e^{\mu} | \, \rho \, | e^{\nu} \rangle \; .
\end{equation*}  
  In this representation, the Louiville - Von Neumann equation becomes
\begin{equation*}
i \, \covdev_t \rho^{\mu\nu} = 
S^{\mu\sigma} H_{\sigma\kappa} \rho^{\kappa\nu} -
\rho^{\mu\sigma} H_{\sigma\kappa} S^{\kappa\nu} \; .
\end{equation*}

  For any of the former equations, the time dependence of the basis 
orbitals may be due to the variation in time of other parameters like
atomic positions $R^i$.
  In such cases, the Christoffel symbols in the covariant derivatives would
satisfy
\begin{equation*}
D^{\mu}_{\phantom{e} \nu t} = v^i D^{\mu}_{\phantom{e} \nu i} \; ,
\end{equation*} 
(or equivalent representations) where $v^i = \partial R^i / \partial t$ 
are the corresponding nuclear velocities.

\subsubsection{Crank-Nicholson integrator}

  In various contexts a time-dependent Schr\"odinger-like equation 
is numerically solved by discretizing time, using adequate
integrator algorithms (for a comparison of the performance and
stability of different options see Refs.~\onlinecite{Koslov,Correa}).
  In the context of non-orthogonal basis sets 
and within mean-field-like
theories for which matrix inversion is affordable, 
the Crank-Nicholson algorithm has been used quite 
successfully.\cite{Tsolakidis2002}  
  The generalization of that procedure to a moving basis set was
achieved by incorporating a L\"owdin orthonormalisation step, 
an idea due to Sankey and collaborators,\cite{Sankey} which will 
be discussed below. 
  The Crank-Nicholson-L\"owdin procedure proved quite successful
in the integration of the Kohn-Sham equations for several studies
of electronic stopping power for ionic projectiles shooting through 
varied materials.\cite{Zeb2013,Correa2013,Ullah2015} 
  Here we define new integrators based on the Crank-Nicholson idea,
inspired by the affine connection defined above, and we compare
them with the Crank-Nicholson-L\"owdin procedure.

  Let us first revise the Crank-Nicholson method in this context
for non-moving bases.\cite{Tsolakidis2002}
  The basics: a state $| \psi \rangle$ evolving according to
$H | \psi \rangle = i \partial_t | \psi \rangle$ can be propagated from $t$ to 
$t+\mathrm{d}t$ by considering the backwards and forwards evolution 
from each of those time points to the one in the middle, as follows:
\begin{align}
\label{CK-abstract}
| \psi (t+\mathrm{d}t/2 ) \rangle 
&= \Big \{ 1 - i \frac{\mathrm{d}t}{2} H (t) \Big \} | \psi (t) \rangle \notag \\
| \psi (t+\mathrm{d}t/2 ) \rangle 
&= \Big \{ 1 + i \frac{\mathrm{d}t}{2} H (t+\mathrm{d}t) \Big \} 
| \psi (t+\mathrm{d}t) \rangle \, ,
\end{align}
which, by eliminating the middle point becomes
\begin{align*}
| \psi (t+\mathrm{d}t) \rangle = \Big [ 1 + i \frac{\mathrm{d}t}{2} 
H (t+\mathrm{d}t) \Big ]^{-1}
\Big \{ 1 - i \frac{\mathrm{d}t}{2} H (t) \Big \} | \psi (t) \rangle
\end{align*}
thereby ensuring, by construction, that the algorithm respects
invariance under time reversal.

  The use of this algorithm is complicated by the dependence
on $H (t+\mathrm{d}t)$, which requires the use of an 
iterative self-consistency procedure.\cite{Kaxiras}
  For practical purposes, however, in many implementations the 
algorithm is simplified by using $H(t)$ in both factors, 
\begin{align}
\label{CK-kets}
| \psi (t+\mathrm{d}t) \rangle = \Big [ 1 + i \frac{\mathrm{d}t}{2} 
H (t) \Big ]^{-1} \Big \{ 1 - i \frac{\mathrm{d}t}{2} H (t) \Big \} | \psi (t) \rangle
\end{align}
which allows a direct evaluation of $| \psi (t+\mathrm{d}t) \rangle$
from information of the previous time step.
  This change is of course of no consequence if the Hamiltonian does
not change with time, but this is hardly the case for any mean-field-like
Hamiltonian (such as that of Kohn and Sham), given the dependence of the 
Hamiltonian on the evolving electron density or wave-functions. 

  For a fixed, not evolving basis set, this is used\cite{Tsolakidis2002} as
\begin{multline}
\label{CK-natural}
\psi^{\mu} (t+\mathrm{d}t) = \\  \Big [ \delta_{\mu}^{\phantom{e}\sigma} + 
i \frac{\mathrm{d}t}{2} H_{\mu}^{\phantom{e}\sigma} (t) \Big ]^{-1}
\Big \{ \delta^{\sigma}_{\phantom{e}\nu}  - i \frac{\mathrm{d}t}{2} 
H^{\sigma}_{\phantom{e}\nu}  (t) \Big \}  \psi^{\nu}(t) \, 
\end{multline}
(please note that the first factor does not indicate an inversion of the
particular $(\mu,\sigma)$ element, but of the tensor as a whole; see 
Appendix~\ref{InverseAppendix} for the notation and relevant definitions 
used in this paper for inverse second-rank tensors).
  In the matrix representation it becomes
\begin{multline}
\label{CK-matrix}
\psi^{\mu} (t+\mathrm{d}t) = \\ \Big [ S_{\mu\sigma} + i \frac{\mathrm{d}t}{2} 
H_{\mu\sigma} (t) \Big ]^{-1}  
\Big \{ S_{\sigma\nu}   - i \frac{\mathrm{d}t}{2} H_{\sigma\nu}  (t) \Big \}
\psi^{\nu}(t) \, ,
\end{multline}
keeping in mind that the lower indices within the inverted tensor become 
upper indices (see Appendix~\ref{InverseAppendix}).

  For fixed bases this algorithm has the enormous virtue of 
being strictly unitary,\cite{Tsolakidis2002}  in the sense that when 
propagating an orthonormal set of states $\psi^{\mu}_m$, the set 
remains orthonormal at $t+\mathrm{d}t$ regardless of the size of d$t$.

\subsubsection{Revising Crank-Nicholson for a moving basis}

  For a moving basis set, the algorithm has to be generalized.
  One straightforward generalisation is achieved by replacing
$H_{\mu\sigma}(t)$ in Eq.~\ref{CK-matrix} by $H_{\mu\sigma}(t) 
- i D_{\mu\sigma t}(t)$ (in the matrix representation, for instance), giving
\begin{multline}
\label{CKmov-matrix}
\psi^{\mu} (t+\mathrm{d}t) =  \Big [ S_{\mu\sigma}  + i \frac{\mathrm{d}t}{2} 
( H_{\mu\sigma}  - i D_{\mu\sigma t} )  \Big ]^{-1}  \\ \times
\Big \{ S_{\sigma\nu} - i \frac{\mathrm{d}t}{2} ( H_{\sigma\nu} - i D_{\sigma\nu t} ) \Big \} 
\psi^{\nu}(t)
\end{multline}
(where we have dropped the $t$-dependence of the tensors for
clarity).
Here, $D_{\mu\sigma t} = \langle e_{\mu} | \partial_t e_{\sigma} \rangle$
is the required temporal Christoffel symbol.
  Eq.~\ref{CKmov-matrix} can be derived from Eq.~\ref{prop} using the 
tensorial time-dependent Schr\"odinger equation in Eq.~\ref{td} and the 
definition of the covariant derivative, Eq.~\ref{covdevdef2}.
  This would be only linearly correct in d$t$ in what concerns both
Hilbert space turning and basis set transformation, 
and therefore the nice 
unitary propagation feature for arbitrary d$t$ is lost.
  Indeed, the loss of hermiticity of the propagated effective Hamiltonian
makes this problem apparent,
albeit that the propagation is correct and well-behaved in the limit of
small d$t$.

   A much more promising  approach is obtained by building on 
Eq.~\ref{affinerot} instead of Eq.~\ref{prop}, 
which exactly accounts 
for basis set change if $\Omega(t+\mathrm{d}t)=\Omega(t)$.
  Within the matrix representation, one obtains
\begin{multline}
\label{CKmov-matrix2}
\psi^{\mu} (t+\mathrm{d}t) = 
\Big [ S_{\mu\lambda}(t+\mathrm{d}t)  + i \frac{\mathrm{d}t}{2} 
H_{\mu\lambda}  (t+\mathrm{d}t) \Big ]^{-1} \\ \times
A_{\lambda}^{\phantom{e}\sigma} (t +\mathrm{d}t : t) 
\Big \{ S_{\sigma\nu}(t) - i \frac{\mathrm{d}t}{2} H_{\sigma\nu} (t) \Big \} 
\psi^{\nu}(t) \; ,
\end{multline}
where $A_{\lambda}^{\phantom{e}\sigma} (t +\mathrm{d}t : t) =
\langle e_{\lambda} (t +\mathrm{d}t) | e^{\sigma} (t) \rangle$ is defined 
analogously to Eq.~\ref{TimeOverlap}.
  The derivation of Eq.~\ref{CKmov-matrix2} is given in 
Appendix~\ref{ModifiedCK-Appendix}.
  The algorithm given by Eq.~\ref{CKmov-matrix2} would again require an
iterative self-consistency procedure for every time step.
  An analog to Eq.~\ref{CK-kets} can also be considered 
with a view to
removing the dependence on $H_{\mu\nu} (t+\mathrm{d}t)$ and
$S_{\mu\nu} (t+\mathrm{d}t)$.
  This is achieved with the following Ansatz:
\begin{multline}
\label{CKmov-matrix3}
\psi^{\mu} (t+\mathrm{d}t) = A^{\mu}_{\phantom{e}\lambda} (t +\mathrm{d}t : t) 
\\ \times \Big [ S_{\lambda\sigma} (t)  + i \frac{\mathrm{d}t}{2} 
H_{\lambda\sigma} (t)  \Big ]^{-1}
\Big \{ S_{\sigma\nu} (t) - i \frac{\mathrm{d}t}{2} H_{\sigma\nu} (t)  \Big \} 
\psi^{\nu}(t) \, .
\end{multline}
  The integrator in Eq.~\ref{CKmov-matrix3} keeps the strict unitary character 
of the algorithm for a transforming basis set in a fixed $\Omega$ space. 
  This can be seen by noticing that the last two transformations in 
Eq.~\ref{CKmov-matrix3} (the ones in curly and square brackets) correspond
to the original Crank-Nicholson scheme for a fixed Hamiltonian, and that
the $A$ tensor transformation is nothing but a change of basis. 
  The latter, however, is only linearly correct in d$t$ if $\Omega$ does turn. 
  Indeed, for an arbitrarily large d$t$, the set 
$P_{\Omega(t + \mathrm{d}t)} | \psi_m \rangle$ 
is not necessarily orthonormal even if the set $P_{\Omega(t)} | 
\psi_m \rangle$ was.
  The advantage of Eq.~\ref{CKmov-matrix3} with respect to 
Eq.~\ref{CKmov-matrix} is important, however, since the space 
turning diminishes with better converged bases, while the basis 
change does not necessarily diminish (think of the perfectly converged limit of 
$\Omega={\cal H}$ for a moving basis: the basis still changes and the 
Hilbert space does not).

  In practice, the tensor $A^{\mu}_{\phantom{e}\lambda} (t +\mathrm{d}t : t)$
of Eq.~\ref{CKmov-matrix3} is somewhat inconvenient for calculations.
  It can be replaced by
\begin{multline}
\label{CKmov-matrix4}
\psi^{\mu} (t+\mathrm{d}t) = S^{\mu\kappa}(t+\mathrm{d}t) A_{\kappa\lambda} 
(t +\mathrm{d}t : t) \\ \times \Big [ S_{\lambda\sigma} (t) + i \frac{\mathrm{d}t}{2} 
H_{\lambda\sigma} (t)  \Big ]^{-1}
\Big \{ S_{\sigma\nu} (t) - i \frac{\mathrm{d}t}{2} H_{\sigma\nu} (t) \Big \} 
\psi^{\nu}(t) \, ,
\end{multline}  
which involves the inversion of the overlap matrix at $t+\mathrm{d}t$
and the calculation of the overlap between the $t$ basis vectors and
the ones at $t+\mathrm{d}t$ (in addition to the Crank-Nicholson operations at $t$).
  Similarly, the practical implementation of the self-consistent procedure
implied by Eq.~\ref{CKmov-matrix2} would actually imply using the
following self-consistent integrator instead:
\begin{multline}
\label{CKmov-matrix5}
\psi^{\mu} (t+\mathrm{d}t) = 
\Big [ S_{\mu\lambda}(t+\mathrm{d}t)  + i \frac{\mathrm{d}t}{2} 
H_{\mu\lambda}  (t+\mathrm{d}t) \Big ]^{-1} \\ \times
A_{\lambda\kappa} (t +\mathrm{d}t \! : \! t) \, \Big \{ \delta^{\kappa}_{\phantom{e}\nu} 
- i \frac{\mathrm{d}t}{2} S^{\kappa\sigma} (t) H_{\sigma\nu} (t) \Big \} 
\psi^{\nu}(t) \, .
\end{multline}

\subsubsection{Procedure based on L\"owdin orthonormalization}
\label{secSankey}
 
  An alternative integrator was proposed\cite{Sankey} and has been 
used\cite{Zeb2013,Correa2013,Ullah2015} for strict unitary propagation
for arbitrary d$t$.
  Following the previous notation, this propagator can be written as
\begin{multline}
\label{Sankey}
\psi^{\mu} (t+\mathrm{d}t) = \{ S^{-1/2} (t+\mathrm{d}t) \}^{\mu l} 
\{ S^{1/2} (t) \}_{l\lambda}  
\\ \times \Big [ S_{\lambda\sigma} (t) + i \frac{\mathrm{d}t}{2} 
H_{\lambda\sigma} (t)  \Big ]^{-1}
\Big \{ S_{\sigma\nu} (t) - i \frac{\mathrm{d}t}{2} H_{\sigma\nu} (t) \Big \} 
\psi^{\nu}(t) \, .
\end{multline}  
  This is analogous to Eq.~\ref{CKmov-matrix4} inasmuch as it first does the
physical (Crank-Nicholson) propagation for the basis at $t$ and within 
$\Omega(t)$, and then 
it transforms to conform to the basis at time $t+\mathrm{d}t$.
  The basis transformation is done differently, however. It can be seen as the
following process: ($i$) It first changes basis 
within $\Omega(t)$ to a L\"owdin orthonormal basis.
  ($ii$) It then assumes that the coefficients in this basis do not change
when changing to the L\"owdin basis of $t+\mathrm{d}t$.   
  ($iii$) It then undoes the L\"owdin transformation in $t+\mathrm{d}t$
obtaining the sought coefficients in the non-orthogonal basis
at $t+\mathrm{d}t$.

  This procedure has the advantage that
the propagation is now strictly unitary by construction for any d$t$,
even considering the turning of the Hilbert space, which should give
larger stability to the method for relatively large values of d$t$.
  The propagated vectors are guaranteed to be orthonormal.
  
  However, the vectors propagated following this L\"owdin procedure
(Eq.~\ref{Sankey}) do not constitute a fair representation of what the 
evolution of the corresponding vectors would be in ${\cal H}$ in the sense
that it does not properly counter the effect of the transforming basis.
  In order to see this, consider the case of a set of vectors at $t=0$, 
$ \{ | \psi_n \rangle \}$, all within $\Omega$ and all initially orthonormal, 
$ {\cal S}_{nm} = \langle \psi_n | \psi_m \rangle = \delta_{nm}$. 
  Consider, as well, that the dynamics is such that the vectors do not 
rotate with time, changing only in phase, as would be the case for
eigenstates of a time-independent Hamiltonian, for instance.
  Take now a basis set for $\Omega$ that does rotate with time, but
keeps orthonormality at all times.
  In this scenario it is clear that the coefficients $\psi^{\mu}_n$ should
transform with time to capture the fact that the non-rotating eigenvectors 
are described in the frame of a rotating basis.
  This is properly taken care of in the integrator proposed in 
Eq.~\ref{CKmov-matrix4} by the $A_{\kappa\lambda} 
(t +\mathrm{d}t : t)$ tensor, which does the needed basis transformation.
   
   For the L\"owdin procedure, however, the coefficients do not change,
 except for the global phase dictated by the Hamiltonian evolution, i.e., 
 \begin{equation*}
 \psi^{\mu}_{\phantom{e}n} (t + \mathrm{d} t) = e^{-i \omega_n 
 \mathrm{d} t}  \psi^{\mu}_{\phantom{e}n} (t), \, ,
 \end{equation*}
as is obvious from the fact that if $S_{\mu\nu} = \delta_{\mu\nu}$ at all
times, and therefore
\begin{equation*}
S^{-1}_{\mu\nu} = S^{-1/2 \, \mu\nu} = S^{1/2}_{\mu\nu} = \delta_{\mu\nu}
\end{equation*} 
at all times. 
  This simple situation clearly illustrates the above assertion on the 
unfair representation by Eq.~\ref{CKmov-matrix4} of the 
states' evolution in a generic situation.
  It is not clear however how significant such discrepancies can be.
  In particular, the case made above is for pure rotations of the basis set.
  We explore this point in more depth in Appendix~\ref{SankeyAppendix}, 
finding interesting dependencies on the rotating versus deforming basis sets.
  In particular, for fixed-shape atomic-like orbitals as basis functions, 
such basis rotations correspond to Galileo transforms in parameter space,
which would suggest no physical significance to the discrepancies
between the solvers in Eqs.~\ref{Sankey} and \ref{CKmov-matrix4} for such 
rotations.
  This could be behind the apparent success of Eq.~\ref{Sankey}.
  A more quantitative analysis should be the focus of further work.

\subsubsection{Strictly unitary propagation}

  If not using Eq.~\ref{Sankey} for propagation, we are left with 
Eq.~\ref{CKmov-matrix4} (or Eq.~\ref{CKmov-matrix5} if using
self-consistency), which does not strictly conserve the orthonormality
of propagated states for finite d$t$ if the space $\Omega$ turns within
${\cal H}$.
  The propagator can still be used as long as d$t$ is small enough such
that the unitary propagation is preserved within a desired tolerance.
  An alternative, however, is re-orthonormalising the states. 
  This can be done with any orthonormalisation procedure, e.g.,
Gram-Schmidt or the iterative procedure used when finding eigenstates
by minimization in electronic structure (see, e.g., 
Ref~\onlinecite{Payne1990}).

  The L\"owdin orthonormalization method described above can
be used for this as well, with the advantage that the orthonormal states
keep closest to the states prior to orthonormalization (we need to remember
we are not just after an orthonormal basis of the evolved occupied space, 
but actually following the evolution of separate states).

  Consider a set of $M$ states, $\{ | \psi_n \rangle , n=1, ... , M \}$, where 
$M < {\cal N}$, being ${\cal N}$ the dimension of $\Omega$, and which are
represented by $\{ \psi^{\mu}_{\phantom{e}n}, n= 1 ... M, \mu = 1, ... , {\cal N} \}$, 
and are all, therefore, within $\Omega$.
  Consider they have been propagated by Eq.~\ref{CKmov-matrix4} and
are not strictly orthonormal, i.e.,
\begin{equation*}
{\cal S}_{nm} = \langle \psi_n | \psi_m \rangle  \ne \delta_{nm} \; ,
\end{equation*}
where 
\begin{equation*}
{\cal S}_{nm} = \psi_{\mu n} \psi^{\mu}_{\phantom{e}m}
\end{equation*}
is an $M\times M$ matrix. 
  By diagonalising ${\cal S}_{mn}$, one can obtain ${\cal S}^{-1/2\phantom{i}nm}$,
whereupon the strictly unitary propagator version of Eq.~\ref{CKmov-matrix4}
becomes
\begin{multline}
\label{CKmov-unitary1}
\psi^{\mu}_{\phantom{e}n} (t+\mathrm{d}t) = {\cal S}^{-1/2 \phantom{i} nm} 
(t+\mathrm{d}t) S^{\mu\kappa}(t+\mathrm{d}t) A_{\kappa\lambda} 
(t +\mathrm{d}t : t) \\ \times \Big [ S_{\lambda\sigma} (t) + i \frac{\mathrm{d}t}{2} 
H_{\lambda\sigma} (t)  \Big ]^{-1}
\Big \{ S_{\sigma\nu} (t) - i \frac{\mathrm{d}t}{2} H_{\sigma\nu} (t) \Big \} 
\psi^{\nu}_{\phantom{e}m} (t) \, .
\end{multline}  
  The self-consistent alternative orthonormalising Eq.~\ref{CKmov-matrix5}
would read:
\begin{multline}
\label{CKmov-unitary2}
\psi^{\mu}_{\phantom{e}n} (t+\mathrm{d}t) = 
{\cal S}^{-1/2 \phantom{i} nm} (t+\mathrm{d}t) \\ \times
\Big [ S_{\mu\lambda}(t+\mathrm{d}t)  + i \frac{\mathrm{d}t}{2} 
H_{\mu\lambda}  (t+\mathrm{d}t) \Big ]^{-1} A_{\lambda\kappa} (t +\mathrm{d}t \! : \! t) 
\, \\ \times \Big \{ \delta^{\kappa}_{\phantom{e}\nu} - i \frac{\mathrm{d}t}{2} 
S^{\kappa\sigma} (t) H_{\sigma\nu} (t) \Big \} \psi^{\nu}_{\phantom{e}m} (t) \, .
\end{multline}

  This orthonormalisation step can be done at every evolution step, 
after a determined number of them, or when the deviation from orthonormality 
reaches some tolerance.

\subsection{Forces}

  Additional terms related to derivatives also appear when calculating
the forces on atoms for geometry relaxation or ab initio molecular
dynamics calculations.
  In the following we restrict the discussion to adiabatic forces, leaving for 
further work the consideration of additional terms that appear for moving 
basis sets in non-adiabatic settings.\cite{Todorov2001}
  Considering for simplicity one single state $| \psi \rangle$, and using 
the language of the Introduction, one calculates quantities like
\begin{align}
\label{PulayOld}
& \frac{\partial E} {\partial R_i} = \partial_i \langle \psi | H | \psi \rangle =
\partial_i \left \{ \sum_{\mu\nu} C_{\mu}^* \langle e_{\mu} | H | e_{\nu} 
\rangle C_{\nu} \right \} \notag \\
&= \sum_{\mu\nu} \bigg \{ C_{\mu}^* \langle e_{\mu} | \partial_iH | e_{\nu} 
\rangle C_{\nu}  \notag \\
& +(\partial_i C_{\mu}^*) \langle e_{\mu} | H | e_{\nu} \rangle C_{\nu} 
+ C_{\mu}^* \langle e_{\mu} | H | e_{\nu} \rangle (\partial_i C_{\nu}) \notag \\ 
& + C_{\mu}^* \langle \partial_i e_{\mu} | H | e_{\nu} \rangle C_{\nu}
+ C_{\mu}^* \langle e_{\mu} | H | \partial_i e_{\nu} \rangle C_{\nu} \bigg \}  \, .
\end{align}  
  The last two terms, called Pulay forces,\cite{Pulay1969} involve again 
basis vector derivatives. 
  We discuss here the relevance of the present formalism for
these forces and related concepts.

\subsubsection{Hellmann-Feynman theorem}

  The Hellman-Feynman theorem states that, given 
$E = \langle \psi | H | \psi \rangle$ (for a normalised $| \psi \rangle$),
the derivative of $E$ with respect to $R^i$ fulfils
\begin{equation}
\label{HF-general}
\partial_i E = \langle \psi | \partial_i H | \psi \rangle
\end{equation}
if $H | \psi \rangle = E | \psi \rangle$ (and $\langle \psi | H = \langle \psi | E$).  
It is easy to see that, using the latter equations in $\Omega$, 
\begin{equation}
\label{schroe}
H^{\mu}_{\phantom{e}\nu} \psi^{\nu} = E \psi^{\mu} \; \, {\rm and} \; \; 
\psi_{\mu} H^{\mu}_{\phantom{e}\nu} = \psi_{\nu} E \; ,
\end{equation}
and the Hellman-Feynman theorem then reads,
\begin{equation*}
\partial_i E = \psi_{\mu} \left ( \partial_i 
H^{\mu}_{\phantom{e}\nu} \right ) \psi^{\nu} \; .
\end{equation*}

  In Eq.~\ref{e_deriv} we saw that the derivative of a scalar
needs no correction. Actually, 
\begin{equation}
\label{forces}
\covdev_i E = \partial_i E = \psi_{\mu} \left ( \partial_i 
H^{\mu}_{\phantom{e}\nu} \right ) \psi^{\nu} 
= \psi_{\mu} \left ( \covdev_i 
H^{\mu}_{\phantom{e}\nu} \right ) \psi^{\nu}  \; ,
\end{equation}
the last identity being easy to check by using 
Eqs.~\ref{covdevop2} and \ref{schroe}.
  This is a very transparent expression of the theorem in 
its general quantum mechanical form in Eq.~\ref{HF-general}.

\subsubsection{Hellman-Feynman theorem in matrix representation}

  The matrix representation gives a less concise expression of the same 
theorem, except when using the covariant derivative of $H$. 
  Starting with the Schr\"odinger equation, instead of Eq.~\ref{schroe}, we have 
\begin{equation}
\label{schroematrix}
H_{\mu\nu} \psi^{\nu} = E S_{\mu\nu} \psi^{\nu} \; \, {\rm and} \; \; 
\psi^{\mu *} H_{\mu\nu} = \psi^{\mu *} S_{\mu\nu} E \; .
\end{equation}
Expanding the derivative $\partial_i E=
\partial_i \left ( \psi^{\mu *} H_{\mu\nu}\psi^{\nu} \right )$ 
and using the fact that
$\langle \psi | \psi \rangle = 1 = \psi^{\mu *} S_{\mu\nu} \psi^{\nu}$,
the Hellman-Feynman theorem is obtained in this representation as
\begin{equation*}
\partial_i E = \psi^{\mu *} \left [ \partial_i  H_{\mu\nu} - 
E \, \partial_i S_{\mu\nu}  \right ] \psi^{\nu} \; ,
\end{equation*}
which can also be expressed as
\begin{equation}
\label{HFmatrix}
\partial_i E = \psi^{\mu *} \left [ \partial_i  H_{\mu\nu}  - 
E ( D_{\mu i\nu} + D_{\mu\nu i} ) \right ] \psi^{\nu} \; .
\end{equation}

  Let us see how it looks using the covariant derivative.
  Introducing its definition for the matrix
representation of $H$ (Eq.~\ref{covdevHmatrix}) into the previous
expression (Eq.~\ref{HFmatrix}), 
\begin{multline}
\label{stepHFmatrixcov}
\partial_i E =  \psi^{\mu *}  [ \covdev_i  H_{\mu\nu}   
- D_{\mu\phantom{e}i}^{\phantom{e}\sigma} H_{\sigma\nu}
- H_{\mu\sigma} D^{\sigma}_{\phantom{e}i\nu} \\
- E ( D_{\mu i\nu} + D_{\mu\nu i} )  ] \psi^{\nu} \; .
\end{multline}
  Using now the Schr\"odinger equation again (Eq.~\ref{schroematrix}),
the following elements of the previous equation become:
\begin{align*}
\psi^{\mu *} D_{\mu\phantom{e}i}^{\phantom{e}\sigma} H_{\sigma\nu} \psi^{\nu} &=
E \, \psi^{\mu *} D_{\mu\phantom{e}i}^{\phantom{e}\sigma} S_{\sigma\nu} \psi^{\nu} 
\, , \quad \mbox{and} \\
\psi^{\mu *} H_{\mu\sigma} D_{\phantom{e}i\nu}^{\sigma} \psi^{\nu} &=
E \psi^{\mu *} S_{\mu\sigma} D_{\phantom{e}i\nu}^{\sigma} \psi^{\nu} \; ,
\end{align*}
whereupon Eq.~\ref{stepHFmatrixcov} becomes
\begin{multline}
\label{step2HFmatrixcov}
\partial_i E =  \psi^{\mu *}  [ \covdev_i  H_{\mu\nu}   
- E ( D_{\mu\phantom{e}i}^{\phantom{e}\sigma} S_{\sigma\nu}
+ S_{\mu\sigma} D^{\sigma}_{\phantom{e}i\nu} \\
+ D_{\mu i\nu} + D_{\mu\nu i} )  ] \psi^{\nu} \; .
\end{multline}
  Using now the relations between Christoffel symbols in 
Eqs.~\ref{ChristRelations}, we find that
\begin{align*}
D_{\mu\phantom{e}i}^{\phantom{e}\sigma} S_{\sigma\nu} &= 
- D_{\mu i}^{\phantom{ee}\sigma} S_{\sigma\nu} = 
- D_{\mu i\nu} \\
S_{\mu\sigma} D^{\sigma}_{\phantom{e}i\nu} &=
- S_{\mu\sigma} D^{\sigma}_{\phantom{e}\nu i} =
- D_{\mu \nu i} \; ,
\end{align*}
and introducing these in Eq.~\ref{step2HFmatrixcov} gives a
quite simple final result for the Hellman-Feynman theorem in the matrix
representation in terms of the Hamiltonian's covariant derivative,
namely,
\begin{equation}
\label{HFmatrixcov}
\partial_i E = \covdev_i E = \psi^{\mu *} ( \covdev_i H_{\mu\nu} ) \psi^{\nu} \; .
\end{equation}

  This last equation can also be derived directly from Eq.~\ref{forces} using the
fact that the covariant derivative conserves the metric, as discussed in
section~\ref{ParallelTransport}, i.e., 
$\covdev_i S_{\mu\nu} = \covdev_i S^{\mu\nu} = 0$.
  From Eq.~\ref{forces}, we have
\begin{align*}
\covdev_i E &= \psi_{\mu} \left ( \covdev_i 
H^{\mu}_{\phantom{e}\nu} \right ) \psi^{\nu} =
\psi^{\mu *} S_{\mu\lambda} (\covdev_i S^{\lambda\sigma} H_{\sigma\nu} )
\psi^{\nu} \\
&= \psi^{\mu *} S_{\mu\lambda} \big \{ S^{\lambda\sigma} 
(\covdev_i  H_{\sigma\nu} ) + (\covdev_i S^{\lambda\sigma} )
H_{\sigma\nu} \big \} \psi^{\nu} \\
&= \psi^{\mu *} \delta_{\mu}^{\phantom{e}\sigma}
(\covdev_i H_{\sigma\nu} ) \psi^{\nu} +
\psi^{\mu *} S_{\mu\lambda} \, 0 \,
H_{\sigma\nu} \psi^{\nu} \\
&= \psi^{\mu *}  (\covdev_i  H_{\mu\nu} ) \psi^{\nu} \;  ,
\end{align*}  
which is nothing but Eq.~\ref{HFmatrixcov}.

\subsubsection{Pulay forces}

  When facing the problem of calculating the forces, $\partial_i E$, 
one still needs to calculate $\partial_i H^{\mu}_{\phantom{e}\nu}$.
  For the time-dependent Schr\"odinger or the von Neumann
equations above, the relevant derivatives were obtained by solving
equations defined in (an evolving) $\Omega$.
  In this case, however, the calculation of  $\partial_i H^{\mu}_{\phantom{e}\nu}$ 
is done by integration, effectively using an auxiliary basis set in $\cal{H}$ 
(analytically with Gaussians, on a real-space grid, etc.).
  The usual procedure follows
\begin{align}
\label{pulay}
\partial_i & H^{\mu}_{\phantom{e}\nu} = \partial_i \langle \psi^{\mu} |
H | \psi_{\nu} \rangle = \\ &= 
\langle e^{\mu} | \partial_i H | e_{\nu} \rangle +
\langle \partial_i e^{\mu} | H | e_{\nu} \rangle + 
\langle e^{\mu} | H | \partial_i e_{\nu} \rangle \; . \notag
\end{align}
  The last two terms give rise to the so-called Pulay terms,\cite{Pulay1969}
as already presented in Eq.~\ref{PulayOld}.
  There is nothing substantially new offered by differential geometry 
in this context.

  It is interesting, however, to separate the terms residing fully within $\Omega$ from
the contributions outside it. 
  Defining $Q_{\Omega}$ as  $P_{\Omega}$'s complement, i.e.,
$P_{\Omega} + Q_{\Omega} = \mathbb{1}$ (the identity operator in $\cal{H}$), 
we can expand
\begin{align*}
\langle \partial_i e^{\mu} | H | e_{\nu} \rangle &=
\langle \partial_i e^{\mu} | ( P_{\Omega} + Q_{\Omega}) H | e_{\nu} \rangle \\ &=
\langle \partial_i e^{\mu} | e_{\sigma} \rangle \langle e^{\sigma} | H | e_{\nu} \rangle 
+ \langle \partial_i e^{\mu} | Q_{\Omega} H | e_{\nu} \rangle  \\ &=
D^{\mu}_{\phantom{e}i\sigma} H^{\sigma}_{\phantom{e}\nu} 
+ \langle \partial_i e^{\mu} | Q_{\Omega} H | e_{\nu} \rangle \; .
\end{align*}
  Doing this for both Pulay terms and using the definition of the covariant
derivative $\covdev_i H^{\mu}_{\phantom{e}\nu}$, one arrives at
\begin{equation*}
\covdev_i H^{\mu}_{\phantom{e}\nu} = 
\langle e^{\mu} |  \partial_i H | e_{\nu} \rangle +
\langle \partial_i e^{\mu} | Q_{\Omega} H | e_{\nu} \rangle +
\langle e^{\mu} | H Q_{\Omega} | \partial_i e_{\nu} \rangle 
\end{equation*}
the last two terms being explicitly built from the components out of
$\Omega$ of both vectors $H | e_{\nu} \rangle$ and 
$| \partial_i e_{\nu} \rangle$ (and their duals/bras).
  Indeed, if neglecting out-of-space components, then
\begin{equation*}
\covdev_i  H^{\mu}_{\phantom{e}\nu}  \simeq
\langle e^{\mu} | \partial_i H | e_{\nu} \rangle \; ,
\end{equation*}
and considering Eq.~\ref{forces}, we arrive upon
\begin{equation*}
\partial_i E \simeq  \psi_{\mu} \langle e^{\mu} | \partial_i H | e_{\nu} \rangle
\psi^{\nu} \; ,
\end{equation*}
or, in the matrix representation,
\begin{equation*}
\partial_i E \simeq  \psi^{\mu *} \langle e_{\mu} | \partial_i H | e_{\nu} \rangle
\psi^{\nu} \; ,
\end{equation*}
where the Pulay terms have disappeared. 
  Neglecting those terms, however, spoil the correspondence between $E$
and $\partial_i E$.
  This just reflects the fact that, in the time-dependent equations above,
the derivatives of the basis vectors were naturally projected onto $\Omega$,
which is not the case here.

\section{Conclusions}

  Covariant derivatives are defined for derivatives of quantum mechanical states
in situations of basis sets varying as a function of external parameters.
  The concepts from differential geometry used to re-formalize dynamical equations
allow for better insights into the meaning of connection terms appearing in
dynamical equations. 
  In addition to relating to the Berry-phase and gauge formalisms, 
these geometrical insights enable the evaluation of existing, and proposal of new,
finite-differences propagators for time evolution equations.

\begin{acknowledgments}
  We would like to thank 
Daniel Sanchez-Portal and Jorge Kohanoff for interesting and intense discussions
on the problem of integrating time-dependent Kohn-Sham equations, Ivo
Souza for discussions on the relation of this work with the Berry formalism,
and Jonathan M. Evans for suggestions on the mathematics.
  D. D. O'R. would like to thank S. M.-M. Dubois, A. A. Mostofi, C.-K. Skylaris, and 
M. C. Payne for helpful comments an early stage of this work.
  Both authors would like to thank the anonymous reviewers for the careful 
reading of the manuscript and for their constructive comments, which have
helped to improve the paper noticeably. 
  E. A. acknowledges funding from the EU through the ElectronStopping Grant 
Number 333813, within the Marie-Curie CIG program, and from MINECO, Spain, 
through Grant Number FIS2015-64886-C5-1-P.
  D. D. O'R. gratefully acknowledges the support of the National University of Ireland 
and the School of Physics at Trinity College Dublin.
\end{acknowledgments}

\appendix

\section{Tensorial notation}
\label{NotationAppendix}

  The tensorial notation for this work presented in Section~\ref{SectForm} 
is different from the one in Ref.~\onlinecite{Artacho1991}, and partly follows 
Ref.~\onlinecite{Head-Gordon1993}. 
  We do specify here an order for the tensor indices,\cite{Head-Gordon1993} 
which defines bra and ket components, instead of using barred indices as 
we did then.\cite{Artacho1991}
  The order of the indices matter as in $\langle e_{\mu} | H | e^{\nu} \rangle = 
H_{\mu}^{\phantom{e}\nu} \ne H^{\nu}_{\phantom{e}\mu}$, and contractions 
are done on two same indices that are up for one and down for the other, and 
left for one and right for the other.
  In order to differentiate between $\langle e_{\mu} | \psi \rangle$ and 
$\langle \psi | e_{\mu} \rangle$, and between
$\langle e^{\mu} | \psi \rangle$ and $\langle \psi | e^{\mu} \rangle$,
Ref.~\onlinecite{Head-Gordon1993} used a dot to indicate order, as
in $\psi_{\mu\bullet}=\langle e_{\mu} | \psi \rangle$ and 
$\psi_{\bullet\mu} = \langle \psi | e_{\mu} \rangle$.
  We choose here to use only one index and explicit complex conjugation, 
with the convention that
\begin{align}
&\psi^{\mu}=\langle e^{\mu} | \psi \rangle \; 
&\psi_{\mu}=\langle \psi | e_{\mu} \rangle \notag \\
&\psi^{\mu *}=\langle \psi | e^{\mu} \rangle \; 
&\psi_{\mu}^*=\langle  e_{\mu} | \psi \rangle \notag
\end{align} 

  With this choice the natural representation becomes very intuitive, with
kets defined with upper index, bras with lower, and an operators with 
upper for the bra (left) and lower for the ket (right).
  It is, however, less transparent for the raising or lowering of indices
for these tensors. 
  They carry an additional complex conjugation and one should 
keep in mind which ones are correct,
\begin{align}
& \psi_{\nu} S^{\nu\mu} = \psi^{\mu *} \;
& S^{\mu\nu} \psi_{\nu}^*= \psi^{\mu} \notag \\
& \psi^{\nu *} S_{\nu\mu} = \psi_{\mu} \;
& S_{\mu\nu} \psi^{\nu} = \psi_{\mu}^* \notag
\end{align}
or, if preferred, go back to any of the other two 
options.\cite{Artacho1991,Head-Gordon1993}
  This is however not an issue for any of the manipulations found in 
this paper.

\section{Transformation under basis change}
\label{BasisChangeAppendix}

  We consider here the behavior under basis change of different tensors 
defined in this work. 
  We first revise the expected behavior of a tensor before getting into derivatives.
  Take $\{ | a_n \rangle \} $ as a new basis set of $\Omega$ that relates 
with the original $\{ | e_{\mu} \rangle \}$ by the transformation tensor 
$A^{\mu}_{\phantom{i} n}$ as follows:
\begin{equation*}
\langle e^{\mu} | = \langle e^{\mu} | a_{n} \rangle \, \langle a^{n} | = 
A^{\mu}_{\phantom{i} n}  \langle a^{n} | \; .
\end{equation*}
  The tensor $\psi^{\mu}$ in the basis $\{ | e_{\mu} \rangle \}$ then transforms 
into the tensor $\psi^{n}$ in the basis $\{ | a_{n} \rangle \}$ as 
\begin{equation*}
\psi^{\mu} = A^{\mu}_{\phantom{i} n} \psi^{n} \; ,
\end{equation*}
and the tensor  $H^{\mu}_{\phantom{i}\nu}$ in the basis $\{ | e_{\mu} \rangle \}$ 
then transforms into the tensor $H^{n}_{\phantom{i}m}$ 
in the basis $\{ | a_{n} \rangle \}$ as 
\begin{equation*}
H^{\mu}_{\phantom{i}\nu} = A^{\mu}_{\phantom{i} n} 
H^{n}_{\phantom{i}m} A^{m}_{\phantom{m}\nu} \; ,
\end{equation*}
where $A^{m}_{\phantom{m}\nu}$ defines the transformation
\begin{equation*}
| e_{\nu} \rangle =  | a_{m} \rangle \, \langle a^{m} | e_{\nu} \rangle = 
| a_{m} \rangle  A^{m}_{\phantom{m} \nu}  \; .
\end{equation*}
  The transformation tensors $A^{\mu}_{\phantom{i}n}$ and 
$A^{m}_{\phantom{m}\nu}$ are interrelated by
\begin{equation*}
\langle e^{\mu} | e_{\nu} \rangle = \langle e^{\mu} | a_{n} \rangle 
\langle a^{n} | e_{\nu} \rangle = 
A^{\mu}_{\phantom{i}n} A^{n}_{\phantom{n}\nu} = \delta^{\mu}_{\phantom{e}\nu} \; ,
\end{equation*}
i.e., they are inverse of each other. 

  In general, lower indices are referred to
as covariant since they transform as the basis set, while  upper indices are
contravariant since they transform as the dual basis.
  Mathematical objects that do not transform following these rules are
non-tensors.
  
  It is easy to show that the straight derivative of a first-rank tensor 
$\partial_i \psi^{\mu}$ is a non-tensor:
\begin{equation}
\label{transfpartial}
\partial_i \psi^{\mu} = \partial_i \left ( A^{\mu}_{\phantom{i}n}  
\psi^{n} \right ) = \left ( \partial_i A^{\mu}_{\phantom{i}n} \right ) 
\psi^{n} + A^{\mu}_{\phantom{i}n} \partial_i \psi^{n} \, ,
\end{equation}
and therefore that $\partial_i \psi^{\mu} \ne A^{\mu}_{\phantom{i}n} 
\partial_i \psi^{n}$,
where there is clearly a term breaking the transformation rule.

  The covariant derivative definition in Eq.~\ref{covdevdef} gives a well-behaved tensor 
since the defined $\covdev_i \psi^{\mu}$ is the tensor representation of the vector
$P_{\Omega} \partial_i \left ( P_{\Omega} | \psi \rangle \right )  \in \Omega$:
\begin{equation*}
\covdev_i \psi^{\mu} = \langle e^{\mu} | \partial_i \left ( P_{\Omega} | \psi \rangle \right )
= A^{\mu}_{\phantom{i}n} \langle a^{n} | \partial_i \left ( P_{\Omega} 
| \psi \rangle \right ) = A^{\mu}_{\phantom{i}n} \covdev_i \psi^{n} \; .
\end{equation*}

  If we want to check this using the definition in Eq.~\ref{covdevdef2}, 
which is a better definition in the sense it only refers to quantities 
defined within $\Omega$, we need to start checking the transformation
of Christoffel symbols under basis change, as follows.
\begin{align*}
D^{\mu}_{\phantom{i} \nu i}  &=  \langle e^{\mu} | \partial_i | e_{\nu} \rangle =
A^{\mu}_{\phantom{i} n}  \langle a^{n} | \left ( \partial_i | a_{m} \rangle
A^{m}_{\phantom{e} \nu}  \right ) \\
&= A^{\mu}_{\phantom{i} n} D^{n}_{\phantom{e} m i} 
A^{m}_{\phantom{e} \nu} +  A^{\mu}_{\phantom{e} n} 
\delta^{n}_{\phantom{e} m} \partial_i A^{m}_{\phantom{e} \nu} \; ,
\end{align*}
giving
\begin{equation}
\label{Christtransf}
D^{\mu}_{\phantom{i} \nu i} = A^{\mu}_{\phantom{i} n} 
D^{n}_{\phantom{e} m i}  A^{m}_{\phantom{e} \nu} + 
A^{\mu}_{\phantom{e} m} \partial_i A^{m}_{\phantom{e} \nu} \; ,
\end{equation}
which shows that the Christoffel symbols are non-tensors, 
as stated in the main text.

  Now, for the transformation of $\covdev_i \psi^{\mu}$ under basis 
set change, we have
\begin{align}
\label{dtransf}
D^{\mu}_{\phantom{i} \nu i}  \psi^{\nu} &= A^{\mu}_{\phantom{i} n} 
D^{n}_{\phantom{e} m i}  A^{m}_{\phantom{e} \nu} \psi^{\nu} +
A^{\mu}_{\phantom{e} m} \left (  \partial_i A^{m}_{\phantom{e} \nu} \right )
\psi^{\nu} \\
&= A^{\mu}_{\phantom{e} n} 
D^{n}_{\phantom{e} m i}  \psi^{m} +
A^{\mu}_{\phantom{e} m} \left (  \partial_i A^{m}_{\phantom{e} \nu} \right )
A^{\nu}_{\phantom{e} l} \psi^{l}  \, .
\end{align}
 The last term contains the derivative of the inverse transformation,
while Eq.~\ref{transfpartial} contains the derivative of the 
direct transformation under basis change.
  In order to relate them, take the fact that $A^{\mu}_{\phantom{i} m}
A^{m}_{\phantom{i} \nu} = \delta^{\mu}_{\phantom{e} \nu}$, and therefore
\begin{equation*}
\partial_i \left ( A^{\mu}_{\phantom{i} m} A^{m}_{\phantom{e} \nu} \right )
=  \left ( \partial_i  A^{\mu}_{\phantom{i} m} \right ) A^{m}_{\phantom{e} \nu} 
+  A^{\mu}_{\phantom{e} m} \left ( \partial_i A^{m}_{\phantom{e} \nu} \right )
 =0 \, .
\end{equation*}
  This gives us the sought after relation
\begin{equation}
\label{partialtransf}
\partial_i  A^{\mu}_{\phantom{i} l} = -  A^{\mu}_{\phantom{i} m} 
\left ( \partial_i A^{m}_{\phantom{e} \nu} \right ) A^{\nu}_{\phantom{i} l}  \, .
\end{equation}
Introducing Eq.~\ref{partialtransf} into Eq.~\ref{dtransf}, it follows that 
\begin{equation}
\label{christtransf2}
D^{\mu}_{\phantom{i} \nu i}  \psi^{\nu} = A^{\mu}_{\phantom{i} n} 
D^{n}_{\phantom{e} m i}  \psi^{m} 
- \left ( \partial_i A^{\mu}_{\phantom{i}n} \right ) \psi^{n} \, .
\end{equation} 
  Finally, the transformed covariant derivative would be the sum
of Eqs.~\ref{transfpartial} and \ref{christtransf2}, giving 
\begin{equation*}
\covdev_i \psi^{\mu} = A^{\mu}_{\phantom{i} n} \partial_i \psi^{n}
+ A^{\mu}_{\phantom{i} n} D^{n}_{\phantom{e} m i}  \psi^{m} 
= A^{\mu}_{\phantom{i} n} \covdev_i \psi^{n} \; ,
\end{equation*}
which confirms the expected tensor character of the covariant derivative.

\section{State propagation}
\label{PropAppendix}

  Here we show the equivalence between Eqs.~\ref{prop} 
and \ref{abstractprop}.
  For that, we may close Eq.~\ref{abstractprop} with $\langle e^{\mu}
(\mathbf{R}+\mathrm{d}\mathbf{R}) |$ from the left, thereby obtaining
\begin{equation*}
\psi^{\mu} (\mathbf{R}+\mathrm{d}\mathbf{R}) =
\langle e^{\mu} (\mathbf{R}+\mathrm{d}\mathbf{R}) |
e_{\nu} (\mathbf{R}) \rangle \{ \psi^{\nu} (\mathbf{R}) +
\covdev_i \psi^{\nu} (\mathbf{R}) \mathrm{d} R^i \}
\end{equation*}
(always keeping in mind that these relations are valid
to linear order for infinitesimal $\mathrm{d} \mathbf{R}$).
  Taking now the bra equivalent of Eq.~\ref{BasisProp}, we have
\begin{equation}
\label{braprop}
\langle e^{\mu} (\mathbf{R} + \mathrm{d}\mathbf{R}) | = 
\langle e^{\mu} (\mathbf{R}) | + D^{\mu}_{\phantom{e}i\nu} 
\mathrm{d} R^i \langle e^{\nu} (\mathbf{R}) | \; .
\end{equation}
  For $D^{\mu}_{\phantom{e}i\nu}$ defined as in Eq.~\ref{AllChristoffels}, 
one then finds that
\begin{equation*}
\psi^{\mu} (\mathbf{R}+\mathrm{d}\mathbf{R}) =
(\delta^{\mu}_{\phantom{e}\nu} + D^{\mu}_{\phantom{e}i\sigma}
\delta^{\sigma}_{\phantom{e}\nu} \mathrm{d} R^i )
\{ \psi^{\nu} + \covdev_i \psi^{\nu} \mathrm{d} R^i \} \, ,
\end{equation*}
where the whole right hand side refers to objects defined at {\bf R}.
  To linear order it gives
\begin{equation*}
\psi^{\mu} (\mathbf{R}+\mathrm{d}\mathbf{R}) =
\psi^{\mu} + (\covdev_i \psi^{\mu} + D^{\mu}_{\phantom{e}i\nu} 
\psi^{\nu}) \mathrm{d} R^i \, .
\end{equation*}
  Using now $D^{\mu}_{\phantom{e}i\nu}=-D^{\mu}_{\phantom{e}\nu i}$
(see Appendix~\ref{ChristAppendix}), and the definition of the
covariant derivative (Eq.~\ref{covdevdef2}), the result is obtained
\begin{equation*}
\psi^{\mu} (\mathbf{R}+\mathrm{d}\mathbf{R}) =
\psi^{\mu} (\mathbf{R}) + \partial_i \psi^{\mu} (\mathbf{R}) \mathrm{d} R^i \, ,
\end{equation*}
which is Eq.~\ref{prop}, and shows how the vector 
$| \psi (\mathbf{R}) \rangle = \psi^{\mu} (\mathbf{R}) | e_{\mu} (\mathbf{R}) \rangle$
evolves into
$| \psi (\mathbf{R}+\mathrm{d}\mathbf{R}) \rangle = 
\psi^{\mu} (\mathbf{R}+\mathrm{d}\mathbf{R}) 
| e_{\mu} (\mathbf{R}+\mathrm{d}\mathbf{R}) \rangle$.

\section{Christoffel symbol relations}
\label{ChristAppendix}

  Christoffel symbols are related to their equivalent in 
representations beyond the natural one introduced in the main text
(in Eqs.~\ref{christoffel} and \ref{christoffelbra}).
  The following can be defined:
\begin{align}
\label{AllChristoffels}
& D^{\mu}_{\phantom{i} \nu i} = \langle e^{\mu} |  \partial_i e_{\nu} \rangle \;
& D^{\mu}_{\phantom{i} i \nu} = \langle \partial_i e^{\mu} |  e_{\nu} \rangle \notag \\
& D_{\mu i}^{\phantom{ai}\nu} =  \langle  \partial_i e_{\mu} | e^{\nu} \rangle \; 
& D_{\mu\phantom{a} i}^{\phantom{a}\nu} =   \langle e_{\mu} |  \partial_i e^{\nu} 
   \rangle \notag \\
& D_{\mu \nu i}  =   \langle e_{\mu} |  \partial_i e_{\nu}  \rangle \; 
& D_{\mu i \nu}  =   \langle  \partial_i e_{\mu} | e_{\nu} \rangle \notag \\
& D^{\mu\phantom{i}\nu}_{\phantom{\mu}i\phantom{\nu}} =  
       \langle \partial_i e^{\mu} | e^{\nu} \rangle \; 
& D^{\mu\nu}_{\phantom{\mu}\phantom{\nu} i} =  
      \langle  e^{\mu} |  \partial_i e^{\nu}\rangle
\end{align}
  A further set of eight would be defined for the contravariant derivatives,
which would have an upper $i$ index, corresponding to 
$\partial^i = \partial / \partial R_i$.
  This is needed if the axes in the parameter space $\Theta$ are oblique. 
  We will not consider that possibility here, although the extension would be 
straightforward using the metric in $\Theta$ space.
  For orthonormal axes, $\partial_i = \partial^i$.

  These magnitudes relate to each other in different ways.
  The shifting identities shift the $i$ derivation from one greek index to the other.
  They are obtained from expanding the derivative of the metric
tensors, $\partial_i S_{\mu\nu}$, $\partial_i S^{\mu\nu}$, and from
$\partial_i \delta^{\mu}_{\phantom{e}\nu} = 
\partial_i \delta_{\mu}^{\phantom{e}\nu} = 0$, and result in 
the following:
\begin{align}
\label{ChristRelations}
D^{\mu}_{\phantom{e}\nu i} &=  - D^{\mu}_{\phantom{e}i\nu}  \notag \\
D_{\mu\phantom{e}i}^{\phantom{e}\nu} &= -D_{\mu i}^{\phantom{ee}\nu} \notag \\
D_{\mu\nu i} &=  - D_{\mu i \nu} + \partial_i S_{\mu\nu} \notag  \\
D^{\mu\nu}_{\phantom{aa} i} &= -
D^{\mu \phantom{a}\nu}_{\phantom{a} i} + \partial_i S^{\mu\nu} \; ,
\end{align}
the first one of them being repeated from Eq.~\ref{christoffelrelation}.

  A further set of relations is obtained simply by complex conjugation, i.e.,
\begin{equation*}
D_{\mu i \nu} = \langle \partial_i e_{\mu} | e_{\nu} \rangle = 
\langle e_{\nu} | \partial_i e_{\mu} \rangle^* = D_{\nu\mu i}^* \; .
\end{equation*}
  This gives rise to the following set:
\begin{align}
& D^{\mu}_{\phantom{e}\nu i} = D_{\nu i}^{\phantom{ee}\mu *} \;
& D^{\mu}_{\phantom{e}i \nu} = D_{\nu\phantom{e}i}^{\phantom{e}\mu *} \notag \\
& D_{\mu\nu i} = D_{\nu i\mu}^* \;
& D^{\mu\phantom{e}\nu}_{\phantom{e}i} = D^{\nu\mu *}_{\phantom{ee}i}  
\end{align}
  
  Metric tensors can be used to raise or lower indices by contraction, but only for
the indices in the symbols that do not refer to a derived basis or dual vector, i.e.,
for Greek indices that do not immediately precede a Latin index. 
  For example:
\begin{equation*}
D_{\mu\nu i} = \langle e_{\mu} | \partial_i e_{\nu} \rangle = 
\langle e_{\mu} | e_{\sigma} \rangle \langle e^{\sigma} | \partial_i e_{\nu} \rangle =
S_{\mu\sigma} D^{\sigma}_{\phantom{e}\nu i} \; .
\end{equation*}
  The following relations follow these raising/lowering rules:
\begin{align}
\label{RaisingLowering}
& D^{\mu}_{\phantom{e}\nu i} =  S^{\mu\sigma} D_{\sigma\nu i} \;
& D^{\mu}_{\phantom{e} i\nu} =  D^{\mu\phantom{e}\sigma}_{\phantom{e}i} S_{\sigma\nu} \notag \\
& D_{\mu\phantom{e}i}^{\phantom{e}\nu} =  S_{\mu\sigma} D^{\sigma\nu}_{\phantom{ee}i} \;
& D_{\mu i}^{\phantom{ee}\nu} =  D_{\mu i\sigma} S^{\sigma\nu} \notag \\
& D_{\mu\nu i} =  S_{\mu\sigma} D^{\sigma}_{\phantom{e}\nu i} \;
& D_{\mu i\nu} =  D_{\mu i} ^{\phantom{ee}\sigma}S_{\sigma\nu} \notag \\
& D^{\mu\nu}_{\phantom{ee}i} =  S^{\mu\sigma} D_{\sigma\phantom{e}i}^{\phantom{e}\nu} \;
& D^{\mu\phantom{e}\nu}_{\phantom{e}i} =  D^{\mu}_{\phantom{e}i\sigma}S^{\sigma\nu} 
\end{align}

  If the Greek index to be lowered or raised is immediately followed by a Latin
index, indicating a derivative, the relations get more complicated.
  See for instance:
\begin{align*}
D_{\mu\nu i} &= \langle e_{\mu} | \partial_i e_{\nu} \rangle = 
\langle e_{\mu} | \partial_i \left \{ | e^{\sigma} \rangle \langle e_{\sigma} | e_{\nu} \rangle \right \} \\ 
&= D_{\mu\phantom{e}i}^{\phantom{e}\sigma} S_{\sigma\nu} + \delta_{\mu}^{\phantom{e}\sigma}
\partial_i S_{\sigma\nu} = D_{\mu\phantom{e}i}^{\phantom{e}\sigma} S_{\sigma\nu} + 
\partial_i S_{\mu\nu}\; .
\end{align*}
This illustrates the raising/lowering derived relations that follow
\begin{align}
\label{DerivedRight}
& D^{\mu\sigma}_{\phantom{ee}i} S_{\sigma\nu} = D^{\mu}_{\phantom{e}\nu i} -
S^{\mu\sigma} \partial_i S_{\sigma\nu} \notag \\
& D_{\mu\sigma i} S^{\sigma\nu} = D_{\mu\phantom{e}i}^{\phantom{e}\nu} -
S_{\mu\sigma} \partial_i S^{\sigma\nu} \notag \\
& D_{\mu\phantom{e}i}^{\phantom{e}\sigma} S_{\sigma\nu} = D_{\mu\nu i} - 
\partial_i S_{\mu\nu} \notag \\
& D^{\mu}_{\phantom{e}\sigma i} S^{\sigma\nu} = D^{\mu\nu}_{\phantom{ee}i} -
\partial_i S^{\mu\nu} \; ,
\end{align}
for derived indices on the right, and 
\begin{align}
\label{DerivedLeft}
& S^{\mu\sigma} D_{\sigma i\nu} = D^{\mu}_{\phantom{e}i\nu} - 
\left ( \partial_i S^{\mu\sigma} \right ) S_{\sigma\nu} \notag \\
& S_{\mu\sigma} D^{\sigma\phantom{e}\nu}_{\phantom{e}i} = D_{\mu i}^{\phantom{ee}\nu} -
\left ( \partial_i S_{\mu\sigma} \right ) S^{\sigma\nu} \notag \\
& S_{\mu\sigma} D^{\sigma}_{\phantom{e}i \nu} = D_{\mu i\nu} -
\partial_i S_{\mu\nu} \notag \\
& S^{\mu\sigma} D_{\sigma i}^{\phantom{ee}\nu} = D^{\mu\phantom{e}\nu}_{\phantom{e}i} -
\partial_i S^{\mu\nu} \; , 
\end{align}
for derived indices on the left. If we now expand the derivative of the overlap,
as in $\partial_i S_{\mu\nu} = D_{\mu i\nu} + D_{\mu\nu i}$, the following 
expressions are obtained from Eqs.~\ref{DerivedRight} and \ref{DerivedLeft}:
\begin{align}
\label{DerivedFinal}
&D^{\mu\sigma}_{\phantom{ee}i} S_{\sigma\nu} = - S^{\mu\sigma} D_{\sigma i\nu} \notag \\
&D_{\mu\sigma i} S^{\sigma\nu} = - S_{\mu\sigma} D^{\sigma\phantom{e}\nu}_{\phantom{e}i} \notag \\
& D_{\mu\phantom{e}i}^{\phantom{e}\sigma} S_{\sigma\nu} = -D_{\mu i\nu} \notag \\
& D^{\mu}_{\phantom{e}\sigma i} S^{\sigma\nu} = -D^{\mu\phantom{e}\nu}_{\phantom{e}i} \notag \\
& S_{\mu\sigma} D^{\sigma}_{\phantom{e}i \nu} = -D_{\mu\nu i} \notag \\
& S^{\mu\sigma} D_{\sigma i}^{\phantom{ee}\nu} = - D^{\mu\nu}_{\phantom{ee}i}  
\end{align}
the first two arising from the first two in both Eqs.~\ref{DerivedRight} and
\ref{DerivedLeft}, third and fourth from the last two in Eqs.~\ref{DerivedRight}, 
and fifth and sixth from the last two in Eqs.~\ref{DerivedLeft}. 
  
  They are further related by metric tensors: 
  Post-multiplying the third and fourth in Eqs.~\ref{DerivedFinal} by $S^{\nu\lambda}$
and $S_{\nu\lambda}$, respectively, (and using the regular raising/lowering of 
Eqs.~\ref{RaisingLowering}) we recover the first two equations in 
Eqs.~\ref{ChristRelations}, as we do by pre-multiplying the fifth and sixth by the 
same metric tensors.
  Pre- or post-multiplying by the appropriate metric tensor the first two
equations in Eqs.~\ref{DerivedFinal} gives the last four equations of the same set.

  The rules for derived raising and lowering are therefore:
  ($i$) Only tensors with one upper and one lower Greek symbols can be
directly contracted.
  ($ii$) They contract with a shift and a corresponding change of sign.
  ($iii$) The other tensors have to be first converted into the former type.
  
  The last four equations in Eqs.~\ref{DerivedFinal} reflect the first two rules 
(and are the only ones actually needed), while the first two reflect the 
third rule as follows.
  Say you need to contract $D^{\mu\sigma}_{\phantom{ee}i} S_{\sigma\nu}$.
  First you convert it to a contractable one, by lowering the first index (a derivativeless
contraction), and then contract it with the second rule:
\begin{equation*}
D^{\mu\sigma}_{\phantom{ee}i} S_{\sigma\nu} = 
S^{\mu\lambda} D_{\lambda\phantom{e}i}^{\phantom{e}\sigma} S_{\sigma\nu}=
- S^{\mu\lambda} D_{\lambda i\nu} \; ,
\end{equation*}
which is the first equation of Eqs.~\ref{DerivedFinal}.

  Finally, we derive within this context the relations $\covdev_i S_{\mu\nu} =
\covdev_i S^{\mu\nu} = 0$.
  Doing it for the latter relation,
\begin{equation*}
\covdev_i S^{\mu\nu} = \partial_i S^{\mu\nu} + D^{\mu}_{\phantom{e}\lambda i}
S^{\lambda\nu} + S^{\mu\lambda}D_{\lambda i}^{\phantom{ee}\nu} \, .
\end{equation*}
  This expression can be obtained from the covariant derivative of the tensor 
corresponding to the unity operator as in Eqs.~\ref{covdevop} and \ref{covdevop2}, 
namely, from 
\begin{equation*}
\covdev_i S^{\mu\nu} = \langle e^{\mu} | \partial_i \{ P_{\Omega} \, 1 \,
P_{\Omega} \} | e^{\nu} \rangle \, .
\end{equation*}
  Expanding now $\partial_i S^{\mu\nu}$ we obtain
\begin{equation*}
\partial_i S^{\mu\nu} = D^{\mu\phantom{e}\nu}_{\phantom{e}i} +
D^{\mu\nu}_{\phantom{ee}i} \, .
\end{equation*}
  Now for the last two terms of $\covdev S^{\mu\nu}$. 
  Using the fourth and sixth equations in \ref{DerivedFinal} we get
\begin{equation*}
D^{\mu}_{\phantom{e}\lambda i} S^{\lambda\nu} = -D^{\mu\phantom{e}\nu}_{\phantom{e}i} 
\; \; \mathrm{and} \; \;
S^{\mu\lambda} D_{\lambda i}^{\phantom{ee}\nu} = - D^{\mu\nu}_{\phantom{ee}i}  \, .
\end{equation*}
  Introducing the last three relations into the above expression for 
$\covdev_i S^{\mu\nu}$ we arrive at the expected result of $\covdev_i S^{\mu\nu}=0$.
  It can be similarly done for $\covdev_i S_{\mu\nu}$.

\section{Chain-rule consistency of tensors}
\label{ChainRuleAppendix}

  Here we verify the following chain rule: 
\begin{equation}
\label{chainApp}
\covdev_i \left ( H^{\mu}_{\phantom{e}\nu} \psi^{\nu} \right ) =
\left ( \covdev_i  H^{\mu}_{\phantom{e}\nu} \right ) \psi^{\nu} +
H^{\mu}_{\phantom{e}\nu} \left ( \covdev_i  \psi^{\nu}  \right ) \; .
\end{equation}
  To see this, on the one hand we have
\begin{align*}
\covdev_i \left ( H^{\mu}_{\phantom{e}\nu} \psi^{\nu} \right ) &=
\partial_i \left ( H^{\mu}_{\phantom{e}\nu} \psi^{\nu} \right ) +
D^{\mu}_{\phantom{e}\lambda i}  \left ( H^{\lambda}_{\phantom{e}\nu} \psi^{\nu}  \right ) \\
&= \left (\partial_i H^{\mu}_{\phantom{e}\nu}  \right ) \psi^{\nu} +
H^{\mu}_{\phantom{e}\nu}  \left ( \partial_i \psi^{\nu} \right )+
D^{\mu}_{\phantom{e}\lambda i}  H^{\lambda}_{\phantom{e}\nu} \psi^{\nu}  \; .
\end{align*}
  On the other hand,
\begin{equation*}
\left ( \covdev_i  H^{\mu}_{\phantom{e}\nu} \right ) \psi^{\nu} =
\left ( \partial_i  H^{\mu}_{\phantom{e}\nu} \right ) \psi^{\nu} +
D^{\mu}_{\phantom{e} \lambda i} H^{\lambda}_{\phantom{e}\nu} \psi^{\nu} 
- H^{\mu}_{\phantom{e}\lambda} D^{\lambda}_{\phantom{e} \nu i}  \psi^{\nu}
\end{equation*}
and
\begin{equation*}
H^{\mu}_{\phantom{e}\nu} \left ( \covdev_i  \psi^{\nu}  \right ) =
H^{\mu}_{\phantom{e}\nu} \left ( \partial_i  \psi^{\nu}  \right ) +
H^{\mu}_{\phantom{e}\nu} D^{\nu}_{\phantom{e}\lambda} \psi^{\lambda} \; .
\end{equation*}
  Summing the last two equations, we see that their respective last terms
cancel, and the sum results equal to the previous one, proving our
original assertion in Eq.~\ref{chainApp}.

  Similarly, for the scalar $E = \psi_{\mu} H^{\mu}_{\phantom{e}\nu} \psi^{\nu}$
it can be shown that
\begin{equation}
\label{e_derivApp}
\covdev_i E = 
\covdev_i \left ( \psi_{\mu} H^{\mu}_{\phantom{e}\nu} \psi^{\nu} \right ) = 
\partial_i \left ( \psi_{\mu} H^{\mu}_{\phantom{e}\nu} \psi^{\nu} \right ) = \partial_i E \; ,
\end{equation}
as expected of a zero-rank tensor. 
  It can be easily seen that
\begin{align*}
\left ( \covdev_i  \psi_{\mu} \right ) & H^{\mu}_{\phantom{e}\nu} \psi^{\nu} + 
\psi_{\mu}  \left ( \covdev_i H^{\mu}_{\phantom{e}\nu} \right ) \psi^{\nu} + 
\psi_{\mu}  H^{\mu}_{\phantom{e}\nu} \left ( \covdev_i \psi^{\nu} \right )  \\
&= \left ( \partial_i  \psi_{\mu} \right ) H^{\mu}_{\phantom{e}\nu} \psi^{\nu} + 
\psi_{\mu}  \left ( \partial_i H^{\mu}_{\phantom{e}\nu} \right ) \psi^{\nu} + 
\psi_{\mu}  H^{\mu}_{\phantom{e}\nu} \left ( \partial_i \psi^{\nu} \right ) \; ,
\end{align*}
whereupon the four terms with Christoffel symbols implied in the left hand side 
cancel each other.

\section{Unitary propagation}
\label{UnitaryAppendix}

  If an orthonormal set of state vectors defined in ${\cal H}$ would propagate in 
$\Theta$ preserving their mutual orthonormality, as under the action
of a unitary propagator in ${\cal H}$, do state vectors defined in 
$\Omega (\mathbf{R})$ do the same? 
  Let us consider time evolution in this case for simplicity in the notation, 
and let us start from the set $\{ | \psi_n (t) \rangle \in \Omega (t) \}$ at
a given time $t$, such that 
\begin{equation*}
\langle \psi_n (t) | \psi_m (t) \rangle = \delta_{nm} \, .
\end{equation*}
  When evolving from $t$ to $t+\mathrm{d}t$, using Eq.~\ref{prop},
\begin{align*}
| \psi_m (t+\mathrm{d}t) \rangle &= (\psi_m^{\mu} + \partial_t \psi_m^{\mu} 
\mathrm{d} t ) | e_{\mu} (t+\mathrm{d}t) \rangle \\
\langle \psi_n (t+\mathrm{d}t) | &= (\psi_{n\nu} + \partial_t \psi_{n\nu} 
\mathrm{d} t ) \langle e^{\nu} (t+\mathrm{d}t) | \, .
\end{align*}
  Their scalar product gives (up to linear terms in $\mathrm{d}t)$
\begin{align*}
\langle \psi_n (t &+\mathrm{d}t) | \psi_m (t+\mathrm{d}t) \rangle = \\
&= (\psi_{n\nu} + \partial_t \psi_{n\nu} \mathrm{d} t ) 
\delta^{\nu}_{\phantom{e}\mu}
(\psi_m^{\mu} + \partial_t \psi_m^{\mu} \mathrm{d} t ) \\
&= \psi_{\mu n} \psi^{\mu}_m +
\{ (\partial_t \psi_{\mu n} ) \psi^{\mu}_m + \psi_{\mu n} (\partial_t \psi^{\mu}_m) \} 
\mathrm{d} t \\
&= \psi_{\mu n} \psi^{\mu}_m + \partial_t (\psi_{\mu n} \psi^{\mu}_m ) \mathrm{d} t \, .
\end{align*}

  If we thus define unitary propagation as the one that keeps
\begin{equation*}
\covdev_t (\psi_{\mu n} \psi^{\mu}_m ) =\partial_t (\psi_{\mu n} \psi^{\mu}_m ) = 0,
\end{equation*}
(it is easy to show that $\partial_t (\psi_{\mu n} \psi^{\mu}_m ) =
\covdev_t (\psi_{\mu n} \psi^{\mu}_m )$, which makes sense being
the scalar products zero-rank tensors) then
\begin{equation*}
\langle \psi_n (t+\mathrm{d}t) | \psi_m (t+\mathrm{d}t) \rangle =
\langle \psi_n (t) | \psi_m (t) \rangle \, ,
\end{equation*}
and, therefore, orthonormal vectors stay orthonormal.

\section{Linearisation of basis transformation}
\label{Geometric-Appendix}

  Here we show that the expressions in Eqs.~\ref{affinerot} and 
\ref{prop} for the propagation of a state vector are linearly equivalent. 
  One can see the latter as a linearization of the former as follows.
  Taking the linear evolution of the bra $\langle e^{\mu} (\mathbf{R} + 
\mathrm{d}\mathbf{R}) |$, as in Eq.~\ref{braprop}, one finds that
\begin{align*}
A^{\mu}_{\phantom{e}\nu} (\mathbf{R} + \mathrm{d}\mathbf{R} : \mathbf{R}) &=
(\delta^{\mu}_{\phantom{e}\sigma} + D^{\mu}_{\phantom{e}i\sigma} \mathrm{d} R^i )
\langle e^{\sigma} (\mathbf{R}) | e_{\nu} (\mathbf{R}) \rangle  \\ &= 
\delta^{\mu}_{\phantom{e}\nu} + D^{\mu}_{\phantom{e}i\nu} \mathrm{d} R^i \, ,
\end{align*}
and, plugging it into Eq.~\ref{affinerot}, that
\begin{align*}
\psi^{\mu} ( \mathbf{R} + \mathrm{d}\mathbf{R})  &=
(\delta^{\mu}_{\phantom{e}\nu} + D^{\mu}_{\phantom{e}i\nu} \mathrm{d} R^i)
\big \{ \psi^{\nu} + ( \covdev_i \psi^{\nu} ) \mathrm{d} R^i \big \} \\
&= \psi^{\mu} + ( \covdev_i \psi^{\mu} + D^{\mu}_{\phantom{e}i\nu} 
\psi^{\nu} ) \mathrm{d} R^i \, 
\end{align*}
(dropping terms in $\mathrm{d} R^i$ beyond linear in the latter equation).
Using that $D^{\mu}_{\phantom{e}i\nu} = -D^{\mu}_{\phantom{e}\nu i}$
(Eq.~\ref{christoffelrelation} and see also Appendix~\ref{ChristAppendix}),
and remembering the definition of the covariant derivative (Eq.~\ref{covdevdef2}),
we recover Eq.~\ref{prop}.

\section{Basis rotation and anti-Hermitian $D_{\mu\nu i}$}
\label{rotation-appendix}

  Here we show how a transformation of the basis corresponding to
a small rotation in $\Omega$ is described by an anti-Hermitian $D_{\mu\nu i}$ 
tensor.
  By a small rotation we mean a basis transformation from the set
$\{ | e_{\mu} ( \mathbf{R} ) \rangle \}$ to the set
$\{ | e_{\mu} ( \mathbf{R} + \mathrm{d} \mathbf{R} ) \rangle \}$, 
such that 
\begin{equation*}
S_{\mu\nu} ( \mathbf{R} + \mathrm{d} \mathbf{R} ) =
S_{\mu\nu} ( \mathbf{R}) \; ,
\end{equation*}
i.e., all the scalar products are maintained.
  Expanding the left hand side, we find that
\begin{align}
\label{S-RdR}
S&_{\mu\nu}  ( \mathbf{R} + \mathrm{d} \mathbf{R} ) =
\langle e_{\mu}  ( \mathbf{R} + \mathrm{d} \mathbf{R} ) |
e_{\nu}  ( \mathbf{R} + \mathrm{d} \mathbf{R} ) \rangle \\
& = \langle e_{\mu}  ( \mathbf{R} + \mathrm{d} \mathbf{R} ) |
e^{\sigma} ( \mathbf{R} ) \rangle  \; S_{\sigma\lambda} ( \mathbf{R} ) 
\; \langle e^{\lambda} ( \mathbf{R} ) | 
 e_{\nu}  ( \mathbf{R} + \mathrm{d} \mathbf{R} ) \rangle \; . \notag
\end{align} 
  Now, to first order in the transformation, we have
\begin{align*}
| e_{\nu}  ( \mathbf{R} + \mathrm{d} \mathbf{R} ) \rangle & \simeq
| e_{\nu} \rangle +  | e_{\kappa} \rangle
\langle  e^{\kappa} | \partial_i e_{\nu} \rangle \mathrm{d} R^i \\
& = | e_{\nu} \rangle +  | e_{\kappa} \rangle
D^{\kappa}_{\phantom{e}\nu i} \mathrm{d} R^i \; ,
\end{align*}
where all the right-hand-side terms are evaluated at {\bf R}. 
  Therefore, 
\begin{equation*}
\langle e^{\lambda} (\mathbf{R}) | e_{\nu}  ( \mathbf{R} + \mathrm{d} \mathbf{R} ) 
\rangle \simeq \delta^{\lambda}_{\phantom{e}\nu} + D^{\lambda}_{\phantom{e}\nu i}
\mathrm{d} R^i \; ,
\end{equation*}
and, similarly, 
\begin{equation*}
\langle  e_{\mu} (\mathbf{R} + \mathrm{d} \mathbf{R}) | e^{\sigma} (\mathbf{R}) \rangle
\simeq \delta_{\mu}^{\phantom{e}\sigma} + D_{\mu i}^{\phantom{e i}\sigma}
\mathrm{d} R^i \; .
\end{equation*}
  Introducing the last two expressions in Eq.~\ref{S-RdR}, we obtain
\begin{equation*}
S_{\mu\nu}(\mathbf{R}+\mathrm{d}\mathbf{R}) \simeq
(\delta_{\mu}^{\phantom{e}\sigma} + D_{\mu i}^{\phantom{ei}\sigma} \mathrm{d} R^i )
\; S_{\sigma\lambda}  \; 
(\delta^{\lambda}_{\phantom{e}\nu} + D^{\lambda}_{\phantom{e}\nu i} \mathrm{d} R^i ) \; ,
\end{equation*}
which, to first order, reads
\begin{equation*}
S_{\mu\nu}(\mathbf{R}+\mathrm{d}\mathbf{R}) \simeq S_{\mu\nu} + 
\left \{ D_{\mu i}^{\phantom{ei}\sigma} S_{\sigma\nu} + S_{\mu\lambda}
D^{\lambda}_{\phantom{e}\nu i}  \right \} \mathrm{d} R^i  \; .
\end{equation*}

  Therefore, the overlap conservation premise for pure basis rotations is 
kept if 
\begin{equation*}
D_{\mu i}^{\phantom{ei}\sigma} S_{\sigma\nu} = -
S_{\mu\lambda} D^{\lambda}_{\phantom{e}\nu i} \; ,
\end{equation*}
or, using the overlaps to lower the indices, 
\begin{equation*}
D_{\mu i \nu} = -D_{\mu\nu i} \; ,
\end{equation*}
which is equivalent to 
\begin{equation*}
D_{\mu \nu i} = -D_{\nu\mu i}^* \; .
\end{equation*}
  The last three expressions are different forms that reflect the fact that
the $D$ tensor should be anti-hermitian for the overlap to be conserved, as
had been proposed.

\section{Inverse second-rank tensors}
\label{InverseAppendix}

  Here, we establish the notation for the inverse of second-rank
tensors, as used in the discussion of the different integrators.
  They conform with inverse (square) matrices for the matrix
representation of such tensors, but they generalise
to other representations. 
  For the sake of completeness we define the inverse from
scratch.

  Drawing from square matrices, if we have a tensor $A_{\mu\nu}$
acting on vectors in $\Omega$, we will define its inverse 
tensor as the tensor $B^{\mu\nu}$ that fulfils
\begin{equation}
\label{inversematrix}
A_{\nu\sigma} B^{\sigma\mu} = B^{\mu\sigma} A_{\sigma\nu}
= \delta^{\mu}_{\nu} \; ,
\end{equation}
where we have chosen to indicate the indices of the inverse
tensor in this representation as both up.
  In addition to being parallel to the metric tensor corresponding
to $S^{-1}$, this choice makes sense if we see this equation as
\begin{equation*}
\langle e_{\nu} | A | e_{\sigma} \rangle 
\langle e^{\sigma} | B | e^{\mu} \rangle =
\langle e^{\mu} | B | e^{\sigma} \rangle 
\langle e_{\sigma} | A | e_{\nu} \rangle = \delta^{\mu}_{\nu} \; ,
\end{equation*}
which is nothing but
\begin{equation}
\label{inverseoperator}
P_{\Omega} A P_{\Omega} B P_{\Omega} =
P_{\Omega} B P_{\Omega} A P_{\Omega} = P_{\Omega} \; ,
\end{equation}
or
\begin{equation*}
(P_{\Omega} A P_{\Omega}) (P_{\Omega} B P_{\Omega}) =
(P_{\Omega} B P_{\Omega}) (P_{\Omega} A P_{\Omega}) = 
P_{\Omega} \; ,
\end{equation*}
which is the definition of the inverse for operators defined as acting 
within $\Omega$.

  Eq.~\ref{inversematrix} allows us to use conventional matrix inversion
procedures to obtain $\underline{B}$ from $\underline{A}$.
  Eq.~\ref{inverseoperator} allows us to extend the definition of 
the inverse tensor to other representations, remembering that
$P_{\Omega}$ is both $| e_{\mu} \rangle \langle e^{\nu} |$ and
$| e^{\mu} \rangle \langle e_{\nu} |$, thus obtaining
\begin{align*}
A^{\mu}_{\phantom{e}\sigma} B^{\sigma}_{\phantom{e}\nu} &=
B^{\mu}_{\phantom{e}\sigma} A^{\sigma}_{\phantom{e}\nu}
= \delta^{\mu}_{\nu} \\
A_{\mu}^{\phantom{e}\sigma} B_{\sigma}^{\phantom{e}\nu} &=
B_{\mu}^{\phantom{e}\sigma} A_{\sigma}^{\phantom{e}\nu}
= \delta_{\mu}^{\nu} \\
A^{\mu\sigma} B_{\sigma\nu} &=
B_{\nu\sigma} A^{\sigma\mu}
= \delta^{\mu}_{\nu} \\
A_{\mu\sigma} B^{\sigma\nu} &=
B^{\nu\sigma} A_{\sigma\mu}
= \delta_{\mu}^{\nu} \; ,
\end{align*}
(the latter being Eq.~\ref{inversematrix}), plus other 
possibilities like
\begin{align*}
A_{\mu}^{\phantom{e}\sigma} S_{\sigma\kappa} 
B^{\kappa}_{\phantom{e}\nu} &= S_{\mu\nu} \\
B_{\mu\sigma} A^{\sigma}_{\phantom{e}\nu} &= S_{\mu\nu} \; .
\end{align*}
Many other expressions can be obtained from Eq.~\ref{inverseoperator},
or derived from each other by using metric tensors to raise or
lower indices, which applies to both $A$ and $B$ tensors, as
representations of well defined operators in the abstract Hilbert space.

  As for notation, we will denote the elements of the inverse 
of $A_{\mu\nu}$ taken as a matrix as
\begin{equation*}
A^{-1\phantom{i}\mu\nu} = \left ( A^{-1} \right ) ^{\mu\nu} =
\left ( A_{\mu\nu} \right )^{-1},
\end{equation*}
corresponding to the matrix representation.
  Note the different position of the indices depending on whether
inside or outside the inversion, a feature which has significant relevance in the 
calculation of energy gradients using non-orthogonal functions.\cite{MHG-gradients}
  For the natural representation, the inverse of 
$A^{\mu}_{\phantom{e}\nu}$ will be represented by
\begin{equation*}
A^{-1\phantom{i}\mu}_{\phantom{-1 ie}\nu} =
\left ( A^{-1} \right )^{\mu}_{\phantom{e}\nu} =
\left ( A_{\mu}^{\phantom{e}\nu} \right ) ^{-1} \; .
\end{equation*}

\section{Modified Crank Nicholson integrator}
\label{ModifiedCK-Appendix}

  Here we derive Eq.~\ref{CKmov-matrix2},
\begin{multline*}
\psi^{\mu} (t+\mathrm{d}t) = 
\Big [ S_{\mu\lambda}(t+\mathrm{d}t)  + i \frac{\mathrm{d}t}{2} 
H_{\mu\lambda}  (t+\mathrm{d}t) \Big ]^{-1} \\ \times
A_{\lambda}^{\phantom{e}\sigma} (t +\mathrm{d}t : t) 
\Big \{ S_{\sigma\nu}(t) - i \frac{\mathrm{d}t}{2} H_{\sigma\nu} (t) \Big \} 
\psi^{\nu}(t) \, .
\end{multline*}
  Beginning with Eq.~\ref{affinerot}, we may first transform to the matrix 
representation 
\begin{align*}
\psi^{\mu} ( t + \mathrm{d}t)  &=
A^{\mu}_{\phantom{e}\nu} (t + \mathrm{d} t : t)
\big \{ \psi^{\nu} + ( \covdev_t \psi^{\nu} ) \mathrm{d} t \big \} \\
&= A^{\mu\sigma} (t + \mathrm{d} t : t) S_{\sigma\nu}
\big \{ \psi^{\nu} + ( \covdev_t \psi^{\nu} ) \mathrm{d} t \big \} \, ,
\end{align*}
and then, using Eq.~\ref{matschroe}, replace the $S_{\sigma\nu}
\covdev_t \psi^{\nu}$ term by  $-i H_{\sigma\nu} \psi^{\nu}$ to solve the 
Schr\"odinger equation.
\begin{equation*}
\psi^{\mu} ( t + \mathrm{d}t)  = A^{\mu\sigma} (t + \mathrm{d} t : t) 
\big \{ S_{\sigma\nu} (t) \psi^{\nu} (t) - i H_{\sigma\nu} (t) \psi^{\nu} (t) 
\mathrm{d} t \big \}  .
\end{equation*}
  Invoking the Crank-Nicholson idea of forward and backward
time-propagation, as in Eq.~\ref{CK-abstract}, we may write 
to first-order that
\begin{align*}
\psi^{\kappa} \bigg (t & +\frac{\mathrm{d}t}{2} \bigg ) = 
A^{\kappa\sigma} \left ( t+\frac{\mathrm{d}t}{2} : t \right ) \\ & \qquad \qquad \ \times
\Big \{ S_{\sigma\nu} (t) - i \frac{\mathrm{d}t}{2} H_{\sigma\nu} (t) \Big \} \psi^{\nu}(t) \, ,
\quad \mbox{and}\\
\psi^{\kappa} \bigg (t & +\frac{\mathrm{d}t}{2} \bigg ) =  
A^{\kappa\lambda} \left ( t+\frac{\mathrm{d}t}{2} : t + \mathrm{d}t \right ) \\ & \times
\Big \{ S_{\lambda\mu} (t+\mathrm{d}t)+ i \frac{\mathrm{d}t}{2} H_{\lambda\mu} 
(t+\mathrm{d}t ) \Big \} \psi^{\mu} (t + \mathrm{d}t) \, .
\end{align*}

  We then get rid of $\psi^{\kappa}(t + \mathrm{d}t/2)$ and use the following relations:
($i$) inversion
\begin{equation*}
A^{\mu\sigma} \left ( t + \frac{\mathrm{d}t}{2} : t \right )
A_{\sigma\nu} \left ( t : t + \frac{\mathrm{d}t}{2} \right ) = \delta^{\mu}_{\phantom{e}\nu}
\end{equation*}
and ($ii$) two-step propagation
\begin{equation*}
A_{\lambda\kappa} \! \! \left ( t + \mathrm{d}t : t+\frac{\mathrm{d}t}{2} \right )
A^{\kappa\sigma} \! \! \left ( t+\frac{\mathrm{d}t}{2} : t \right ) =
A_{\lambda}^{\phantom{e}\sigma} \! \! \left ( t + \mathrm{d}t : t \right )
\end{equation*}
which lead to Eq.~\ref{CKmov-matrix2}.

\section{Adequacy of L\"owdin-based propagation}
\label{SankeyAppendix}

\subsection{General considerations}

  The key question is to what extent the propagation given by the L\"owdin
orthogonalisation procedure, Eq.~\ref{Sankey}, faithfully integrates the 
time-dependent Schr\"odinger equation of motion in Eq.~\ref{td}.
  Section~\ref{secSankey} shows a clear counter-example, demonstrating that 
it is not the case in general. 
  Here we get a bit deeper into the question.
    
  The correct time propagation is defined by 
\begin{align}
\label{goodpropag}
\psi^{\mu} ( t+ \mathrm{d}t)  &=  \psi^{\mu} (t) + \partial_t \psi^{\mu}(t) \mathrm{d}t  \\
&= \psi^{\mu} (t) + \big \{ \covdev_t \psi^{\mu}(t) 
- D^{\mu}_{\phantom{e}\nu t} (t) \psi^{\nu}(t) \big \} \mathrm{d}t \; , \notag
\end{align}
the latter being a linearisation of 
\begin{equation*}
\psi^{\mu} ( t+ \mathrm{d}t)  = 
A^{\mu}_{\phantom{e}\nu} (t + \mathrm{d}t : t)
\big \{ \psi^{\nu} +  \covdev_t \psi^{\nu}  \mathrm{d} t \big \} \, ,
\end{equation*}
which accounts for the basis set transformation between $t$ and $t+\mathrm{d}t$, 
as in Eq.~\ref{affinerot}.
  The question is now whether the above coincides with 
\begin{equation}
\label{testsankey}
\psi^{\mu} (t+\mathrm{d}t) = S^{-1/2\phantom{i}\mu \lambda} (t+\mathrm{d}t) \,
S^{1/2}_{\phantom{ee}\lambda\nu} (t)  \big \{ \psi^{\nu} + \covdev_t \psi^{\nu}  \mathrm{d}t \big \}
\end{equation}
to linear order in d$t$. To that order we can expand $S^{-1/2\phantom{i}\mu \lambda} 
(t+\mathrm{d}t)$ as follows
\begin{equation*}
S^{-1/2\phantom{i}\mu \lambda} (t+\mathrm{d}t) =
S^{-1/2\phantom{i}\mu \lambda} (t) + \partial_t S^{-1/2\phantom{i}\mu \lambda} (t)
\, \mathrm{d} t \; ,
\end{equation*}
thereby obtaining, also to linear order
\begin{multline*}
S^{-1/2\phantom{i}\mu \lambda} (t+\mathrm{d}t) \,
S^{1/2}_{\phantom{ee}\lambda\nu} (t) = \\ = \delta^{\mu}_{\phantom{e}\nu} +
\partial_t S^{-1/2\phantom{i}\mu \lambda} (t) \, S^{1/2}_{\phantom{ee}\lambda\nu} 
(t) \, \mathrm{d}t \; ,
\end{multline*}
and Eq.~\ref{testsankey} becomes, again to linear order,
\begin{equation}
\label{badpropag}
\psi^{\mu} (t+\mathrm{d}t) = \psi^{\mu} (t) + \big \{ \covdev_t \psi^{\mu}(t) 
- G^{\mu}_{\phantom{e}\nu t} (t) \psi^{\nu}(t) \big \} \mathrm{d}t \; ,
\end{equation}
where we have defined 
\begin{equation}
\label{gdef}
G^{\mu}_{\phantom{e}\nu t} \equiv 
- \; \partial_t S^{-1/2\phantom{i}\mu \lambda} \, \, S^{1/2}_{\phantom{ee}\lambda\nu}  \; .
\end{equation}
  Comparing then Eqs.~\ref{goodpropag} and \ref{badpropag}, it becomes clear
that the desired evolution of $\psi^{\mu}$ is recovered if
\begin{equation*}
G^{\mu}_{\phantom{e}\nu t} = D^{\mu}_{\phantom{e}\nu t} \; ,
\end{equation*}
which is generally not the case, as inferred by the orthonormal evolution
explained above, and is clear is some (not all) of the examples we present 
below.

\subsection{Examples}

  Consider in the following a two-dimensional $\Omega$ Hilbert space, 
spanned by two basis vectors that we take as reference for $t=0$
\begin{equation*}
| e_1 (0) \rangle = |e_1 \rangle \; \; \mathrm{and} \; \; 
| e_2 (0) \rangle = |e_2 \rangle \; .
\end{equation*}
Starting with the illustration of the quite generic argument at the end of 
Section~\ref{secSankey}, we consider the evolving basis set
\begin{align*}
| e_1 (t) \rangle &= \phantom{-} \cos \theta (t) \, | e_1 \rangle + \sin \theta (t) \, | e_2 \rangle \\ 
| e_2 (t) \rangle &= - \sin \theta(t) \, | e_1 \rangle + \cos \theta(t) \, | e_2 \rangle  \; ,
\end{align*} 
which is orthonormal at all times if it is for $t=0$.
  It is easy to check that 
\begin{equation*}
D^{\bullet}_{\phantom{e}\bullet t} = \left ( 
\begin{array}{cc}
0  & - \partial_t \theta \\
\partial_t \theta & 0
\end{array} \right )  \ne
G^{\bullet}_{\phantom{e}\bullet t} = \left ( 
\begin{array}{cc}
0  & 0 \\ 0 & 0
\end{array} \right ) 
\end{equation*}
the latter resulting from $S_{\bullet\bullet} = S^{-1\phantom{i}\bullet\bullet} = 
S^{1/2}_{\phantom{ee}\bullet\bullet} = S^{-1/2\phantom{i}\bullet\bullet} = 
\mathbb{1}$ and $\partial_t S^{-1/2\phantom{i}\bullet\bullet} = \mathbb{0}$ 
at all times.

  A similarly simple example is one in which the basis vectors are orthogonal,
and only change norm with time, a particular case of what this 
formalism should cope with.
  Take then
\begin{align*}
| e_1 (t) \rangle &= \alpha_1 (t) \, | e_1 \rangle \\ 
| e_2 (t) \rangle &= \alpha_2 (t) \, | e_2 \rangle  \; .
\end{align*} 
  Then,
\begin{equation*}
\begin{array}{cc}
S_{\bullet\bullet}(t) = \left ( 
\begin{array}{cc}
\alpha_1^2 & 0 \\ 0 & \alpha_2^2 
\end{array} \right ) &
S^{-1\phantom{i}\bullet\bullet}(t) = \left ( 
\begin{array}{cc}
\alpha_1^{-2} & 0 \\ 0 & \alpha_2^{-2} 
\end{array} \right ) \\ \phantom{a} & \phantom{a} \\
S^{1/2}_{\bullet\bullet}(t) = \left ( 
\begin{array}{cc}
\alpha_1 & 0 \\ 0 & \alpha_2
\end{array} \right ) &
S^{-1/2\phantom{i}\bullet\bullet}(t) = \left ( 
\begin{array}{cc}
\alpha_1^{-1} & 0 \\ 0 & \alpha_2^{-1} 
\end{array} \right ) 
\end{array}
\end{equation*} 
and
\begin{align*}
\langle e^1 (t) | &= \alpha_1^{-1} (t) \, \langle e^1 | \\ 
\langle e^2 (t) | &= \alpha_2^{-1} (t) \, \langle e^2 | \; .
\end{align*} 
The tensor $D^{\mu}_{\phantom{e}\nu t}$ then becomes
\begin{equation*}
D^{\bullet}_{\phantom{e}\bullet t} = \left (
\begin{array}{cc} 
{ \partial_t \alpha_1 \over  \alpha_1 } & 0 \\
0 & { \partial_t \alpha_2 \over  \alpha_2 }
\end{array} \right )\; ,
\end{equation*}
and $G^{\mu}_{\phantom{e}\nu t} = - \; \partial_t 
S^{-1/2\phantom{i}\mu l} \, \, S^{1/2}_{\phantom{ee}l\nu}$ becomes
\begin{equation*}
G^{\bullet}_{\phantom{e}\bullet t} =  - \left (
\begin{array}{cc} 
{ - \partial_t \alpha_1 \over  \alpha_1^2 } & 0 \\
0 & { - \partial_t \alpha_2 \over  \alpha_2^2 }
\end{array} \right ) \left (
\begin{array}{cc} 
\alpha_1 & 0 \\ 0 &  \alpha_2 
\end{array} \right )\;  = D^{\bullet}_{\phantom{e}\bullet t} \; .
\end{equation*}
  In this case, the condition is fulfilled, and the propagator
would thus be faithful.

  Finally, let us consider both basis vectors normalised but with 
a mutual (real) overlap, $s(t)$, so that 
\begin{equation*}
S_{\bullet\bullet} =  \left (
\begin{array}{cc} 
1 & s \\ s & 1 
\end{array} \right ) \;  \mathrm{and} \; S^{-1\phantom{i}\bullet\bullet} =
{1 \over 1- s^2} \left ( 
\begin{array}{rr}
1 & -s \\ -s & 1 \end{array} \right ) \; .
\end{equation*}
  By diagonalising the metric as
\begin{equation*}
S_{\bullet\bullet} =  {1\over 2} \left (
\begin{array}{rr} 
1 & 1 \\ 1 & -1 
\end{array} \right ) \; \left (
\begin{array}{cc} 
1 + s & 0 \\ 0 & 1 -s
\end{array} \right ) \;  \left (
\begin{array}{rr} 
1 & 1 \\ 1 & - 1 
\end{array} \right ) \;  ,
\end{equation*}
one readily obtains
\begin{align*}
S^{1/2}_{\bullet\bullet} &=  {1\over 2} \left (
\begin{array}{rrr} 
a+b & & a-b \\ a-b & & a+b 
\end{array} \right ) \\ \phantom{a} \\
S^{-1/2\phantom{i}\bullet\bullet} &=  {1\over 2} \left (
\begin{array}{ccc} 
a^{-1}+b^{-1} & & a^{-1}-b^{-1} \\ a^{-1}-b^{-1} & & a^{-1}+b^{-1} 
\end{array} \right ) \\ \phantom{a} \\
\partial_t S^{-1/2\phantom{i}\bullet\bullet} &= - {\partial_t s \over 4} \left (
\begin{array}{ccc} 
a^{-3}-b^{-3} & & a^{-3}+b^{-3} \\ a^{-3}+b^{-3} & & a^{-3}-b^{-3} 
\end{array} \right ) \; ,
\end{align*}
where $a \equiv \sqrt{1+s}$ and $b \equiv \sqrt{1-s}$ . Using the definition 
of $G^{\mu}_{\phantom{e}\nu t}$  of Eq.~\ref{gdef} and operating, one gets
\begin{equation}
\label{gmodel}
G^{\bullet}_{\phantom{e}\bullet t} = - {\partial_t s \over 2} {1\over 1-s^2} \left (
\begin{array}{rr} s & -1 \\ -1 & s \end{array} \right ) \; ,
\end{equation}
which incidentally is equal to $S^{-1\phantom{i}\mu\sigma} \partial_t 
S_{\sigma\nu} / 2$, the expression we obtain by differentiating 
Eq.~\ref{gdef} as if $S$ were a scalar.
  
  We turn now to $D^{\mu}_{\phantom{e}\nu t}$, and we build it from
$D_{\mu\nu t} = \langle e_{\mu} | \partial_t | e_{\nu} \rangle$.
  Let us assume that our two basis states correspond to, say, 
two $s$-like atomic orbitals as we would have for a minimal-basis
description of H$_2$.
  In that case $D_{11t}=D_{22t}= 0$, since the derivative of an $s$
orbital is a function of $p$ character, orthogonal to the original $s$.
  Remembering now that $s=\langle e_1 | e_2 \rangle$,
\begin{equation*}
\partial_t s = \langle e_1 | \partial_t e_2 \rangle
+ \langle \partial_t e_1 | e_2 \rangle = D_{12t}+D_{21t} \; .
\end{equation*}
  Let us now make a final assumption, which will turn out to be
important: let us assume that the motion respects the 
centre of symmetry, such that both states move equally towards 
or away from each other.
  In this case $D_{12t}=D_{21t}$. 
  Putting it all together we have
\begin{equation}
\label{D2a}
D_{\bullet\bullet t} = \left ( \begin{array}{cc} 0 & \partial_t s /2 \\ \partial_t s /2 & 0
\end{array} \right ) \; ,
\end{equation}
and since $D^{\mu}_{\phantom{e}\nu t}=S^{\mu\sigma}D_{\sigma\nu t}$,
\begin{equation*}
D^{\bullet}_{\phantom{e}\bullet t} = {\partial_t s \over 2}
{1\over 1-s^2}  \left ( \begin{array}{rr} 1 & -s \\ -s & 1 \end{array} \right ) 
\left ( \begin{array}{cc} 0 & 1 \\ 1 & 0 \end{array} \right ) \; ,
\end{equation*}
which is exactly the same as found for  $G^{\bullet}_{\phantom{e}\bullet t}$
in Eq.~\ref{gmodel} above, finding thus another example for the 
L\"owdin orthogonalisation procedure working fine.

  Interestingly, however, if instead of considering the symmetric motion
we have orbital 1 moving towards a fixed orbital 2, while 
keeping the same $s(t)$, then
\begin{equation}
\label{D2b}
D_{\bullet\bullet t} = \left ( \begin{array}{cc} 0 & 0 \\ \partial_t s & 0
\end{array} \right ) \; ,
\end{equation}
and 
\begin{equation*}
D^{\bullet}_{\phantom{e}\bullet t} = {\partial_t s \over 1-s^2}
\left ( \begin{array}{rr} -s & 0 \\ 1 & 0 \end{array} \right ) \ne
G^{\bullet}_{\phantom{e}\bullet t}  \; ,
\end{equation*}
since $G^{\mu}_{\phantom{e}\nu t}$ does not change from Eq.~\ref{gmodel}.

  It is interesting to note that the difference between Eqs.~\ref{D2a} and \ref{D2b}
is an anti-Hermitian $D_{\bullet\bullet t}$ tensor corresponding to a pure
rotation of the basis in $\Omega$ (see Appendix~\ref{rotation-appendix}), which 
corresponds to a change of reference frame (Galilean transform) in real space 
for the moving atoms.
  
  For an $N$-dimensional $\Omega$ space there are $N^2$ degrees of
freedom in the set of anti-Hermitian matrices (corresponding to the dimension
of the $U(N)$ group of unitary matrices).
  There would then be $N^2$ possible ways to rotate in $\Omega$ without
altering the metric tensors. 
  Interestingly, however, for a basis of atomic-like orbitals moving with atoms,
there are only four such dimensions, that correspond to the three translations
in real space plus a global phase. 
  All other possible rotations in $\Omega$ are not compatible with maintaining
the shape of the atomic-like orbitals.
  (Only in the case of a basis made purely of $s$ orbitals, the three possible 
rotations in real (3D) space would also keep the metric tensors invariant).
  Such a translation is the difference between Eqs.~\ref{D2a} and \ref{D2b}.
  Two key questions remain: ($i$) how relevant is the difference for the physical
evolution, and ($ii$) can such a translation make the $G^{\mu}_{\phantom{e}\nu t}$
and $D^{\mu}_{\phantom{e}\nu t}$ tensors agree in general (as they do 
in the example above). 
  The answer to the latter seems likely to be negative since $4 < N^2$ for systems
with more than two basis states.
  These points should be studied in future work.


\begin{thebibliography}{54}%
\makeatletter
\providecommand \@ifxundefined [1]{%
 \@ifx{#1\undefined}
}%
\providecommand \@ifnum [1]{%
 \ifnum #1\expandafter \@firstoftwo
 \else \expandafter \@secondoftwo
 \fi
}%
\providecommand \@ifx [1]{%
 \ifx #1\expandafter \@firstoftwo
 \else \expandafter \@secondoftwo
 \fi
}%
\providecommand \natexlab [1]{#1}%
\providecommand \enquote  [1]{``#1''}%
\providecommand \bibnamefont  [1]{#1}%
\providecommand \bibfnamefont [1]{#1}%
\providecommand \citenamefont [1]{#1}%
\providecommand \href@noop [0]{\@secondoftwo}%
\providecommand \href [0]{\begingroup \@sanitize@url \@href}%
\providecommand \@href[1]{\@@startlink{#1}\@@href}%
\providecommand \@@href[1]{\endgroup#1\@@endlink}%
\providecommand \@sanitize@url [0]{\catcode `\\12\catcode `\$12\catcode
  `\&12\catcode `\#12\catcode `\^12\catcode `\_12\catcode `\%12\relax}%
\providecommand \@@startlink[1]{}%
\providecommand \@@endlink[0]{}%
\providecommand \url  [0]{\begingroup\@sanitize@url \@url }%
\providecommand \@url [1]{\endgroup\@href {#1}{\urlprefix }}%
\providecommand \urlprefix  [0]{URL }%
\providecommand \Eprint [0]{\href }%
\providecommand \doibase [0]{http://dx.doi.org/}%
\providecommand \selectlanguage [0]{\@gobble}%
\providecommand \bibinfo  [0]{\@secondoftwo}%
\providecommand \bibfield  [0]{\@secondoftwo}%
\providecommand \translation [1]{[#1]}%
\providecommand \BibitemOpen [0]{}%
\providecommand \bibitemStop [0]{}%
\providecommand \bibitemNoStop [0]{.\EOS\space}%
\providecommand \EOS [0]{\spacefactor3000\relax}%
\providecommand \BibitemShut  [1]{\csname bibitem#1\endcsname}%
\let\auto@bib@innerbib\@empty
\bibitem [{\citenamefont {Frisch}\ \emph {et~al.}()\citenamefont {Frisch},
  \citenamefont {Trucks}, \citenamefont {Schlegel}, \citenamefont {Scuseria},
  \citenamefont {Robb}, \citenamefont {Cheeseman}, \citenamefont {Scalmani},
  \citenamefont {Barone}, \citenamefont {Mennucci}, \citenamefont {Petersson},
  \citenamefont {Nakatsuji}, \citenamefont {Caricato}, \citenamefont {Li},
  \citenamefont {Hratchian}, \citenamefont {Izmaylov}, \citenamefont {Bloino},
  \citenamefont {Zheng}, \citenamefont {Sonnenberg}, \citenamefont {Hada},
  \citenamefont {Ehara}, \citenamefont {Toyota}, \citenamefont {Fukuda},
  \citenamefont {Hasegawa}, \citenamefont {Ishida}, \citenamefont {Nakajima},
  \citenamefont {Honda}, \citenamefont {Kitao}, \citenamefont {Nakai},
  \citenamefont {Vreven}, \citenamefont {Montgomery}, \citenamefont {Peralta},
  \citenamefont {Ogliaro}, \citenamefont {Bearpark}, \citenamefont {Heyd},
  \citenamefont {Brothers}, \citenamefont {Kudin}, \citenamefont {Staroverov},
  \citenamefont {Kobayashi}, \citenamefont {Normand}, \citenamefont
  {Raghavachari}, \citenamefont {Rendell}, \citenamefont {Burant},
  \citenamefont {Iyengar}, \citenamefont {Tomasi}, \citenamefont {Cossi},
  \citenamefont {Rega}, \citenamefont {Millam}, \citenamefont {Klene},
  \citenamefont {Knox}, \citenamefont {Cross}, \citenamefont {Bakken},
  \citenamefont {Adamo}, \citenamefont {Jaramillo}, \citenamefont {Gomperts},
  \citenamefont {Stratmann}, \citenamefont {Yazyev}, \citenamefont {Austin},
  \citenamefont {Cammi}, \citenamefont {Pomelli}, \citenamefont {Ochterski},
  \citenamefont {Martin}, \citenamefont {Morokuma}, \citenamefont {Zakrzewski},
  \citenamefont {Voth}, \citenamefont {Salvador}, \citenamefont {Dannenberg},
  \citenamefont {Dapprich}, \citenamefont {Daniels}, \citenamefont {Farkas},
  \citenamefont {Foresman}, \citenamefont {Ortiz}, \citenamefont {Cioslowski},\
  and\ \citenamefont {Fox}}]{Gaussian}%
  \BibitemOpen
  \bibfield  {author} {\bibinfo {author} {\bibfnamefont {M.~J.}\ \bibnamefont
  {Frisch}}, \bibinfo {author} {\bibfnamefont {G.~W.}\ \bibnamefont {Trucks}},
  \bibinfo {author} {\bibfnamefont {H.~B.}\ \bibnamefont {Schlegel}}, \bibinfo
  {author} {\bibfnamefont {G.~E.}\ \bibnamefont {Scuseria}}, \bibinfo {author}
  {\bibfnamefont {M.~A.}\ \bibnamefont {Robb}}, \bibinfo {author}
  {\bibfnamefont {J.~R.}\ \bibnamefont {Cheeseman}}, \bibinfo {author}
  {\bibfnamefont {G.}~\bibnamefont {Scalmani}}, \bibinfo {author}
  {\bibfnamefont {V.}~\bibnamefont {Barone}}, \bibinfo {author} {\bibfnamefont
  {B.}~\bibnamefont {Mennucci}}, \bibinfo {author} {\bibfnamefont {G.~A.}\
  \bibnamefont {Petersson}}, \bibinfo {author} {\bibfnamefont {H.}~\bibnamefont
  {Nakatsuji}}, \bibinfo {author} {\bibfnamefont {M.}~\bibnamefont {Caricato}},
  \bibinfo {author} {\bibfnamefont {X.}~\bibnamefont {Li}}, \bibinfo {author}
  {\bibfnamefont {H.~P.}\ \bibnamefont {Hratchian}}, \bibinfo {author}
  {\bibfnamefont {A.~F.}\ \bibnamefont {Izmaylov}}, \bibinfo {author}
  {\bibfnamefont {J.}~\bibnamefont {Bloino}}, \bibinfo {author} {\bibfnamefont
  {G.}~\bibnamefont {Zheng}}, \bibinfo {author} {\bibfnamefont {J.~L.}\
  \bibnamefont {Sonnenberg}}, \bibinfo {author} {\bibfnamefont
  {M.}~\bibnamefont {Hada}}, \bibinfo {author} {\bibfnamefont {M.}~\bibnamefont
  {Ehara}}, \bibinfo {author} {\bibfnamefont {K.}~\bibnamefont {Toyota}},
  \bibinfo {author} {\bibfnamefont {R.}~\bibnamefont {Fukuda}}, \bibinfo
  {author} {\bibfnamefont {J.}~\bibnamefont {Hasegawa}}, \bibinfo {author}
  {\bibfnamefont {M.}~\bibnamefont {Ishida}}, \bibinfo {author} {\bibfnamefont
  {T.}~\bibnamefont {Nakajima}}, \bibinfo {author} {\bibfnamefont
  {Y.}~\bibnamefont {Honda}}, \bibinfo {author} {\bibfnamefont
  {O.}~\bibnamefont {Kitao}}, \bibinfo {author} {\bibfnamefont
  {H.}~\bibnamefont {Nakai}}, \bibinfo {author} {\bibfnamefont
  {T.}~\bibnamefont {Vreven}}, \bibinfo {author} {\bibfnamefont {J.~A.}\
  \bibnamefont {Montgomery}, \bibfnamefont {{Jr.}}}, \bibinfo {author}
  {\bibfnamefont {J.~E.}\ \bibnamefont {Peralta}}, \bibinfo {author}
  {\bibfnamefont {F.}~\bibnamefont {Ogliaro}}, \bibinfo {author} {\bibfnamefont
  {M.}~\bibnamefont {Bearpark}}, \bibinfo {author} {\bibfnamefont {J.~J.}\
  \bibnamefont {Heyd}}, \bibinfo {author} {\bibfnamefont {E.}~\bibnamefont
  {Brothers}}, \bibinfo {author} {\bibfnamefont {K.~N.}\ \bibnamefont {Kudin}},
  \bibinfo {author} {\bibfnamefont {V.~N.}\ \bibnamefont {Staroverov}},
  \bibinfo {author} {\bibfnamefont {R.}~\bibnamefont {Kobayashi}}, \bibinfo
  {author} {\bibfnamefont {J.}~\bibnamefont {Normand}}, \bibinfo {author}
  {\bibfnamefont {K.}~\bibnamefont {Raghavachari}}, \bibinfo {author}
  {\bibfnamefont {A.}~\bibnamefont {Rendell}}, \bibinfo {author} {\bibfnamefont
  {J.~C.}\ \bibnamefont {Burant}}, \bibinfo {author} {\bibfnamefont {S.~S.}\
  \bibnamefont {Iyengar}}, \bibinfo {author} {\bibfnamefont {J.}~\bibnamefont
  {Tomasi}}, \bibinfo {author} {\bibfnamefont {M.}~\bibnamefont {Cossi}},
  \bibinfo {author} {\bibfnamefont {N.}~\bibnamefont {Rega}}, \bibinfo {author}
  {\bibfnamefont {J.~M.}\ \bibnamefont {Millam}}, \bibinfo {author}
  {\bibfnamefont {M.}~\bibnamefont {Klene}}, \bibinfo {author} {\bibfnamefont
  {J.~E.}\ \bibnamefont {Knox}}, \bibinfo {author} {\bibfnamefont {J.~B.}\
  \bibnamefont {Cross}}, \bibinfo {author} {\bibfnamefont {V.}~\bibnamefont
  {Bakken}}, \bibinfo {author} {\bibfnamefont {C.}~\bibnamefont {Adamo}},
  \bibinfo {author} {\bibfnamefont {J.}~\bibnamefont {Jaramillo}}, \bibinfo
  {author} {\bibfnamefont {R.}~\bibnamefont {Gomperts}}, \bibinfo {author}
  {\bibfnamefont {R.~E.}\ \bibnamefont {Stratmann}}, \bibinfo {author}
  {\bibfnamefont {O.}~\bibnamefont {Yazyev}}, \bibinfo {author} {\bibfnamefont
  {A.~J.}\ \bibnamefont {Austin}}, \bibinfo {author} {\bibfnamefont
  {R.}~\bibnamefont {Cammi}}, \bibinfo {author} {\bibfnamefont
  {C.}~\bibnamefont {Pomelli}}, \bibinfo {author} {\bibfnamefont {J.~W.}\
  \bibnamefont {Ochterski}}, \bibinfo {author} {\bibfnamefont {R.~L.}\
  \bibnamefont {Martin}}, \bibinfo {author} {\bibfnamefont {K.}~\bibnamefont
  {Morokuma}}, \bibinfo {author} {\bibfnamefont {V.~G.}\ \bibnamefont
  {Zakrzewski}}, \bibinfo {author} {\bibfnamefont {G.~A.}\ \bibnamefont
  {Voth}}, \bibinfo {author} {\bibfnamefont {P.}~\bibnamefont {Salvador}},
  \bibinfo {author} {\bibfnamefont {J.~J.}\ \bibnamefont {Dannenberg}},
  \bibinfo {author} {\bibfnamefont {S.}~\bibnamefont {Dapprich}}, \bibinfo
  {author} {\bibfnamefont {A.~D.}\ \bibnamefont {Daniels}}, \bibinfo {author}
  {\bibfnamefont {O.}~\bibnamefont {Farkas}}, \bibinfo {author} {\bibfnamefont
  {J.~B.}\ \bibnamefont {Foresman}}, \bibinfo {author} {\bibfnamefont {J.~V.}\
  \bibnamefont {Ortiz}}, \bibinfo {author} {\bibfnamefont {J.}~\bibnamefont
  {Cioslowski}}, \ and\ \bibinfo {author} {\bibfnamefont {D.~J.}\ \bibnamefont
  {Fox}},\ }\href@noop {} {\enquote {\bibinfo {title} {Gaussian-09 {R}evision
  {E}.01},}\ }\bibinfo {note} {Gaussian Inc. Wallingford CT 2009}\BibitemShut
  {NoStop}%
\bibitem [{\citenamefont {Ahlrichs}\ \emph {et~al.}(1989)\citenamefont
  {Ahlrichs}, \citenamefont {B\"ar}, \citenamefont {H\"aser}, \citenamefont
  {Horn},\ and\ \citenamefont {K\"olmel}}]{Turbomol}%
  \BibitemOpen
  \bibfield  {author} {\bibinfo {author} {\bibfnamefont {R.}~\bibnamefont
  {Ahlrichs}}, \bibinfo {author} {\bibfnamefont {M.}~\bibnamefont {B\"ar}},
  \bibinfo {author} {\bibfnamefont {M.}~\bibnamefont {H\"aser}}, \bibinfo
  {author} {\bibfnamefont {H.}~\bibnamefont {Horn}}, \ and\ \bibinfo {author}
  {\bibfnamefont {C.}~\bibnamefont {K\"olmel}},\ }\href@noop {} {\bibfield
  {journal} {\bibinfo  {journal} {Chem. Phys. Lett.}\ }\textbf {\bibinfo
  {volume} {162}},\ \bibinfo {pages} {165} (\bibinfo {year}
  {1989})}\BibitemShut {NoStop}%
\bibitem [{\citenamefont {Gordon}\ and\ \citenamefont
  {Schmidt}(2005)}]{Gamess}%
  \BibitemOpen
  \bibfield  {author} {\bibinfo {author} {\bibfnamefont {M.~S.}\ \bibnamefont
  {Gordon}}\ and\ \bibinfo {author} {\bibfnamefont {M.~W.}\ \bibnamefont
  {Schmidt}},\ }in\ \href@noop {} {\emph {\bibinfo {booktitle} {Theory and
  Applications of Computational Chemistry: the first forty years}}},\ \bibinfo
  {editor} {edited by\ \bibinfo {editor} {\bibfnamefont {C.~E.}\ \bibnamefont
  {Dykstra}}, \bibinfo {editor} {\bibfnamefont {G.}~\bibnamefont {Frenking}},
  \bibinfo {editor} {\bibfnamefont {K.~S.}\ \bibnamefont {Kim}}, \ and\
  \bibinfo {editor} {\bibfnamefont {G.~E.}\ \bibnamefont {Scuseria}}}\
  (\bibinfo  {publisher} {Elsevier},\ \bibinfo {address} {Amsterdam},\ \bibinfo
  {year} {2005})\ pp.\ \bibinfo {pages} {1167--1189}\BibitemShut {NoStop}%
\bibitem [{\citenamefont {van Gisbergen}\ \emph {et~al.}(1999)\citenamefont
  {van Gisbergen}, \citenamefont {Snijders},\ and\ \citenamefont
  {Baerends}}]{ADF}%
  \BibitemOpen
  \bibfield  {author} {\bibinfo {author} {\bibfnamefont {S.}~\bibnamefont {van
  Gisbergen}}, \bibinfo {author} {\bibfnamefont {J.}~\bibnamefont {Snijders}},
  \ and\ \bibinfo {author} {\bibfnamefont {E.}~\bibnamefont {Baerends}},\
  }\href@noop {} {\bibfield  {journal} {\bibinfo  {journal} {Comp. Phys.
  Commun.}\ }\textbf {\bibinfo {volume} {118}},\ \bibinfo {pages} {119}
  (\bibinfo {year} {1999})}\BibitemShut {NoStop}%
\bibitem [{\citenamefont {Soler}\ \emph {et~al.}(2002)\citenamefont {Soler},
  \citenamefont {Artacho}, \citenamefont {Gale}, \citenamefont {Garcia},
  \citenamefont {Junquera}, \citenamefont {Ordejon},\ and\ \citenamefont
  {Sanchez-Portal}}]{Siesta}%
  \BibitemOpen
  \bibfield  {author} {\bibinfo {author} {\bibfnamefont {J.~M.}\ \bibnamefont
  {Soler}}, \bibinfo {author} {\bibfnamefont {E.}~\bibnamefont {Artacho}},
  \bibinfo {author} {\bibfnamefont {J.~D.}\ \bibnamefont {Gale}}, \bibinfo
  {author} {\bibfnamefont {A.}~\bibnamefont {Garcia}}, \bibinfo {author}
  {\bibfnamefont {J.}~\bibnamefont {Junquera}}, \bibinfo {author}
  {\bibfnamefont {P.}~\bibnamefont {Ordejon}}, \ and\ \bibinfo {author}
  {\bibfnamefont {D.}~\bibnamefont {Sanchez-Portal}},\ }\href@noop {}
  {\bibfield  {journal} {\bibinfo  {journal} {J. Phys. Condens. Matter}\
  }\textbf {\bibinfo {volume} {14}},\ \bibinfo {pages} {2745} (\bibinfo {year}
  {2002})}\BibitemShut {NoStop}%
\bibitem [{\citenamefont {VandeVondele}\ \emph {et~al.}(2005)\citenamefont
  {VandeVondele}, \citenamefont {Krack}, \citenamefont {Mohamed}, \citenamefont
  {Parrinello}, \citenamefont {Chassaing},\ and\ \citenamefont
  {Hutter}}]{CP2K}%
  \BibitemOpen
  \bibfield  {author} {\bibinfo {author} {\bibfnamefont {J.}~\bibnamefont
  {VandeVondele}}, \bibinfo {author} {\bibfnamefont {M.}~\bibnamefont {Krack}},
  \bibinfo {author} {\bibfnamefont {F.}~\bibnamefont {Mohamed}}, \bibinfo
  {author} {\bibfnamefont {M.}~\bibnamefont {Parrinello}}, \bibinfo {author}
  {\bibfnamefont {T.}~\bibnamefont {Chassaing}}, \ and\ \bibinfo {author}
  {\bibfnamefont {J.}~\bibnamefont {Hutter}},\ }\href@noop {} {\bibfield
  {journal} {\bibinfo  {journal} {Comp. Phys. Commun.}\ }\textbf {\bibinfo
  {volume} {167}},\ \bibinfo {pages} {107} (\bibinfo {year}
  {2005})}\BibitemShut {NoStop}%
\bibitem [{\citenamefont {Dovesi}\ \emph {et~al.}(2005)\citenamefont {Dovesi},
  \citenamefont {Civalleri}, \citenamefont {Orlando}, \citenamefont {Roetti},\
  and\ \citenamefont {Saunders}}]{Crystal}%
  \BibitemOpen
  \bibfield  {author} {\bibinfo {author} {\bibfnamefont {R.}~\bibnamefont
  {Dovesi}}, \bibinfo {author} {\bibfnamefont {B.}~\bibnamefont {Civalleri}},
  \bibinfo {author} {\bibfnamefont {R.}~\bibnamefont {Orlando}}, \bibinfo
  {author} {\bibfnamefont {C.}~\bibnamefont {Roetti}}, \ and\ \bibinfo {author}
  {\bibfnamefont {V.~R.}\ \bibnamefont {Saunders}},\ }\href@noop {} {\bibfield
  {journal} {\bibinfo  {journal} {Rev. Comput. Chem.}\ }\textbf {\bibinfo
  {volume} {21}},\ \bibinfo {pages} {1} (\bibinfo {year} {2005})}\BibitemShut
  {NoStop}%
\bibitem [{\citenamefont {Ozaki}(2003)}]{OpenMX}%
  \BibitemOpen
  \bibfield  {author} {\bibinfo {author} {\bibfnamefont {T.}~\bibnamefont
  {Ozaki}},\ }\href@noop {} {\bibfield  {journal} {\bibinfo  {journal} {Phys.
  Rev. B}\ }\textbf {\bibinfo {volume} {67}},\ \bibinfo {pages} {155108}
  (\bibinfo {year} {2003})}\BibitemShut {NoStop}%
\bibitem [{\citenamefont {Blum}\ \emph {et~al.}(2009)\citenamefont {Blum},
  \citenamefont {Gehrke}, \citenamefont {Hanke}, \citenamefont {Havu},
  \citenamefont {Havu}, \citenamefont {Ren}, \citenamefont {Reuter},\ and\
  \citenamefont {Scheffler}}]{FHI-AIMS}%
  \BibitemOpen
  \bibfield  {author} {\bibinfo {author} {\bibfnamefont {V.}~\bibnamefont
  {Blum}}, \bibinfo {author} {\bibfnamefont {R.}~\bibnamefont {Gehrke}},
  \bibinfo {author} {\bibfnamefont {F.}~\bibnamefont {Hanke}}, \bibinfo
  {author} {\bibfnamefont {P.}~\bibnamefont {Havu}}, \bibinfo {author}
  {\bibfnamefont {V.}~\bibnamefont {Havu}}, \bibinfo {author} {\bibfnamefont
  {X.}~\bibnamefont {Ren}}, \bibinfo {author} {\bibfnamefont {K.}~\bibnamefont
  {Reuter}}, \ and\ \bibinfo {author} {\bibfnamefont {M.}~\bibnamefont
  {Scheffler}},\ }\href@noop {} {\bibfield  {journal} {\bibinfo  {journal}
  {Comp. Phys. Commun.}\ }\textbf {\bibinfo {volume} {180}},\ \bibinfo {pages}
  {2175} (\bibinfo {year} {2009})}\BibitemShut {NoStop}%
\bibitem [{\citenamefont {Skylaris}\ \emph {et~al.}(2005)\citenamefont
  {Skylaris}, \citenamefont {Haynes}, \citenamefont {Mostofi},\ and\
  \citenamefont {Payne}}]{Onetep}%
  \BibitemOpen
  \bibfield  {author} {\bibinfo {author} {\bibfnamefont {C.-K.}\ \bibnamefont
  {Skylaris}}, \bibinfo {author} {\bibfnamefont {P.~D.}\ \bibnamefont
  {Haynes}}, \bibinfo {author} {\bibfnamefont {A.~A.}\ \bibnamefont {Mostofi}},
  \ and\ \bibinfo {author} {\bibfnamefont {M.~C.}\ \bibnamefont {Payne}},\
  }\href@noop {} {\bibfield  {journal} {\bibinfo  {journal} {J. Chem. Phys.}\
  }\textbf {\bibinfo {volume} {122}},\ \bibinfo {pages} {084119} (\bibinfo
  {year} {2005})}\BibitemShut {NoStop}%
\bibitem [{\citenamefont {Bowler}\ and\ \citenamefont
  {Miyazaki}(2012)}]{Conquest}%
  \BibitemOpen
  \bibfield  {author} {\bibinfo {author} {\bibfnamefont {D.~R.}\ \bibnamefont
  {Bowler}}\ and\ \bibinfo {author} {\bibfnamefont {T.}~\bibnamefont
  {Miyazaki}},\ }\href@noop {} {\bibfield  {journal} {\bibinfo  {journal} {Rep.
  Prog. Phys.}\ }\textbf {\bibinfo {volume} {75}},\ \bibinfo {pages} {036503}
  (\bibinfo {year} {2012})}\BibitemShut {NoStop}%
\bibitem [{\citenamefont {Mohr}\ \emph {et~al.}(2015)\citenamefont {Mohr},
  \citenamefont {Ratcliff}, \citenamefont {Genovese}, \citenamefont {Caliste},
  \citenamefont {Boulanger}, \citenamefont {Goedecker},\ and\ \citenamefont
  {Deutsch}}]{BigDFT}%
  \BibitemOpen
  \bibfield  {author} {\bibinfo {author} {\bibfnamefont {S.}~\bibnamefont
  {Mohr}}, \bibinfo {author} {\bibfnamefont {L.~E.}\ \bibnamefont {Ratcliff}},
  \bibinfo {author} {\bibfnamefont {L.}~\bibnamefont {Genovese}}, \bibinfo
  {author} {\bibfnamefont {D.}~\bibnamefont {Caliste}}, \bibinfo {author}
  {\bibfnamefont {P.}~\bibnamefont {Boulanger}}, \bibinfo {author}
  {\bibfnamefont {S.}~\bibnamefont {Goedecker}}, \ and\ \bibinfo {author}
  {\bibfnamefont {T.}~\bibnamefont {Deutsch}},\ }\href@noop {} {\bibfield
  {journal} {\bibinfo  {journal} {Phys. Chem. Chem. Phys.}\ }\textbf {\bibinfo
  {volume} {17}},\ \bibinfo {pages} {31360} (\bibinfo {year}
  {2015})}\BibitemShut {NoStop}%
\bibitem [{\citenamefont {L\"owdin}(1950)}]{Lowdin1950}%
  \BibitemOpen
  \bibfield  {author} {\bibinfo {author} {\bibfnamefont {P.~O.}\ \bibnamefont
  {L\"owdin}},\ }\href@noop {} {\bibfield  {journal} {\bibinfo  {journal} {J.
  Chem. Phys.}\ }\textbf {\bibinfo {volume} {18}},\ \bibinfo {pages} {365}
  (\bibinfo {year} {1950})}\BibitemShut {NoStop}%
\bibitem [{\citenamefont {Runge}\ and\ \citenamefont
  {Gross}(1984)}]{RungeGross}%
  \BibitemOpen
  \bibfield  {author} {\bibinfo {author} {\bibfnamefont {E.}~\bibnamefont
  {Runge}}\ and\ \bibinfo {author} {\bibfnamefont {E.~K.~U.}\ \bibnamefont
  {Gross}},\ }\href@noop {} {\bibfield  {journal} {\bibinfo  {journal} {Phys.
  Rev. Lett.}\ }\textbf {\bibinfo {volume} {52}},\ \bibinfo {pages} {997}
  (\bibinfo {year} {1984})}\BibitemShut {NoStop}%
\bibitem [{\citenamefont {Yabana}\ and\ \citenamefont
  {Bertsch}(1996)}]{Yabana}%
  \BibitemOpen
  \bibfield  {author} {\bibinfo {author} {\bibfnamefont {K.}~\bibnamefont
  {Yabana}}\ and\ \bibinfo {author} {\bibfnamefont {G.~F.}\ \bibnamefont
  {Bertsch}},\ }\href@noop {} {\bibfield  {journal} {\bibinfo  {journal} {Phys.
  Rev. B}\ }\textbf {\bibinfo {volume} {54}},\ \bibinfo {pages} {4484}
  (\bibinfo {year} {1996})}\BibitemShut {NoStop}%
\bibitem [{\citenamefont {Tsolakidis}\ \emph {et~al.}(2002)\citenamefont
  {Tsolakidis}, \citenamefont {Sanchez-Portal},\ and\ \citenamefont
  {Martin}}]{Tsolakidis2002}%
  \BibitemOpen
  \bibfield  {author} {\bibinfo {author} {\bibfnamefont {A.}~\bibnamefont
  {Tsolakidis}}, \bibinfo {author} {\bibfnamefont {D.}~\bibnamefont
  {Sanchez-Portal}}, \ and\ \bibinfo {author} {\bibfnamefont {R.~M.}\
  \bibnamefont {Martin}},\ }\href@noop {} {\bibfield  {journal} {\bibinfo
  {journal} {Phys. Rev. B}\ }\textbf {\bibinfo {volume} {66}},\ \bibinfo
  {pages} {235416} (\bibinfo {year} {2002})}\BibitemShut {NoStop}%
\bibitem [{\citenamefont {Andrade}\ \emph {et~al.}(2015)\citenamefont
  {Andrade}, \citenamefont {Strubbe}, \citenamefont {Giovannini}, \citenamefont
  {Larsen}, \citenamefont {Oliveira}, \citenamefont {Alberdi-Rodriguez},
  \citenamefont {Varas}, \citenamefont {Theophilou}, \citenamefont {Helbig},
  \citenamefont {Verstraete}, \citenamefont {Stella}, \citenamefont {Nogueira},
  \citenamefont {Aspuru-Guzik}, \citenamefont {Castro}, \citenamefont
  {Marques},\ and\ \citenamefont {Rubio}}]{Octopus}%
  \BibitemOpen
  \bibfield  {author} {\bibinfo {author} {\bibfnamefont {X.}~\bibnamefont
  {Andrade}}, \bibinfo {author} {\bibfnamefont {D.}~\bibnamefont {Strubbe}},
  \bibinfo {author} {\bibfnamefont {U.~D.}\ \bibnamefont {Giovannini}},
  \bibinfo {author} {\bibfnamefont {A.~H.}\ \bibnamefont {Larsen}}, \bibinfo
  {author} {\bibfnamefont {M.~J.~T.}\ \bibnamefont {Oliveira}}, \bibinfo
  {author} {\bibfnamefont {J.}~\bibnamefont {Alberdi-Rodriguez}}, \bibinfo
  {author} {\bibfnamefont {A.}~\bibnamefont {Varas}}, \bibinfo {author}
  {\bibfnamefont {I.}~\bibnamefont {Theophilou}}, \bibinfo {author}
  {\bibfnamefont {N.}~\bibnamefont {Helbig}}, \bibinfo {author} {\bibfnamefont
  {M.~J.}\ \bibnamefont {Verstraete}}, \bibinfo {author} {\bibfnamefont
  {L.}~\bibnamefont {Stella}}, \bibinfo {author} {\bibfnamefont
  {F.}~\bibnamefont {Nogueira}}, \bibinfo {author} {\bibfnamefont
  {A.}~\bibnamefont {Aspuru-Guzik}}, \bibinfo {author} {\bibfnamefont
  {A.}~\bibnamefont {Castro}}, \bibinfo {author} {\bibfnamefont {M.~A.~L.}\
  \bibnamefont {Marques}}, \ and\ \bibinfo {author} {\bibfnamefont
  {A.}~\bibnamefont {Rubio}},\ }\href@noop {} {\bibfield  {journal} {\bibinfo
  {journal} {Phys. Chem. Chem. Phys.}\ }\textbf {\bibinfo {volume} {17}},\
  \bibinfo {pages} {31371} (\bibinfo {year} {2015})}\BibitemShut {NoStop}%
\bibitem [{\citenamefont {Yam}\ \emph {et~al.}(2003)\citenamefont {Yam},
  \citenamefont {Yokojima},\ and\ \citenamefont {Chen}}]{ONTDDFT}%
  \BibitemOpen
  \bibfield  {author} {\bibinfo {author} {\bibfnamefont {C.~Y.}\ \bibnamefont
  {Yam}}, \bibinfo {author} {\bibfnamefont {S.}~\bibnamefont {Yokojima}}, \
  and\ \bibinfo {author} {\bibfnamefont {G.~H.}\ \bibnamefont {Chen}},\
  }\href@noop {} {\bibfield  {journal} {\bibinfo  {journal} {Phys. Rev. B}\
  }\textbf {\bibinfo {volume} {68}},\ \bibinfo {pages} {153105} (\bibinfo
  {year} {2003})}\BibitemShut {NoStop}%
\bibitem [{\citenamefont {O'Rourke}\ and\ \citenamefont
  {Bowler}(2015)}]{Bowler}%
  \BibitemOpen
  \bibfield  {author} {\bibinfo {author} {\bibfnamefont {C.}~\bibnamefont
  {O'Rourke}}\ and\ \bibinfo {author} {\bibfnamefont {D.~R.}\ \bibnamefont
  {Bowler}},\ }\href@noop {} {\bibfield  {journal} {\bibinfo  {journal} {J.
  Chem. Phys.}\ }\textbf {\bibinfo {volume} {143}},\ \bibinfo {pages} {102801}
  (\bibinfo {year} {2015})}\BibitemShut {NoStop}%
\bibitem [{\citenamefont {Todorov}(2001)}]{Todorov2001}%
  \BibitemOpen
  \bibfield  {author} {\bibinfo {author} {\bibfnamefont {T.~N.}\ \bibnamefont
  {Todorov}},\ }\href@noop {} {\bibfield  {journal} {\bibinfo  {journal} {J.
  Phys. Condens. Matter}\ }\textbf {\bibinfo {volume} {13}},\ \bibinfo {pages}
  {10125} (\bibinfo {year} {2001})}\BibitemShut {NoStop}%
\bibitem [{\citenamefont {Kunert}\ and\ \citenamefont
  {Schmidt}(2003)}]{Rudiger2002}%
  \BibitemOpen
  \bibfield  {author} {\bibinfo {author} {\bibfnamefont {T.}~\bibnamefont
  {Kunert}}\ and\ \bibinfo {author} {\bibfnamefont {R.}~\bibnamefont
  {Schmidt}},\ }\href@noop {} {\bibfield  {journal} {\bibinfo  {journal} {Euro.
  Phys. J. D}\ }\textbf {\bibinfo {volume} {25}},\ \bibinfo {pages} {15}
  (\bibinfo {year} {2003})}\BibitemShut {NoStop}%
\bibitem [{\citenamefont {Kolesov}\ \emph {et~al.}(2016)\citenamefont
  {Kolesov}, \citenamefont {Gran\"as}, \citenamefont {Hoyt}, \citenamefont
  {Vinichenko},\ and\ \citenamefont {Kaxiras}}]{Kaxiras2015}%
  \BibitemOpen
  \bibfield  {author} {\bibinfo {author} {\bibfnamefont {G.}~\bibnamefont
  {Kolesov}}, \bibinfo {author} {\bibfnamefont {O.}~\bibnamefont {Gran\"as}},
  \bibinfo {author} {\bibfnamefont {R.}~\bibnamefont {Hoyt}}, \bibinfo {author}
  {\bibfnamefont {D.}~\bibnamefont {Vinichenko}}, \ and\ \bibinfo {author}
  {\bibfnamefont {E.}~\bibnamefont {Kaxiras}},\ }\href@noop {} {\bibfield
  {journal} {\bibinfo  {journal} {J. Chem. Th. Comp.}\ }\textbf {\bibinfo
  {volume} {12}},\ \bibinfo {pages} {466} (\bibinfo {year} {2016})}\BibitemShut
  {NoStop}%
\bibitem [{\citenamefont {Meng}\ and\ \citenamefont {Kaxiras}(2008)}]{Kaxiras}%
  \BibitemOpen
  \bibfield  {author} {\bibinfo {author} {\bibfnamefont {S.}~\bibnamefont
  {Meng}}\ and\ \bibinfo {author} {\bibfnamefont {E.}~\bibnamefont {Kaxiras}},\
  }\href@noop {} {\bibfield  {journal} {\bibinfo  {journal} {J. Chem. Phys.}\
  }\textbf {\bibinfo {volume} {129}},\ \bibinfo {pages} {054110} (\bibinfo
  {year} {2008})}\BibitemShut {NoStop}%
\bibitem [{\citenamefont {Cui}\ \emph {et~al.}(2010)\citenamefont {Cui},
  \citenamefont {Fang},\ and\ \citenamefont {Yang}}]{WeitaoYang2010}%
  \BibitemOpen
  \bibfield  {author} {\bibinfo {author} {\bibfnamefont {G.}~\bibnamefont
  {Cui}}, \bibinfo {author} {\bibfnamefont {W.}~\bibnamefont {Fang}}, \ and\
  \bibinfo {author} {\bibfnamefont {W.}~\bibnamefont {Yang}},\ }\href@noop {}
  {\bibfield  {journal} {\bibinfo  {journal} {Phys. Chem. Chem. Phys.}\
  }\textbf {\bibinfo {volume} {12}},\ \bibinfo {pages} {416} (\bibinfo {year}
  {2010})}\BibitemShut {NoStop}%
\bibitem [{\citenamefont {Pulay}(1969)}]{Pulay1969}%
  \BibitemOpen
  \bibfield  {author} {\bibinfo {author} {\bibfnamefont {P.}~\bibnamefont
  {Pulay}},\ }\href@noop {} {\bibfield  {journal} {\bibinfo  {journal} {Mol.
  Phys.}\ }\textbf {\bibinfo {volume} {17}},\ \bibinfo {pages} {197} (\bibinfo
  {year} {1969})}\BibitemShut {NoStop}%
\bibitem [{\citenamefont {Vanderbilt}\ and\ \citenamefont
  {Joannopoulos}(1980)}]{Vanderbilt1984}%
  \BibitemOpen
  \bibfield  {author} {\bibinfo {author} {\bibfnamefont {D.}~\bibnamefont
  {Vanderbilt}}\ and\ \bibinfo {author} {\bibfnamefont {J.~D.}\ \bibnamefont
  {Joannopoulos}},\ }\href@noop {} {\bibfield  {journal} {\bibinfo  {journal}
  {Phys. Rev. B}\ }\textbf {\bibinfo {volume} {22}},\ \bibinfo {pages} {2927}
  (\bibinfo {year} {1980})}\BibitemShut {NoStop}%
\bibitem [{\citenamefont {Ballantine}\ and\ \citenamefont
  {Kol\'a}(1986)}]{Ballantine1986}%
  \BibitemOpen
  \bibfield  {author} {\bibinfo {author} {\bibfnamefont {L.~E.}\ \bibnamefont
  {Ballantine}}\ and\ \bibinfo {author} {\bibfnamefont {M.}~\bibnamefont
  {Kol\'a}},\ }\href@noop {} {\bibfield  {journal} {\bibinfo  {journal} {J.
  Phys. C}\ }\textbf {\bibinfo {volume} {19}},\ \bibinfo {pages} {981}
  (\bibinfo {year} {1986})}\BibitemShut {NoStop}%
\bibitem [{\citenamefont {Artacho}\ and\ \citenamefont {Mil\'ans~del
  Bosch}(1991)}]{Artacho1991}%
  \BibitemOpen
  \bibfield  {author} {\bibinfo {author} {\bibfnamefont {E.}~\bibnamefont
  {Artacho}}\ and\ \bibinfo {author} {\bibfnamefont {L.}~\bibnamefont
  {Mil\'ans~del Bosch}},\ }\href@noop {} {\bibfield  {journal} {\bibinfo
  {journal} {Phys. Rev. A}\ }\textbf {\bibinfo {volume} {43}},\ \bibinfo
  {pages} {5770} (\bibinfo {year} {1991})}\BibitemShut {NoStop}%
\bibitem [{\citenamefont {Head-Gordon}\ \emph {et~al.}(1998)\citenamefont
  {Head-Gordon}, \citenamefont {Maslen},\ and\ \citenamefont
  {White}}]{Head-Gordon1993}%
  \BibitemOpen
  \bibfield  {author} {\bibinfo {author} {\bibfnamefont {M.}~\bibnamefont
  {Head-Gordon}}, \bibinfo {author} {\bibfnamefont {P.~E.}\ \bibnamefont
  {Maslen}}, \ and\ \bibinfo {author} {\bibfnamefont {C.~A.}\ \bibnamefont
  {White}},\ }\href@noop {} {\bibfield  {journal} {\bibinfo  {journal} {J.
  Chem. Phys.}\ }\textbf {\bibinfo {volume} {108}},\ \bibinfo {pages} {616}
  (\bibinfo {year} {1998})}\BibitemShut {NoStop}%
\bibitem [{\citenamefont {Hiscock}\ and\ \citenamefont
  {Thom}(2014)}]{Holomorphic}%
  \BibitemOpen
  \bibfield  {author} {\bibinfo {author} {\bibfnamefont {H.~G.}\ \bibnamefont
  {Hiscock}}\ and\ \bibinfo {author} {\bibfnamefont {A.~J.~W.}\ \bibnamefont
  {Thom}},\ }\href@noop {} {\bibfield  {journal} {\bibinfo  {journal} {J. Chem.
  Theory Comput.}\ }\textbf {\bibinfo {volume} {10}},\ \bibinfo {pages} {4795}
  (\bibinfo {year} {2014})}\BibitemShut {NoStop}%
\bibitem [{\citenamefont {O'Regan}\ \emph {et~al.}(2011)\citenamefont
  {O'Regan}, \citenamefont {Payne},\ and\ \citenamefont
  {Mostofi}}]{ORegan2011}%
  \BibitemOpen
  \bibfield  {author} {\bibinfo {author} {\bibfnamefont {D.~D.}\ \bibnamefont
  {O'Regan}}, \bibinfo {author} {\bibfnamefont {M.~C.}\ \bibnamefont {Payne}},
  \ and\ \bibinfo {author} {\bibfnamefont {A.~A.}\ \bibnamefont {Mostofi}},\
  }\href@noop {} {\bibfield  {journal} {\bibinfo  {journal} {Phys. Rev. B}\
  }\textbf {\bibinfo {volume} {83}},\ \bibinfo {pages} {245124} (\bibinfo
  {year} {2011})}\BibitemShut {NoStop}%
\bibitem [{\citenamefont {Soriano}\ and\ \citenamefont
  {Palacios}(2014)}]{Palacios2014}%
  \BibitemOpen
  \bibfield  {author} {\bibinfo {author} {\bibfnamefont {M.}~\bibnamefont
  {Soriano}}\ and\ \bibinfo {author} {\bibfnamefont {J.~J.}\ \bibnamefont
  {Palacios}},\ }\href@noop {} {\bibfield  {journal} {\bibinfo  {journal}
  {Phys. Rev. B}\ }\textbf {\bibinfo {volume} {90}},\ \bibinfo {pages} {075128}
  (\bibinfo {year} {2014})}\BibitemShut {NoStop}%
\bibitem [{\citenamefont {Jacob}(2015)}]{Jacob2015}%
  \BibitemOpen
  \bibfield  {author} {\bibinfo {author} {\bibfnamefont {D.}~\bibnamefont
  {Jacob}},\ }\href@noop {} {\bibfield  {journal} {\bibinfo  {journal} {J.
  Phys. Condens. Matter}\ }\textbf {\bibinfo {volume} {27}},\ \bibinfo {pages}
  {245606} (\bibinfo {year} {2015})}\BibitemShut {NoStop}%
\bibitem [{\citenamefont {Weeks}\ \emph {et~al.}(1973)\citenamefont {Weeks},
  \citenamefont {Anderson},\ and\ \citenamefont {Davidson}}]{Weeks}%
  \BibitemOpen
  \bibfield  {author} {\bibinfo {author} {\bibfnamefont {J.~D.}\ \bibnamefont
  {Weeks}}, \bibinfo {author} {\bibfnamefont {P.~W.}\ \bibnamefont {Anderson}},
  \ and\ \bibinfo {author} {\bibfnamefont {A.~G.~H.}\ \bibnamefont
  {Davidson}},\ }\href@noop {} {\bibfield  {journal} {\bibinfo  {journal} {J.
  Chem. Phys.}\ }\textbf {\bibinfo {volume} {58}},\ \bibinfo {pages} {1388}
  (\bibinfo {year} {1973})}\BibitemShut {NoStop}%
\bibitem [{\citenamefont {Bullett}(1975)}]{Bullett}%
  \BibitemOpen
  \bibfield  {author} {\bibinfo {author} {\bibfnamefont {D.~W.}\ \bibnamefont
  {Bullett}},\ }\href@noop {} {\bibfield  {journal} {\bibinfo  {journal} {J.
  Phys. C}\ }\textbf {\bibinfo {volume} {17}},\ \bibinfo {pages} {2695}
  (\bibinfo {year} {1975})}\BibitemShut {NoStop}%
\bibitem [{\citenamefont {Anderson}(1984)}]{Anderson}%
  \BibitemOpen
  \bibfield  {author} {\bibinfo {author} {\bibfnamefont {P.~W.}\ \bibnamefont
  {Anderson}},\ }\href@noop {} {\bibfield  {journal} {\bibinfo  {journal}
  {Phys. Rep.}\ }\textbf {\bibinfo {volume} {110}},\ \bibinfo {pages} {311}
  (\bibinfo {year} {1984})}\BibitemShut {NoStop}%
\bibitem [{\citenamefont {White}\ \emph {et~al.}(1997)\citenamefont {White},
  \citenamefont {Maslen}, \citenamefont {Lee},\ and\ \citenamefont
  {Head-Gordon}}]{MHG-gradients}%
  \BibitemOpen
  \bibfield  {author} {\bibinfo {author} {\bibfnamefont {C.~A.}\ \bibnamefont
  {White}}, \bibinfo {author} {\bibfnamefont {P.}~\bibnamefont {Maslen}},
  \bibinfo {author} {\bibfnamefont {M.~S.}\ \bibnamefont {Lee}}, \ and\
  \bibinfo {author} {\bibfnamefont {M.}~\bibnamefont {Head-Gordon}},\
  }\href@noop {} {\bibfield  {journal} {\bibinfo  {journal} {Chem. Phys.
  Lett.}\ }\textbf {\bibinfo {volume} {276}},\ \bibinfo {pages} {133} (\bibinfo
  {year} {1997})}\BibitemShut {NoStop}%
\bibitem [{\citenamefont {Head-Gordon}\ \emph {et~al.}(2003)\citenamefont
  {Head-Gordon}, \citenamefont {Shao}, \citenamefont {Saravanan},\ and\
  \citenamefont {White}}]{MHG-curvy}%
  \BibitemOpen
  \bibfield  {author} {\bibinfo {author} {\bibfnamefont {M.}~\bibnamefont
  {Head-Gordon}}, \bibinfo {author} {\bibfnamefont {Y.}~\bibnamefont {Shao}},
  \bibinfo {author} {\bibfnamefont {C.}~\bibnamefont {Saravanan}}, \ and\
  \bibinfo {author} {\bibfnamefont {C.~A.}\ \bibnamefont {White}},\ }\href@noop
  {} {\bibfield  {journal} {\bibinfo  {journal} {Mol. Phys.}\ }\textbf
  {\bibinfo {volume} {101}},\ \bibinfo {pages} {37} (\bibinfo {year}
  {2003})}\BibitemShut {NoStop}%
\bibitem [{\citenamefont {O'Regan}(2012)}]{ORegan}%
  \BibitemOpen
  \bibfield  {author} {\bibinfo {author} {\bibfnamefont {D.~D.}\ \bibnamefont
  {O'Regan}},\ }\href@noop {} {\emph {\bibinfo {title} {Optimised Projections
  for the Ab Initio Simulation of Large and Strongly Correlated Systems,}}},\
  Springer Theses XVI\ (\bibinfo  {publisher} {Springer Verlag, Berlin,
  Heidelberg},\ \bibinfo {year} {2012})\BibitemShut {NoStop}%
\bibitem [{\citenamefont {Lebedev}\ \emph {et~al.}(2010)\citenamefont
  {Lebedev}, \citenamefont {Cloud},\ and\ \citenamefont {Eremeyev}}]{Algebra}%
  \BibitemOpen
  \bibfield  {author} {\bibinfo {author} {\bibfnamefont {L.~P.}\ \bibnamefont
  {Lebedev}}, \bibinfo {author} {\bibfnamefont {M.~J.}\ \bibnamefont {Cloud}},
  \ and\ \bibinfo {author} {\bibfnamefont {V.~A.}\ \bibnamefont {Eremeyev}},\
  }\href@noop {} {\emph {\bibinfo {title} {Tensor Analysis With Applications to
  Mechanics}}}\ (\bibinfo  {publisher} {World scientific, Singapore},\ \bibinfo
  {year} {2010})\ p.~\bibinfo {pages} {12}\BibitemShut {NoStop}%
\bibitem [{\citenamefont {Einstein}(1916)}]{Einstein1916}%
  \BibitemOpen
  \bibfield  {author} {\bibinfo {author} {\bibfnamefont {A.}~\bibnamefont
  {Einstein}},\ }\href@noop {} {\bibfield  {journal} {\bibinfo  {journal}
  {Annalen der Physik}\ }\textbf {\bibinfo {volume} {354}},\ \bibinfo {pages}
  {769} (\bibinfo {year} {1916})}\BibitemShut {NoStop}%
\bibitem [{\citenamefont {Kohn}\ and\ \citenamefont {Sham}(1965)}]{Kohn-Sham}%
  \BibitemOpen
  \bibfield  {author} {\bibinfo {author} {\bibfnamefont {W.}~\bibnamefont
  {Kohn}}\ and\ \bibinfo {author} {\bibfnamefont {L.~J.}\ \bibnamefont
  {Sham}},\ }\href@noop {} {\bibfield  {journal} {\bibinfo  {journal} {Phys.
  Rev.}\ }\textbf {\bibinfo {volume} {140}},\ \bibinfo {pages} {A1133}
  (\bibinfo {year} {1965})}\BibitemShut {NoStop}%
\bibitem [{\citenamefont {Husem\"oller}(1994)}]{FibreBundle}%
  \BibitemOpen
  \bibfield  {author} {\bibinfo {author} {\bibfnamefont {D.}~\bibnamefont
  {Husem\"oller}},\ }\href@noop {} {\emph {\bibinfo {title} {Fibre Bundles}}}\
  (\bibinfo  {publisher} {Springer Verlag},\ \bibinfo {year}
  {1994})\BibitemShut {NoStop}%
\bibitem [{\citenamefont {Berry}(1984)}]{Berry-orig}%
  \BibitemOpen
  \bibfield  {author} {\bibinfo {author} {\bibfnamefont {M.~V.}\ \bibnamefont
  {Berry}},\ }\href@noop {} {\bibfield  {journal} {\bibinfo  {journal} {Proc.
  Royal Soc. A}\ }\textbf {\bibinfo {volume} {392}},\ \bibinfo {pages} {45}
  (\bibinfo {year} {1984})}\BibitemShut {NoStop}%
\bibitem [{\citenamefont {Xiao}\ \emph {et~al.}(2010)\citenamefont {Xiao},
  \citenamefont {Chang},\ and\ \citenamefont {Niu}}]{Berry}%
  \BibitemOpen
  \bibfield  {author} {\bibinfo {author} {\bibfnamefont {D.}~\bibnamefont
  {Xiao}}, \bibinfo {author} {\bibfnamefont {M.~C.}\ \bibnamefont {Chang}}, \
  and\ \bibinfo {author} {\bibfnamefont {Q.}~\bibnamefont {Niu}},\ }\href@noop
  {} {\bibfield  {journal} {\bibinfo  {journal} {Rev. Mod. Phys.}\ }\textbf
  {\bibinfo {volume} {82}},\ \bibinfo {pages} {1959} (\bibinfo {year}
  {2010})}\BibitemShut {NoStop}%
\bibitem [{\citenamefont {Marzari}\ and\ \citenamefont
  {Vanderbilt}(1997)}]{Marzari1997}%
  \BibitemOpen
  \bibfield  {author} {\bibinfo {author} {\bibfnamefont {N.}~\bibnamefont
  {Marzari}}\ and\ \bibinfo {author} {\bibfnamefont {D.}~\bibnamefont
  {Vanderbilt}},\ }\href@noop {} {\bibfield  {journal} {\bibinfo  {journal}
  {Phys. Rev. B}\ }\textbf {\bibinfo {volume} {56}},\ \bibinfo {pages} {12847}
  (\bibinfo {year} {1997})}\BibitemShut {NoStop}%
\bibitem [{\citenamefont {Mead}(1992)}]{Mead1992}%
  \BibitemOpen
  \bibfield  {author} {\bibinfo {author} {\bibfnamefont {M.~A.}\ \bibnamefont
  {Mead}},\ }\href@noop {} {\bibfield  {journal} {\bibinfo  {journal} {Rev.
  Mod. Phys.}\ }\textbf {\bibinfo {volume} {64}},\ \bibinfo {pages} {51}
  (\bibinfo {year} {1992})}\BibitemShut {NoStop}%
\bibitem [{\citenamefont {Leforestier}\ \emph {et~al.}(1991)\citenamefont
  {Leforestier}, \citenamefont {Bisseling}, \citenamefont {Cerjan},
  \citenamefont {Feit}, \citenamefont {Friesner}, \citenamefont {Guldberg},
  \citenamefont {Hammerich}, \citenamefont {Jolicard}, \citenamefont
  {Karrlein}, \citenamefont {Meyer}, \citenamefont {Lipkin}, \citenamefont
  {Roncero},\ and\ \citenamefont {Kosloff}}]{Koslov}%
  \BibitemOpen
  \bibfield  {author} {\bibinfo {author} {\bibfnamefont {C.}~\bibnamefont
  {Leforestier}}, \bibinfo {author} {\bibfnamefont {R.~H.}\ \bibnamefont
  {Bisseling}}, \bibinfo {author} {\bibfnamefont {C.}~\bibnamefont {Cerjan}},
  \bibinfo {author} {\bibfnamefont {M.~D.}\ \bibnamefont {Feit}}, \bibinfo
  {author} {\bibfnamefont {R.}~\bibnamefont {Friesner}}, \bibinfo {author}
  {\bibfnamefont {A.}~\bibnamefont {Guldberg}}, \bibinfo {author}
  {\bibfnamefont {A.}~\bibnamefont {Hammerich}}, \bibinfo {author}
  {\bibfnamefont {G.}~\bibnamefont {Jolicard}}, \bibinfo {author}
  {\bibfnamefont {W.}~\bibnamefont {Karrlein}}, \bibinfo {author}
  {\bibfnamefont {H.~D.}\ \bibnamefont {Meyer}}, \bibinfo {author}
  {\bibfnamefont {N.}~\bibnamefont {Lipkin}}, \bibinfo {author} {\bibfnamefont
  {O.}~\bibnamefont {Roncero}}, \ and\ \bibinfo {author} {\bibfnamefont
  {R.}~\bibnamefont {Kosloff}},\ }\href@noop {} {\bibfield  {journal} {\bibinfo
   {journal} {J. Comp. Phys.}\ }\textbf {\bibinfo {volume} {94}},\ \bibinfo
  {pages} {59} (\bibinfo {year} {1991})}\BibitemShut {NoStop}%
\bibitem [{\citenamefont {Schleife}\ \emph {et~al.}(2012)\citenamefont
  {Schleife}, \citenamefont {Draeger}, \citenamefont {Kanai},\ and\
  \citenamefont {Correa}}]{Correa}%
  \BibitemOpen
  \bibfield  {author} {\bibinfo {author} {\bibfnamefont {A.}~\bibnamefont
  {Schleife}}, \bibinfo {author} {\bibfnamefont {E.~W.}\ \bibnamefont
  {Draeger}}, \bibinfo {author} {\bibfnamefont {Y.}~\bibnamefont {Kanai}}, \
  and\ \bibinfo {author} {\bibfnamefont {A.~A.}\ \bibnamefont {Correa}},\
  }\href@noop {} {\bibfield  {journal} {\bibinfo  {journal} {J. Chem. Phys.}\
  }\textbf {\bibinfo {volume} {137}},\ \bibinfo {pages} {22A546} (\bibinfo
  {year} {2012})}\BibitemShut {NoStop}%
\bibitem [{\citenamefont {Tomfohr}\ and\ \citenamefont
  {Sankey}(2001)}]{Sankey}%
  \BibitemOpen
  \bibfield  {author} {\bibinfo {author} {\bibfnamefont {J.~K.}\ \bibnamefont
  {Tomfohr}}\ and\ \bibinfo {author} {\bibfnamefont {O.~F.}\ \bibnamefont
  {Sankey}},\ }\href@noop {} {\bibfield  {journal} {\bibinfo  {journal} {Phys.
  Stat. Sol. (b)}\ }\textbf {\bibinfo {volume} {226}},\ \bibinfo {pages} {115}
  (\bibinfo {year} {2001})}\BibitemShut {NoStop}%
\bibitem [{\citenamefont {Zeb}\ \emph {et~al.}(2012)\citenamefont {Zeb},
  \citenamefont {Kohanoff}, \citenamefont {Sanchez-Portal}, \citenamefont
  {Arnau}, \citenamefont {Juaristi},\ and\ \citenamefont {Artacho}}]{Zeb2013}%
  \BibitemOpen
  \bibfield  {author} {\bibinfo {author} {\bibfnamefont {M.}~\bibnamefont
  {Zeb}}, \bibinfo {author} {\bibfnamefont {K.}~\bibnamefont {Kohanoff}},
  \bibinfo {author} {\bibfnamefont {D.}~\bibnamefont {Sanchez-Portal}},
  \bibinfo {author} {\bibfnamefont {A.}~\bibnamefont {Arnau}}, \bibinfo
  {author} {\bibfnamefont {J.~I.}\ \bibnamefont {Juaristi}}, \ and\ \bibinfo
  {author} {\bibfnamefont {E.}~\bibnamefont {Artacho}},\ }\href@noop {}
  {\bibfield  {journal} {\bibinfo  {journal} {Phys. Rev. Lett.}\ }\textbf
  {\bibinfo {volume} {108}},\ \bibinfo {pages} {225504} (\bibinfo {year}
  {2012})}\BibitemShut {NoStop}%
\bibitem [{\citenamefont {Correa}\ \emph {et~al.}(2012)\citenamefont {Correa},
  \citenamefont {Kohanoff}, \citenamefont {Artacho}, \citenamefont
  {Sanchez-Portal},\ and\ \citenamefont {Caro}}]{Correa2013}%
  \BibitemOpen
  \bibfield  {author} {\bibinfo {author} {\bibfnamefont {A.~A.}\ \bibnamefont
  {Correa}}, \bibinfo {author} {\bibfnamefont {J.}~\bibnamefont {Kohanoff}},
  \bibinfo {author} {\bibfnamefont {E.}~\bibnamefont {Artacho}}, \bibinfo
  {author} {\bibfnamefont {D.}~\bibnamefont {Sanchez-Portal}}, \ and\ \bibinfo
  {author} {\bibfnamefont {A.}~\bibnamefont {Caro}},\ }\href@noop {} {\bibfield
   {journal} {\bibinfo  {journal} {Phys. Rev. Lett.}\ }\textbf {\bibinfo
  {volume} {108}},\ \bibinfo {pages} {213201} (\bibinfo {year}
  {2012})}\BibitemShut {NoStop}%
\bibitem [{\citenamefont {Ullah}\ \emph {et~al.}(2015)\citenamefont {Ullah},
  \citenamefont {Corsetti}, \citenamefont {Sanchez-Portal},\ and\ \citenamefont
  {Artacho}}]{Ullah2015}%
  \BibitemOpen
  \bibfield  {author} {\bibinfo {author} {\bibfnamefont {R.}~\bibnamefont
  {Ullah}}, \bibinfo {author} {\bibfnamefont {F.}~\bibnamefont {Corsetti}},
  \bibinfo {author} {\bibfnamefont {D.}~\bibnamefont {Sanchez-Portal}}, \ and\
  \bibinfo {author} {\bibfnamefont {E.}~\bibnamefont {Artacho}},\ }\href@noop
  {} {\bibfield  {journal} {\bibinfo  {journal} {Phys. Rev. B}\ }\textbf
  {\bibinfo {volume} {91}},\ \bibinfo {pages} {125203} (\bibinfo {year}
  {2015})}\BibitemShut {NoStop}%
\bibitem [{\citenamefont {Payne}\ \emph {et~al.}(1992)\citenamefont {Payne},
  \citenamefont {Teter}, \citenamefont {Allan}, \citenamefont {Arias},\ and\
  \citenamefont {Joannopoulos}}]{Payne1990}%
  \BibitemOpen
  \bibfield  {author} {\bibinfo {author} {\bibfnamefont {M.~C.}\ \bibnamefont
  {Payne}}, \bibinfo {author} {\bibfnamefont {M.~P.}\ \bibnamefont {Teter}},
  \bibinfo {author} {\bibfnamefont {D.~C.}\ \bibnamefont {Allan}}, \bibinfo
  {author} {\bibfnamefont {T.~A.}\ \bibnamefont {Arias}}, \ and\ \bibinfo
  {author} {\bibfnamefont {J.~D.}\ \bibnamefont {Joannopoulos}},\ }\href@noop
  {} {\bibfield  {journal} {\bibinfo  {journal} {Rev. Mod. Phys.}\ }\textbf
  {\bibinfo {volume} {64}},\ \bibinfo {pages} {1045} (\bibinfo {year}
  {1992})}\BibitemShut {NoStop}%
\end{thebibliography}

%

\end{document}